\documentclass[twocolumn]{aastex631}
\usepackage{amsmath}
\usepackage{mathrsfs}

\begin{document}

\title{How High-Specific-Energy Winds Regulate the Circumgalactic Medium of Dwarf Galaxies}

\author[0009-0003-3855-2708]{Michael Messere}
\affiliation{Department of Astronomy, Columbia University, New York, NY 10027, USA}

\author[0000-0003-2630-9228]{Greg L. Bryan}
\affiliation{Department of Astronomy, Columbia University, New York, NY 10027, USA}

\correspondingauthor{Michael Messere}
\email{mam2645@columbia.edu}



\begin{abstract}

We investigate the role of ejective and preventive feedback in $\mathrm{\sim10^{10}-10^{11}\,M_\odot}$ dwarf halos using cosmological zoom-in simulations. These simulations use adaptive mesh refinement to capture high-specific-energy outflows, together with an implementation of discrete supernovae (SNe). We show that episodic, SNe-driven shock heating sustains the circumgalactic medium (CGM) at $\mathrm{\sim T_{vir}}$. This process also increases the ratio $\mathrm{t_{cool}/t_{ff} > 10}$ in the outer CGM and intergalactic medium (IGM), placing the gas in a radiatively stable regime. Hot outflows ($\mathrm{\gtrsim10^5\, K}$) dominate the energy budget, and their high specific energy allows them to traverse the CGM, escape the halo, and heat the IGM. In contrast, warm outflows ($\mathrm{\lesssim10^5\, K}$) dominate the mass budget and are largely recycled back into the interstellar medium (ISM), where they fuel future star formation. We identify a gradual transition at $\mathrm{\sim 5\, Gyr}$ that marks a shift in the balance between ejective and preventive feedback. At early times ($\mathrm{< 5\, Gyr}$), although the CGM cooling rate dominates for a larger fraction of time, the infrequent yet powerful SNe energy injection into the CGM is able to quickly dominate the cumulative energy balance. These outflows and their high specific energy are able to \lq sweep' up mass in the CGM and IGM. At late times ($\mathrm{> 5\, Gyr}$), the CGM baryon fraction is only $\mathrm{\sim0.1}$, leading to a transition toward a preventive feedback mode in which SNe maintain $\mathrm{t_{cool}/t_{ff} > 10}$ and prevent $\mathrm{\sim75\%}$ of the expected baryon accretion rate. 

\end{abstract}



\section{Introduction} \label{sec:intro}

\emph{How} did the gas in and around galaxies get there? This simple question has become central to one of the most important (yet elusive) topics in galaxy formation and evolution –– feedback. In order to form a galaxy, the rate of radiative cooling ($\mathrm{t_{cool}}$) for some fraction of the baryon content must be less than the age of the Universe \citep{Spitzer_1956}. Yet, early models of galaxy formation revealed that, without feedback, gas cools too efficiently, leading to excessive star formation -- an issue known as the “overcooling” problem \citep{Dekel_Silk_1986,White_Frenk_1991,Katz1996}. Feedback, which encompasses processes such as stellar winds, supernova (SNe), radiation, and active galactic nucleus (AGN) outflows, plays a critical role in regulating star formation. A direct consequence of feedback is that gas does not always trace dark matter. Instead, the difference between the expected cosmic baryon fraction ($\mathrm{\mathit{f}_b = \Omega_b/\Omega_m}$) and the observed gas fraction within the CGM ($f_\mathrm{CGM}$) reflects the extent to which feedback has redistributed the gas \citep[see review articles from][]{Crain_VanDeVoort_2023,Wright_2024}. 

The advent of the Cosmic Origins Spectrograph (COS) aboard the Hubble Space Telescope (HST) and large-scale galaxy surveys has allowed astronomers to study the circumgalactic medium (CGM) and its connection to the interstellar medium (ISM) and intergalactic medium (IGM) in unprecedented detail \citep{Chen1998,Tumlinson_2011,Tumlinson2013,Werk_2014}. However, the sightline approach of HST/COS has made it difficult to directly link these observations to feedback-driven processes.  For example, absorption lines probe individual lines of sight which may not trace the gas phase carrying most of the energy from feedback. To help, astronomers have turned to recent advances in computation, enabling the detailed study of gas at various scales: from the interaction between the cold and hot phase at the parsec level \citep{Gronke2020,McCourt2018,Abruzzo2024} to cosmological simulations tracking large-scale gas accretion \citep{Nelson_2013,Nelson_2016}. These cosmological simulations now routinely reproduce many properties of galaxies, including a star-formation rate that correctly places them on the stellar-to-halo mass relation  \citep[SHMR;][]{Moster_2010,Behroozi_2019}. However, there are two very different ways this can be done, resulting in very distinct gas distributions and fraction of gas retained in the halo: ejective feedback and preventive feedback.

\subsection{The Importance of Energy}

In ejective feedback models, strong galactic winds remove gas from the ISM and deposit it into the CGM and IGM. In order to avoid too much star formation in the ISM, these models require a high ($\gg 1$) mass-loading factor ($\mathrm{\eta_M}$): 
\vspace{0cm}
\begin{gather}
\mathrm{\dot{M}_{out} = \eta_M \dot{M}_*}
\end{gather}
\noindent
which quantifies the rate of mass ejection ($\mathrm{\dot{M}_{out}}$) through galactic winds just outside the galactic disk relative to the star-formation rate ($\mathrm{\dot{M}_{\ast}}$). Ejective feedback typically results in gas-rich CGMs, since the gas that doesn't form stars in the ISM is ejected back to the CGM. However, recent observations indicate that dwarf galaxies \citep{Chisholm_2017,McQuinn_2019,Concas_2022,Marasco_2023,KadoFong_2024} tend to have low $\mathrm{\eta_M}$, often an order of magnitude lower than what is measured in IllustrisTNG \citep{Pillepich_2018,Nelson_2019} or FIRE and FIRE-2 \citep{Muratov2015,Angles-Alcazar2017,Pandya_2021,Porter2024}. In addition, idealized galaxy and  tall-box (or \lq\lq patch'') simulation work that explored the low mass halo regime has independently found that $\mathrm{\eta_M}$ is low ($\sim1$; in agreement with observational work) and driven by cool-warm outflow \citep{Schneider_Robertson_2018,Kim_2020,Steinwandel_2024}. In comparison with a typical cosmological simulation where the gas cell resolution is low (e.g., $\sim10^5\, \mathrm{M}_\odot$; TNG50-1) and an effective equation of state is assumed, idealized galaxy and patch simulations are able to better resolve the interaction between the cold and hot phase since their gas cell resolution is high \citep[e.g., $\mathrm{\sim4\, M_\odot}$;][]{Smith2021,Steinwandel_2024}.  For example, in an idealized simulation of an isolated $\mathrm{M_h\sim10^{11}\,M_\odot}$ galaxy, \cite{Steinwandel_2024} find $\mathrm{\eta_M=0.1-1}$ at 10 kpc ($\mathrm{\sim0.1\,R_{vir}}$) and 1 kpc above the disk, respectively, where most of the mass is locked in the \lq warm' phase ($\mathrm{<5\times10^5\,K}$). 

In contrast to the mass removal (high $\mathrm{\eta_M}$) in ejective feedback, the focus of preventive feedback is galactic wind \emph{energy} transport. Therefore, the emphasis is on the energy-loading factor ($\mathrm{\eta_E}$) in preventive feedback:
\vspace{0cm}
\begin{gather}
\mathrm{\dot{E}_{out} = \eta_E \frac{\dot{M}_* E_{SN}}{m_*}}
\end{gather}
\noindent
where $\mathrm{E_{SN}}$ is the energy available from each core-collapse SNe and $\mathrm{m_*}$ is the mass required to produce one core-collapse SNe. $\mathrm{\eta_E}$ is therefore a measure of the amount of energy available from SNe to enter the CGM and IGM. In a review, \cite{Donahue_Voit_2022} discuss the importance of energy in heating and expanding the CGM, which in turn will prevent the further accretion of gas (see their Section 5). As a result, preventive feedback will often leave the CGM gas-depleted. The prevention of gas accretion can occur at both the CGM-scale (IGM $\rightarrow$ CGM) and ISM-scale (CGM $\rightarrow$ ISM).  This is explored in \cite{Wright_2024}, where they calculated the mass inflow rate at the ISM ($\mathrm{\sim0.25\,R_{vir}}$) and CGM ($\mathrm{\sim1\,R_{vir}}$) in IllustrisTNG and EAGLE. In IllustrisTNG, the accretion rate is \emph{high} at both the CGM- and ISM-scale. In contrast, in EAGLE, the accretion rate is \emph{low} at both the CGM- and ISM-scale, likely in part due to the prevention of gas accretion and therefore low $f_\mathrm{CGM}$ in EAGLE at $\mathrm{z\sim0}$.    
 
In the same idealized galaxy and patch simulation work listed above \citep{Schneider_Robertson_2018,Kim_2020,Steinwandel_2024}, they found that $\eta_E$ can be high and is driven by the hot outflows. For example, \cite{Kim_2020} found that $\mathrm{\eta_E\sim0.1}$ with a star-formation surface density of $\mathrm{\log_{10} \left( \Sigma_{SFR,40}/M_\odot\, kpc^{-2}\,yr^{-1} \right)} \sim -4$. 
They also found that $\mathrm{\eta_E}$ is dominated by the outflow in the hot phase ($\mathrm{>5\times 10^5\,K}$).

In order to further investigate the role of $\mathrm{\eta_E}$, \cite{Carr_2023} developed a simple regulator model that followed the flow of  mass and energy in to and out of the CGM. In their halo mass range ($\mathrm{M_h\lesssim10^{12}\,M_\odot}$), \cite{Carr_2023} demonstrated that the SHMR is most sensitive to $\mathrm{\eta_E}$, not $\mathrm{\eta_M}$ (see their Figure 3). A similar conclusion is found in \cite{Pandya2023,Pandya2026,Voit2024b,Voit2024}, where the latter found that the specific energy of the outflow ($\mathrm{e_{s,fb}}$) compared to the specific energy of the gas accreting onto the halo ($\mathrm{e_{s,acc}}$) is important in driving either a \lq coupled' or \lq uncoupled' feedback flow (see Section 2 in \cite{Voit2024} for further discussion).

The specific energy of the outflow is defined as 

\begin{gather}\label{equation:specific_energy}
\mathrm{e_s \equiv \frac{\dot{E}_{out}}{\dot{M}_{out}} = \frac{\eta_E E_{SN}}{\eta_M m_*}}
\end{gather}

\cite{Voit2024b} find that low mass halos ($\mathrm{\sim10^{11}\,M_\odot
}$) in IllustrisTNG have $\mathrm{e_{s,fb}\sim e_{s,acc}}$ such that some of the CGM gas is lifted out of the halo, but will likely result in gas recycling. In comparison, at the same halo mass in EAGLE \citep{Schaye_2015,Crain_2015}, $\mathrm{e_{s,fb}\gg e_{s,acc}}$, resulting in CGM gas easily removed and heated. The discrepancy between the $\mathrm{e_s}$ manifests itself in $f_\mathrm{CGM}$, where $f_\mathrm{CGM}$ in IllustrisTNG is a factor of $\sim 5$ greater than $f_\mathrm{CGM}$ in EAGLE at $\mathrm{M_h\sim10^{11}\,M_\odot}$ \citep{Crain_VanDeVoort_2023,Wright_2024}. EAGLE has a more \lq preventive-like' feedback model, where not only is the outflow high specific energy ($\mathrm{e_{s,fb}\gg e_{s,acc}}$), but it has been shown that gas is prevented from accreting and re-accreting at the halo-scale \citep{Mitchell2020b,Mitchell_2020,Mitchell2022}. 

This stark contrast in $f_\mathrm{CGM}$ is also seen in the Arkenstone, a novel cosmological simulation using a two-phase galactic wind model that is able to capture the high-specific energy winds required in preventive feedback  \citep{Fielding_Bryan2022,Smith_2024b,Smith_2024a,Bennett2025,Sullivan2026}. \cite{Bennett2025} found that compared to the fiducial IllustrisTNG run (high $\mathrm{\eta_M}$), Arkenstone \lq hot' model is able to produce a gas-depleted CGM while still remaining on the SHMR (see their Figure 3).
 
\subsection{Low Mass Halos}

Given the observed low $\eta_M$ and clear evidence for the necessity for energy-loaded, high-specific energy winds, the idea of SNe-driven preventive feedback regulating star-formation in the dwarf galaxy regime is promising. In the halo mass range below $\sim10^{11.5}\,\mathrm{M}_\odot$, the role of SNe feedback can be isolated from AGN feedback \citep[although this is an open question, see][]{Koudmani_2019,Barai_2019,Sharma_2020,Arjona2024} and stable virial shock-heating of the halo \citep{Birnboim_Dekel_2003,Dekel_Birnboim_2006}. In addition, low mass halos ($\sim10^{10}-10^{11}\,\mathrm{M}_\odot$) have a virial temperature of $\mathrm{T}_{\rm vir} \sim 10^5\,\rm K$ and are therefore uniquely positioned at the peak of the cooling curve \citep{Mo_2010}.

In considering whether SNe-driven preventive feedback can regulate star formation, two distinct but related questions arise. The first concerns the thermal state of the CGM: in the traditional view of a dwarf galaxy, we might naively expect SNe heating of the halo to cool rapidly \citep{White_Frenk_1991}, driving $\mathrm{T}_{\rm CGM} \ll \mathrm{T}_{\rm vir}$. The second concerns the radiative stability of the CGM: even if SNe wind-induced shock heating sustains $\mathrm{T}_{\rm CGM} \sim \mathrm{T}_{\rm vir}$, the CGM may still be radiatively unstable with $\mathrm{t}_{\rm cool}/\mathrm{t}_{\rm ff} \lesssim 10$ \citep{McCourt_2012, Sharma2012, Voit_2017}, allowing continued gas accretion and thus requiring an unrealistically high $\eta_M$ to keep the galaxy on the SHMR. 

The notion of SNe wind-induced shock-heating the CGM has been discussed in recent literature \citep{Voit_2015,Fielding_2017,Pandya2023,Carr_2023}. For example, in an idealized setup of a $\mathrm{M_h\sim10^{11}\,M_\odot}$ halo, \cite{Fielding_2017} demonstrated that energy-loaded winds produced a shock-heated CGM and mass-loaded winds produced a cool, clumpy CGM. \cite{Voit_2015} argue that low mass halo CGMs should be similar to the behavior observed and simulated at the galaxy cluster scale, where the heating and cooling of gas surrounding the central black hole regulates star-formation \citep{Li_2015,Donahue_Voit_2022}. In recent work, the time evolution of $\mathrm{t_{cool} / t_{ff}}$ was measured in $\mathrm{M_h\sim10^{12}\,M_\odot}$ halos in FIRE-2 as part of a process described as \lq outside-in virialization' \citep{Stern2019,Stern2020,Stern2021}. In FIRE-2, \cite{Stern2021} found that SNe feedback is (eventually) able to increase the inner CGM temperature to $\mathrm{\sim T_{vir}}$ and $\mathrm{t_{cool} / t_{ff}}\sim10$, which is interestingly correlated with the transition from bursty to steady star-formation and disk-formation. 

Lastly, feedback can be traced in gas observables, such as the kinetic and thermal Sunyaev-Zel'dovich effect \citep[SZ;][]{Moser2022}, X-ray surface brightness \citep[XSB;][]{Lau2025}, fast-radio bursts \cite[FRBs;][]{Connor2025}, and HST/COS absorption. However, the SZ effect and XSB are currently used to trace the ionized gas in massive halos ($\mathrm{M_h}\gtrsim10^{12}\,\mathrm{M}_\odot$) and there has been little work exploring the role of the electron dispersion measure contribution from the low mass halo population. HST/COS absorption, on the other hand, could be used to help distinguish between ejective and preventive feedback, where the temperature and baryon content of the CGM will manifest itself in absorption (although there will likely be some degeneracy). For example, neutral hydrogen ({\sc Hi}) is ubiquitously detected to large radii and low metal ions (e.g., C\,\textsc{ii}, C\,\textsc{iii}, C\,\textsc{iv}, Si\,\textsc{ii}, Si\,\textsc{iii}, Si\,\textsc{iv}) do not extend beyond $0.5\, \mathrm{R_{vir}}$  \citep{Bordoli_2014,Bordoloi_2018,Zheng_2019,Zheng_2020,Zheng_2024}. In addition, \cite{Mishra_2024} detected highly ionized {\sc Ovi} extending out to $1-2\, \mathrm{R_{vir}}$ in a low-redshift, star-forming dwarf galaxy sample. However, the origin of {\sc Ovi} is unclear, but could be used to better understand and constrain the underlying SNe-driven feedback in $\mathrm{\sim10^{10}-10^{11}\,M_\odot}$ halos. 

There has been considerable progress simulating $\mathrm{M_h \sim 10^{10} - 10^{11}\,M_\odot}$ halos in a cosmological context. A number of zoom-in suites have characterized the star formation histories, structural properties, and scaling relations of field dwarfs in this mass range \citep{Munshi2021, Tremmel2017, Shen2014, Applebaum2021, Tomaru2025}. Others have focused more directly on the CGM, including ion column density distributions \citep{Mina2021, Baumschlager2025} and the thermal structure of the halo gas \citep{Tung2025, Cook2025}.
However, comparatively little attention has been paid to the time evolution of the CGM thermal structure in this mass regime --- in particular, whether SNe-driven outflows can sustain $\mathrm{T_{CGM} \sim T_{vir}}$, drive $\mathrm{t_{cool}/t_{ff} \gtrsim 10}$ in the outer CGM, and prevent gas accretion into the halo in the first place. We use this paper to quantify several outstanding questions regarding 
feedback and the flow of energy in the low-mass halo regime. In particular,

\begin{enumerate}
  \item Can the CGM and nearby IGM be sustained at $\mathrm{\sim T_{vir}}$ through SNe wind-induced shocks?
  \item What is the specific energy of the outflow and more broadly, what are the energy outflow and inflow rates at the ISM boundary and CGM boundary? 
  \item Can these outflows entrain mass in the CGM (and beyond) and prevent future gas accretion? 
  \item Can feedback drive the dwarf CGM to a radiatively stable regime, where $\mathrm{t_{cool} / t_{ff} \gtrsim 10}$, similar to what is observed in more massive systems? 
  \item How can this model be tested through observations? For example, the column density profile of highly ionized {\sc Ovi}.  
\end{enumerate}

In this paper, we explore these five key questions using cosmological zoom-in  simulations using the adaptive mesh refinement code \textsc{enzo} \citep{Bryan_2014} and a discrete SNe feedback model where high-specific energy winds penetrate a high-resolution CGM \citep{Forbes_2016,Goldbaum_2015,Goldbaum_2016}.

The paper is organized as follows. We first briefly introduce the simulation setup and feedback model in Section \ref{sec:methodology}. In Section \ref{sec:results}, we introduce the dwarf galaxy sample and present our main results. We then discuss these findings in the context of other work in Section \ref{sec:discussion}. Finally, we conclude in Section \ref{sec:summary and conclusion} and discuss the future work needed to build upon our findings. Throughout this paper, we adopt the cosmological parameters presented in \cite{Planck2018}, where $\mathrm{H_0 = 67.4\ km/s/Mpc}$, $\mathrm{\Omega_\Lambda = 0.685}$, $\mathrm{\Omega_m = 0.315}$, and $\mathrm{\Omega_b = 0.049}$. This corresponds to a cosmic baryon fraction of $\mathrm{f_b \sim 0.16}$.

\section{Methodology} \label{sec:methodology}

In order to understand the role of preventive feedback in low-mass $\mathrm{M_h \sim 10^{10}-10^{11}\, M_\odot}$ halos, we are interested in a high resolution CGM where the SNe-driven high specific energy winds are able to propagate. We choose to use the structured adaptive mesh-refinement (AMR) code \textsc{enzo} \citep{Bryan_2014}. \textsc{enzo} is a Cartesian code designed to model astrophysical fluid flow, and can be used in any dimension and scale. The collisionless matter (e.g., dark matter and stars) is treated as a discrete particle on the Cartesian grid that interacts with the surrounding baryon content via gravity.  AMR refers to the use of an adaptive hierarchy of grids that can vary spatial resolution based on the user-defined threshold; therefore, \textsc{enzo} can achieve a large spatial dynamical range. In the context of our simulation suite, we aim to model both the small-scale star-formation and SNe and large-scale cosmological gas flow. In \textsc{enzo}, the fluid equations are solved using a modified Godunov third-order accurate piecewise parabolic method \citep{Bryan_1995}.
In addition, compared to a purely Lagrangian method using a fixed particle mass, \textsc{enzo}, although in general cells are refined based on baryon and dark matter overdensities, low mass cells with high specific energy are permitted. Therefore, \textsc{enzo} may be more suitable to better resolve both the high specific energy flows we expect from SNe-driven feedback \citep[see also the discussion in][]{Smith_2024a}.

\textsc{enzo} also enables the implementation of heating and cooling in our simulation suite using the standardized chemistry and cooling library, \textsc{grackle} \citep{Bryan_2014,Grackle2017}. \textsc{grackle} contains the heating and cooling rates for primordial species, metal species, and UV background photo-heating and photo-ionization rates. These are calculated for the temperature and density range found in our simulation. We use a six-species model: H, H$^+$, He, He$^+$, He$^{++}$, e$^-$ and model other ionization species (e.g., {\sc Ovi}) using a photo-ionization model (\S\ref{subsec:metal distribution}). We also use a time-dependent meta-galactic radiation field with self-shielding \citep{Haardt_Madau_2012}, beginning at $\mathrm{z = 9}$ and ending at $\mathrm{z = 0}$. 

\subsection{Simulation Setup} \label{sec:simulation setup}

We generate our initial conditions using MUlti-Scale Initial Conditions \citep[\textsc{\textsc{music}};][]{Hahn_Abel_2011}, adopting the best-fit cosmological parameters from  \cite{Planck2018} and the transfer function of \cite{Eisenstein_Hu_1998}. We perform cosmological zoom-in simulations within both $(10\ \mathrm{Mpc},h^{-1})^3$ and $(20\, \mathrm{Mpc\,h^{-1}})^3$ volumes, following a workflow similar to that described in \cite{Onorbe_2014}.

We begin with a low-resolution, dark matter–only simulation evolved from $z = 99$ to $z = 0$. Halos are identified at $z=0$ using the HOP halo-finding algorithm \citep{Eisenstein_Hut_1998}, as implemented in \textsc{yt} \citep{Turk_2011}. From this catalog, we select systems with halo masses in the range $(3$–$5) \times 10^{10}\ \mathrm{M_\odot}$. We then compute the distance to each halo’s nearest neighbor and select those that are relatively isolated at $z \sim 0$. This selection is designed to ensure that the CGM is relatively unperturbed at $z \sim 0$. As discussed in Section~\ref{subsec:closure radius}, some systems nevertheless host infalling satellites at low redshift.

We select 10 halos in the mass range $(3$–$5) \times 10^{10}\ \mathrm{M_\odot}$ for our final sample. For each system, we identify all dark matter particles located within $3\, \mathrm{R_{vir}}$ of the present-day host halo and record their initial ($z = 99$) simulation positions (since each particle has a unique, time-independent ID). These initial positions are then supplied to \textsc{music} to define the Lagrangian zoom-in region. 

\textsc{music} constructs a bounding convex hull that encloses only the \lq must-refine' particles. These are a specialized class of dark matter particles that force additional mesh refinement in their local surroundings (specifically, the eight nearest cells), up to a prescribed level beyond the base zoom-in refinement. We set the minimum refinement level of the must-refine particles equal to the highest level of the initial zoom-in refinement. Compared to the traditional approach of refining a fixed Eulerian zoom-in volume, the must-refine particle method offers two key advantages. First, it prevents artificial collapse and excessive refinement of structures outside the primary halo of interest. Second, it allows the refinement region to follow the target halo as it moves through the cosmological volume. This is particularly important for low-mass halos, which can drift out of (and later re-enter) a fixed zoom-in region due to large-scale velocity flows.

Cells are refined based on both baryonic and dark matter overdensity criteria. To ensure adequate resolution in the CGM  -- which is significantly lower density than the ISM -- we increase the number of initial refinement levels to 3. For the $20\ \mathrm{Mpc\, h^{-1}}$ cosmological box, the root grid has a resolution of $256^3$ ($128^3$ for the $10\ \mathrm{Mpc\, h^{-1}}$ box). Therefore, in the 3 initial level region, the initial spatial resolution is $\sim 15\ \mathrm{kpc}$. However, beyond this initial refinement configuration, gas cells are allowed to refine up to 8 levels. This yields a maximum spatial resolution of $\sim 450\ \mathrm{pc}$. We repeat one zoom-in simulation with 4 initial levels of refinement, corresponding to a CGM spatial resolution a factor of $\sim2$ smaller than our fiducial sample. Throughout this paper, we refer to the high resolution simulation run as M8-h (\S\ref{sec:results}). 

\subsection{Feedback Model} \label{sec:feedback}

The other important component of the simulation outside of the refinement scheme is the feedback model. We adopt the same feedback model that was recently used in an idealized MW mass system \citep{Goldbaum_2015,Goldbaum_2016} and compared to other code in the Assembling Galaxies Of Resolved Anatomy (AGORA) project \citep{Kim_2016}. The model parameters match a \textsc{starburst99} \citep{Leitherer_1999,Vazquez_Leitherer_2005,Leitherer_2014} model of a young stellar population that fully samples the Chabrier initial mass function \citep{Chabrier_2003}. See \cite{Forbes_2016} for a detailed description of the model; we repeat the most important aspects in the context of our work below.  

The feedback model first requires the formation of a star particle, where the star-formation density in a cell is calculated using:
\begin{gather}
\mathrm{\frac{\rm{d \rho_*}}{dt} = \left\{\begin{matrix}
 f_* \frac{\rho}{\rm{t_{ff}}} & \rho > \rho_{\rm{thresh}} \\ \\
 0 & \rho \leq \rho_{\rm{thresh}}
\end{matrix}\right.} 
\end{gather}
where $f_*$ is the star-formation efficiency, $\mathrm{t_{ff}}$ is the local free-fall time, and $\mathrm{\rho_{thresh} = \mu\ m_H\ n_{thresh}}$ is the star-formation density threshold. We adopt $\mathrm{f_* = 1 \%}$ and $\mathrm{n_{thresh} = 50 cm^{-3}}$, based on recent observational work \citep{Krumholz_Tan_2007,Krumholz_2012}. We also fix the mean molecular weight $\mu  = 1.4$. If the gas (1) reaches this star-formation density threshold and (2) is located on a maximum refinement level grid, then the gas is converted into a star particle, representing a stellar population. In the model, this is implemented in a stochastic scheme in order to avoid the production of too many low-mass stars. The probability to form a star particle $\mathrm{m_{sf}}$ is given by:
\begin{gather}
\mathrm{P_* = \left\{\begin{matrix}
 f_* \frac{\rho \Delta x^3}{\rm{m_{sf}}} \frac{\rm{dt}}{\rm{t_{ff}}} & \rho > \rho_{\rm{thresh}} \\ \\
 0 & \rho \leq \rho_{\rm{thresh}}
\end{matrix}\right.} 
\end{gather}
where $\mathrm{dt}$ is the timestep on the maximum level of refinement and $\mathrm{m_{sf}} = 10^3\,\mathrm{M}_{\odot}$.

There are three sources of feedback in our simulation: {\sc Hii} regions, massive stars, and core-collapse SNe. The ionizing radiation from the massive star at the center of the {\sc Hii} region can heat the surrounding medium and prevent gravitational collapse and star-formation. {\sc Hii} region heating will only occur in a gas cell that contains at least one massive star where $\mathrm{T < 10^4\ K}$. The amount of heat is determined by first calculating the ratio of the Str\"omgren sphere volume (determined from the star particle ionizing radiation and gas cell density) to the cell volume. If the Str\"omgren sphere volume ($\mathrm{V_s}$) is larger than then cell volume ($\mathrm{V_c}$), then the cell is heated to $\mathrm{10^4\ K}$. If the Str\"omgren sphere volume is less than the cell volume, then the cell is heated to $\mathrm{10^4\ K\ (V_s/V_c)}$. Since there are often more than one massive star in a star particle, each massive star will contribute to the heating of the gas cell, but the gas cell will not exceed $\mathrm{10^4\ K}$. Stellar winds from young, massive stars have a similar effect of suppressing star-formation. \textsc{starburst99} is used to calculate the mass and internal energy that should be injected into the host gas cell as a function of time for a stellar population. This is computed until the last SN occurs.      

In the core-collapse SNe model, each SN has $\mathrm{10^{51}\ erg.}$ of energy available, which can be divided into momentum feedback and energy feedback. At a given timestep, if a SN occurs within a gas cell, then the explosion is approximated at the center of the grid cell. In order to reproduce the SNe extent of a SN bubble, momentum is injected equally into each of the nearest 26 cells using: 
\begin{gather}
\mathrm{\Delta p_{SN} = 1.2 \times 10^4\ M_\odot\ km\ s^{-1}\ \hat{r}} 
\end{gather}
where the total momentum deposited ($\mathrm{3 \times 10^5\ km\ s^{-1}}$) is based on recent high-resolution simulations of SNe in a realistic inhomogeneous ISM \citep{Kim_2011,Kim_Ostriker_2015,Martizzi_2015}. This result is not too different from the simple physical model presented in \cite{Cioffi_1988} and \cite{Kim_Ostriker_2015} found that this is only slightly less ($\sim 5 \%$) than the total momentum from SNe placed in a homogeneous ISM with the same mean density. Once the momentum injection is complete, the resulting net change in kinetic energy of the surrounding 26 cells (with respect to the simulation rest frame) is subtracted from the $\mathrm{10^{51}\ erg}$ energy budget. The remaining non-kinetic energy is then deposited into the SN host cell as thermal energy.         

\section{Results} \label{sec:results}

This simulation framework is applied to each of our ten $\mathrm{\sim 10^{10} M_\odot}$ cosmological zoom-in systems.  We first briefly comment on the final resolution of our sample, including the highest resolution run (\S\ref{sec:present day resolution}). We then discuss the $\mathrm{z=0}$ properties of our sample (\S\ref{sec:present day properties}), including the stellar mass-halo mass relationship. We then present the time evolution of the halo temperature (\S\ref{subsec:sustaining virial temperature}), energy flow (\S\ref{subsec:flow of mass and energy}), mass flow (\S\ref{subsec:flow of mass}), and radiative stability of the CGM (\S\ref{subsec:radiative stability}). Lastly, we present the CGM baryon fraction (\S\ref{subsec:baryon fraction}), closure radius (\S\ref{subsec:closure radius}), and ion absorption (\S\ref{subsec:metal distribution}).     

\subsection{Present-Day Resolution} \label{sec:present day resolution}     

\begin{figure}
    \centering
    \includegraphics[width=0.48\textwidth]{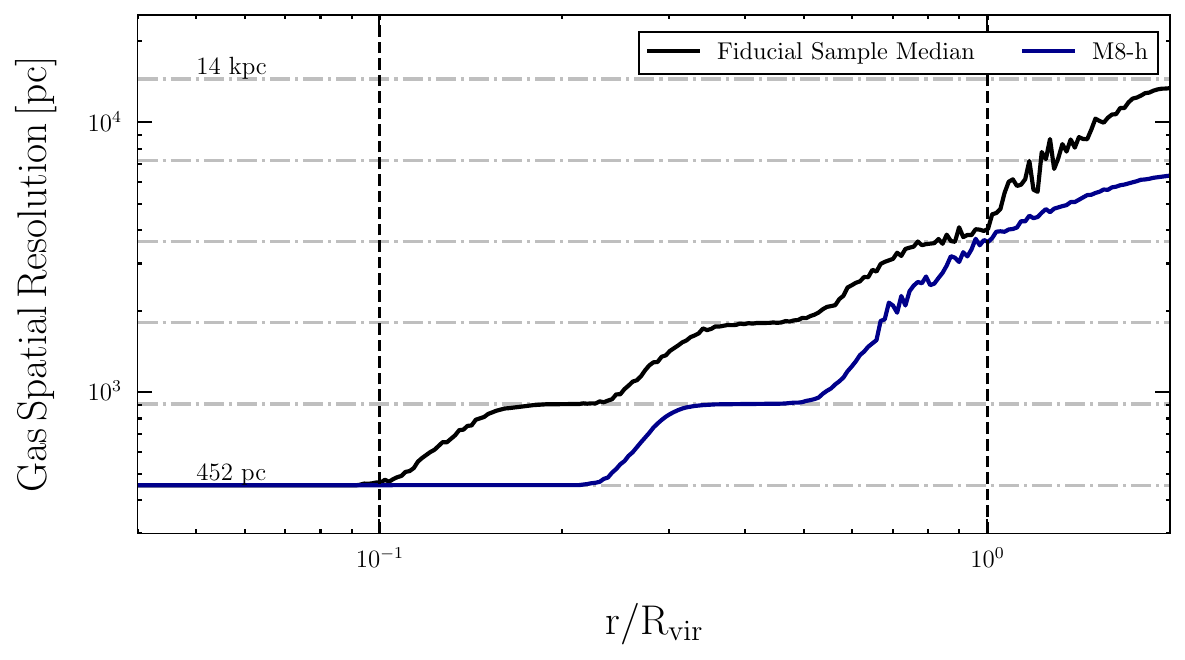} 
    \caption{The gas spatial resolution as a function of distance from the center of the halo. We present both M8-h and the median of the fiducial sample. For each individual halo, we calculate the volume-weighted average gas spatial resolution between $\mathrm{0.04\,R_{vir}}$ and $\mathrm{2\,R_{vir}}$. The highest spatial resolution in both M8-h and the fiducial sample is in the ISM ($\mathrm{\sim452\,pc}$).        
}\label{fig:gas spatial resolution}
\end{figure}

We briefly describe the resolution of our simulation suite. In Figure 
\ref{fig:gas spatial resolution}, we show the $\mathrm{z\sim0}$ spatial resolution of the gas as a function of distance from the center of the halo. In particular, we plot the volume-weighted average spatial resolution for M8-h (solid blue) and the median volume-weighted average across the fiducial sample (solid black). In addition, we include the minimum CGM resolution ($\mathrm{14\,kpc}$), minimum simulation resolution ($\mathrm{452\,pc}$), and every level of refinement in-between as a horizontal line. The minimum CGM resolution is set by the initial level of refinement at the simulation start and the minimum simulation resolution is five additional levels refinement beyond the maximum level of initial refinement (see \S\ref{sec:methodology}). In the M8-h simulation, there is an additional level of initial refinement, hence the CGM spatial resolution is a factor of $\mathrm{\sim2}$ smaller than the fiducial sample. The typical dark matter mass resolution in the halo in the fiducial sample is $\mathrm{\sim10^5\,M_\odot}$. In comparison, the dark matter mass resolution in M8-h is $\mathrm{\sim10^4\,M_\odot}$. 

\subsection{Present-Day Properties} \label{sec:present day properties}

The present-day galaxy properties are presented in Table \ref{table:properties}, where each dwarf is identified using the letter M (for \lq Messere' instead of \lq Messier' object) followed by a number corresponding to $\mathrm{M_{vir}}$ (ordered from most massive to least massive). In Table \ref{table:properties}, we include $\mathrm{M_{vir}}$, $\mathrm{R_{vir}}$, $\mathrm{M_{ISM}}$, $\mathrm{M_{CGM}}$, and $\mathrm{M_{*}}$. We define the outer boundary of the ISM using $\mathrm{0.1\, R_{vir}(t)}$, which is the definition often adopted in other simulation work \citep[e.g.,][]{Pandya_2020,Pandya2023}. In addition, we define the CGM as the region contained between $\mathrm{0.1 \, R_{vir}(t)}$ and $\mathrm{1 \, R_{vir}(t)}$. In particular, we use the time-dependent halo radius, $\mathrm{R_{vir}(t)}$, which is calculated in our implementation of \textsc{rockstar} \citep{Behroozi2013} and \textsc{consistent-trees} \citep{Behroozi2013b}, which is based on the definition presented in \cite{Bryan_Norman1998}. The merger tree enables us to smoothly track the time evolution of the largest progenitor branch of our present-day low mass halo. We output 331 snapshots per zoom-in simulation, corresponding to $\sim42$ Myr timestep in our analysis. In our time series analysis (e.g., temperature evolution, energy flow rate) we exclude the first 14 snapshots, corresponding to the first $\sim500$ Myr of halo growth, since this is approximately the earliest output when the entire sample was present in the merger tree.

\begin{figure} 
    \centering
    \includegraphics[width=0.47\textwidth]{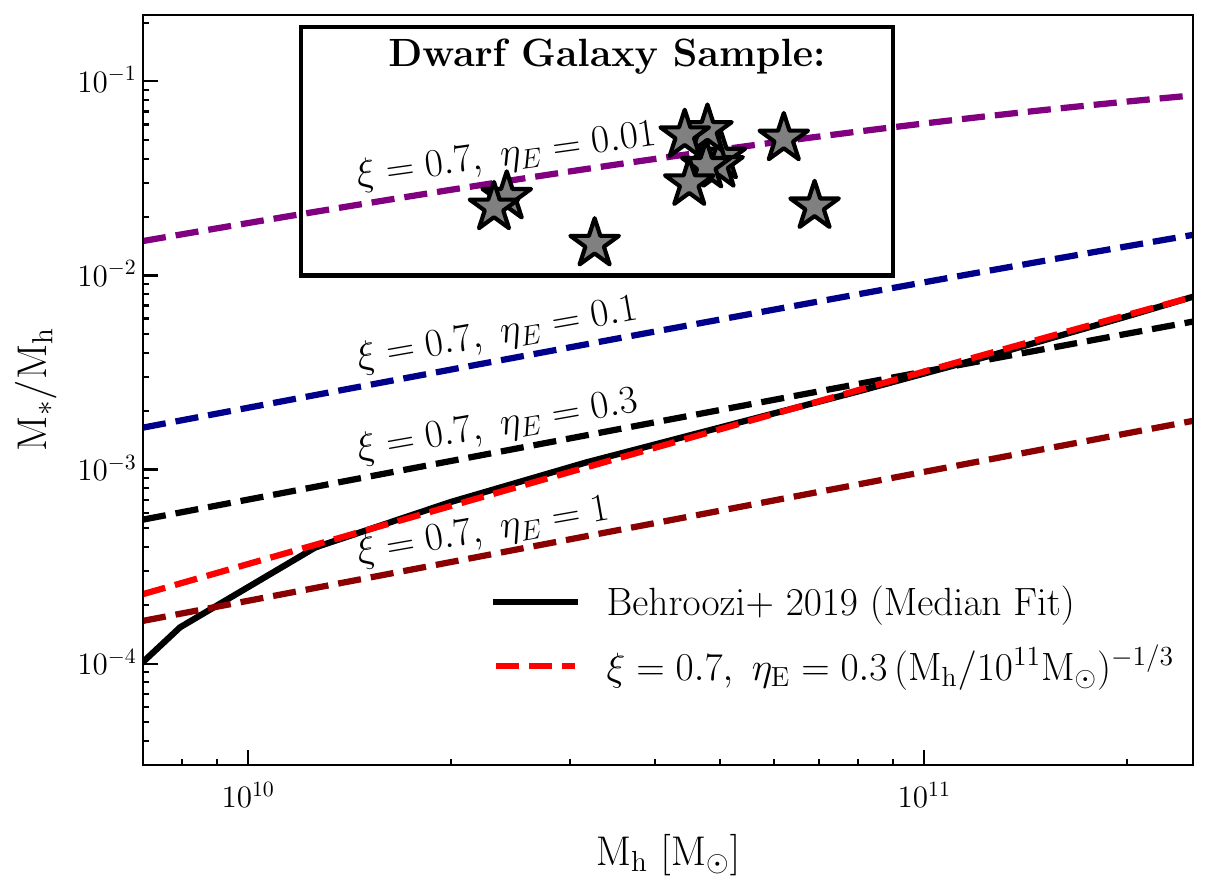} 
    \caption{The z = 0 stellar mass--halo mass relationship for our dwarf galaxy sample compared to \cite{Behroozi_2019} and \cite{Voit2024b,Voit2024}. In comparison with the minimalist regulator model presented in \cite{Voit2024b} (where radiative loss is assumed to be negligible), we find that our high stellar mass is likely driven by  $\mathrm{\eta_E}=0.01-0.05$.  
    \label{fig:SHMR}}
\end{figure}

It is important to note that the range of stellar mass in our dwarf sample is high compared to the observed SHMR. We show a direct comparison to \cite{Behroozi_2019} in Figure \ref{fig:SHMR}, denoted using the solid black line. This discrepancy is likely due to the inability of the feedback model to effectively heat the ISM gas at high redshift (i.e. due to overcooling), resulting in too much early star-formation. We put this into context using the regulator model presented in \cite{Voit2024b,Voit2024}. In their model, \cite{Voit2024b} find that low mass halos self-regulate through the expansion of their atmosphere (i.e., CGM), where the SNe energy is able to couple with the CGM gas. In agreement with the conclusion from \cite{Carr_2023}, \cite{Voit2024b}  find that the SHMR is not sensitive to the mass loading of the outflow and instead driven by the SNe energy, or energy loading factor.   

In Figure \ref{fig:SHMR}, we include the asymptotic stellar baryon fraction from \cite{Voit2024b} (see their Equation 14):
\begin{gather} \label{equation:voit}
\mathrm{\mathit{f}_{\ast,asy} = \frac{\xi \mathit{v}^2_c}{\eta_E\varepsilon_{SN}+ \xi \mathit{v}^2_c
}}
\end{gather}
In our calculation, we ignore the CGM radiative loss term (since \cite{Voit2024b} assumes it is small) in the denominator $\left[ \mathrm{\varepsilon_{acc}-(1+\eta_M)\varepsilon_{loss}}\right]$, where $\varepsilon_\mathrm{loss}$ is the specific energy of the CGM gas cooling onto the ISM. $\xi$ in Equation \ref{equation:voit} is the difference between the equilibrium specific energy ($\varepsilon_\mathrm{eq}$) and the gas accretion specific energy ($\varepsilon_\mathrm{acc}$). $v_\mathrm{c}$ is the maximum circular velocity of the halo. See \cite{Voit2024b,Voit2024} for additional model details.   

In Figure \ref{fig:SHMR}, we show four different $\mathrm{M_\ast/M_h}$ from Equation \ref{equation:voit}, where we assume a different $\mathrm{\eta_E}$ and $\xi=0.7$. Note that we convert $f_\mathrm{\ast}$ to $\mathrm{M_\ast/M_h}$, where $f_\mathrm{\ast}=\mathrm{M_\ast/\mathit{f}_b M_h}$. In addition, we include a simple halo mass dependence of $\mathrm{\eta_E}$, where $\mathrm{\eta_E=0.3(M_h/10^{11}\,M_\odot)}^{-1/3}$ and $\xi=0.7$.

Assuming our present-day $\mathrm{M_h\sim5\times10^{10}\,M_\odot}$, the mass dependent $\mathrm{\eta_E}$ suggests that $\sim38\%$ of the available energy from SNe is needed to couple with the CGM in order for our dwarf sample to reside along the SMHM relationship. However, we find that given our high $\mathrm{M_\ast}$, only $\sim 1\%-5\%$ of the available energy from SNe is coupled with the gaseous atmosphere. This difference is most likely due to excessive radiative losses in the simulated ISM. 


\begin{deluxetable}{cccccc}
\tablecaption{Properties of the dwarf galaxy sample. We label each dwarf using the notation M$
X$, where $
X$ is the mass ranking of the dwarf galaxy (from most to least massive). M8-h is M8 with one additional level of initial refinement.}
\label{table:properties}
\tablenum{1}
\tablehead{\colhead{ID} & \colhead{$\mathrm{log_{}\frac{M_{vir}}{M_\odot}}$} & \colhead{$\mathrm{\frac{R_{vir}}{kpc}}$} & \colhead{$\mathrm{log_{}\frac{M_{ISM}}{M_\odot}}$} & \colhead{$\mathrm{log_{}\frac{M_{CGM}}{M_\odot}}$} &
\colhead{$\mathrm{log_{}\frac{M_{\ast}}{M_\odot}}$}}
\startdata
M1 & 10.84 & 107.88 & 8.85 & 8.12  & 9.19 \\
M2 & 10.82 & 106.09 & 9.24 & 9.21 & 9.50  \\
M3 & 10.73 & 98.48 & 8.68 & 9.06 & 9.31 \\
M4 & 10.72 & 98.08 & 9.24 & 8.82 & 9.27\\
M5 & 10.71 & 97.45 & 8.98 & 8.75 & 9.43  \\
M6 & 10.70 & 96.75 & 8.90 & 9.06 & 9.24  \\
M7 & 10.67 & 93.83 & 9.22 & 8.96 & 9.13 \\
M8 & 10.67 & 94.95 & 8.84 & 7.98 & 9.37  \\
M9 & 10.53 & 84.54 & 8.65 & 8.78 & 8.68  \\
M8-h & 10.40 & 76.86 & 8.78 & 7.97 & 8.79  \\
M10 & 10.38 & 73.60 & 8.85 & 8.54 & 8.71  
\enddata
\end{deluxetable}

\subsection{Sustaining the Virial Temperature} \label{subsec:sustaining virial temperature}

The first question we address is whether SNe wind-induced shock heating can sustain the CGM at $\mathrm{\sim T_{vir}}$:

\begin{gather} \label{equation:tvir}
\mathrm{T_{vir} = \frac{\mu m_p}{2 k_B}\frac{G M_{vir}}{R_{vir}}}
\end{gather}
where we ignore external pressure \citep{Lochhaas_2021} and assume the mean molecular weight $\mu = 0.59$.

As discussed in \S\ref{sec:intro}, the present-day halo masses of our sample lie near the peak of the cooling curve. Therefore, we naively expect the CGM to cool on a short timescale compared to the halo free-fall time ($\mathrm{t_{cool} \ll t_{ff}}$) \citep{White_Frenk_1991}.

\subsubsection{SNe-Induced Shock Heating in M8-h} \label{subsubsec:M8-h}

\begin{figure*} 
    \centering
    \includegraphics[width=1\textwidth]{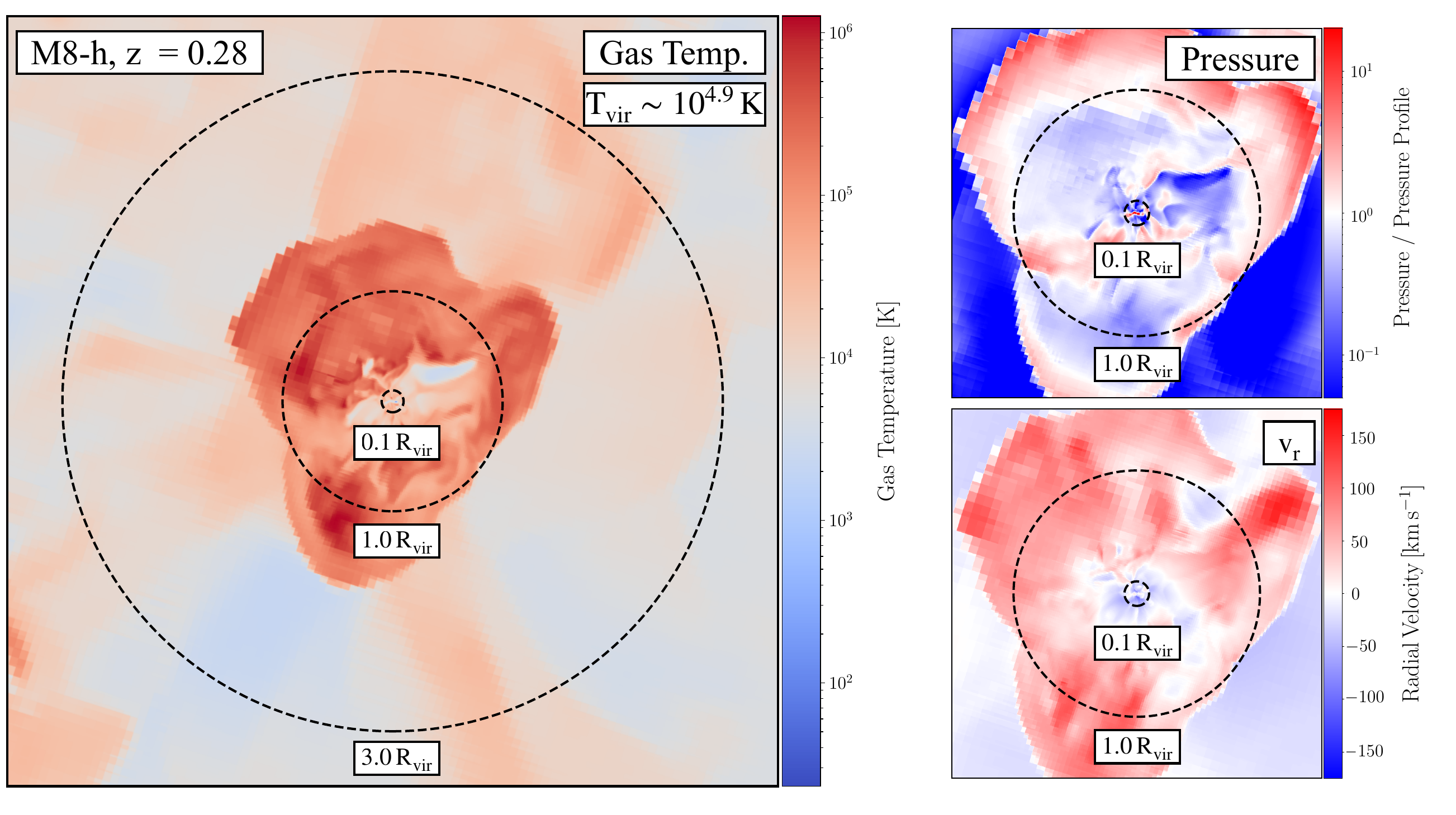} 
    \caption{\textbf{Left Panel:} M8-h (z = 0.28) temperature slice out to $\sim3\,\mathrm{R_{vir}}$. \textbf{Upper-Right Panel}: M8-h pressure slice normalized by the volume-weighted three dimensional pressure profile out to $\sim1\, \mathrm{R_{vir}}$. \textbf{Lower-Right Panel}: M8-h radial velocity slice out to $\sim1\, \mathrm{R_{vir}}$, where $\mathrm{v_r>0\,km\,s^{-1}}$ is outflow away from the galaxy center. The M8-h image in each panel is viewed edge-on (calculated from the inner galaxy angular momentum vector). In the ISM ($\mathrm{<0.1\,R_{vir}}$), the shock-driven outflow appears preferentially along the minor axis and is able to engulf the entire halo at the CGM-scale ($\mathrm{\sim1\,R_{vir}}$). The temperature within this outer shock envelope tends to exceed $\mathrm{T_{vir}\sim10^{4.9}\,K}$.   
    }\label{fig:temp_slice_two}
\end{figure*}

To demonstrate an example of SNe-induced shock heating, we show our high-resolution dwarf galaxy (M8-h) at $\mathrm{z=0.28}$ in Figure \ref{fig:temp_slice_two}. The left panel shows an edge-on temperature slice \citep[using \textsc{yt};][]{Turk_2011} extending out to $\mathrm{\sim3\,R_{vir}}$. We define the edge-on orientation using the angular momentum vector of gas in the inner halo ($\mathrm{r/R_{vir}<0.25}$). For reference, the virial temperature of M8-h at $\mathrm{z=0.28}$ is $\mathrm{T_{vir}\sim10^{4.9}\,K}$ (Equation \ref{equation:tvir}). The upper-right panel shows the halo gas pressure out to $\mathrm{\sim1\,R_{vir}}$. Instead of plotting the physical pressure, we normalize the pressure slice by the volume-weighted three-dimensional radial pressure profile in order to better visualize the pressure contrast across the shock. The lower-right panel shows the gas radial velocity relative to the galaxy center. The radial velocity is calculated after correcting for the bulk motion of the ISM ($\mathrm{r/R_{vir} <0.1}$).

In M8-h, the ISM ($\mathrm{r/R_{vir} <0.1}$) is cold, with the mass-weighted temperature below $\mathrm{10^4\,K}$ (dark blue). Outside the ISM, hot gas vents along the minor axis of the star-forming disk into the surrounding CGM. This hot gas is driven out of the ISM by SNe feedback and produces multiple shock fronts at small radii. The shocks are most clearly visible at larger radii; for example, a previous episode of star formation and subsequent SNe produced the shock at $\mathrm{\sim1\,R_{vir}}$, where the CGM interior to the shock has been heated to $\mathrm{\sim T_{vir}}$ and above (dark red). Although the shock has likely cooled since first breaking through the star-forming disk and reaching its current radial extent, it remains well above $\mathrm{T_{vir}}$ ($\mathrm{\gtrsim10^6\,K}$).

The same shock front at $\mathrm{\sim1\,R_{vir}}$ can be seen in the upper-right panel in Figure~\ref{fig:temp_slice_two}, where a pressure enhancement appears along the shock edge, surrounded by the lower-pressure CGM and IGM. The lower-right panel further shows that the radial velocity at the shock front is directed away from the galaxy center ($50$ to $100\,\mathrm{km\,s^{-1}}$). Interestingly, the inner CGM ($\lesssim0.3\,\mathrm{R_{vir}}$) is dominated by net inflow ($-10$ to $-50\,\mathrm{km\,s^{-1}}$), likely fueling future star formation as it reaches the ISM.

\subsubsection{Volume-Weighted Temperature Evolution} \label{subsubsec:temperature evolution}

\begin{figure*}
    \centering
    \includegraphics[width=1\textwidth]{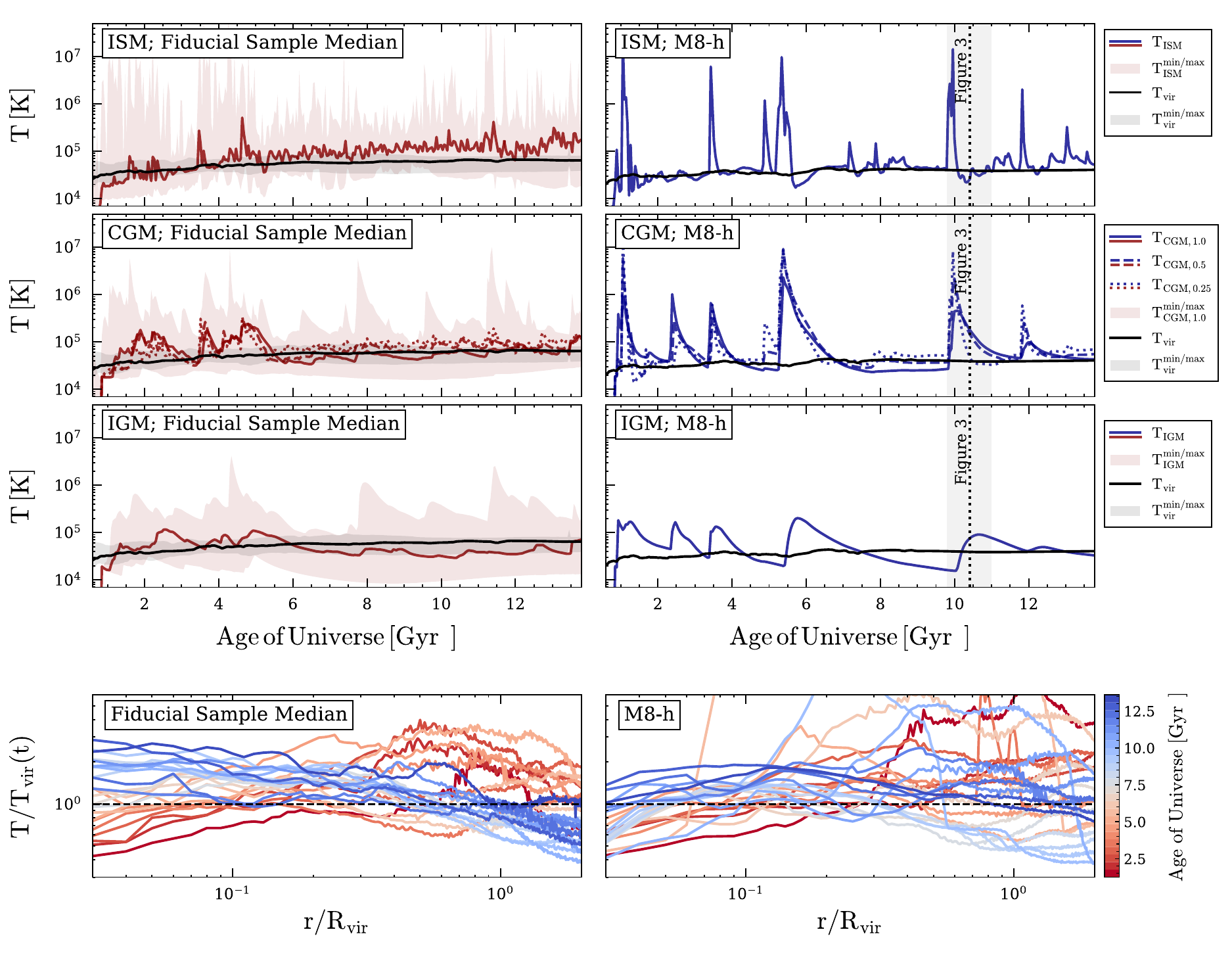} 
    \caption{\textbf{Upper-Top Panel:} The ISM volume-weighted temperature evolution in both the fiducial sample (red; left) and M8-h (blue; right). We also include the full range (minimum to maximum) temperature in the fiducial sample. It is important to note that this isn't the traditional ISM temperature definition. Since the temperature is volume-weighted, it is probing the outflow temperature close to $\mathrm{\sim0.1\,R_{vir}}$ (the outer boundary of our ISM definition). \textbf{Upper-Middle Panel:} The same temperature time evolution, but for the CGM. In addition to the volume-weighted temperature of the entire CGM ($\mathrm{T_{CGM,1.0}}$), we also include the temperature of the inner half ($\mathrm{T_{CGM,0.5}}$) and inner quarter ($\mathrm{T_{CGM,0.25}}$) of the CGM. \textbf{Upper-Bottom Panel:} The volume-weighted IGM ($\mathrm{1-2\,R_{vir}}$) temperature time evolution. In all panels showing the temperature time evolution, we include the fiducial sample median $\mathrm{T_{vir}}$ and M8-h $\mathrm{T_{vir}}$ (solid black). The full range of $\mathrm{T_{vir}}$ in the fiducial sample is shown using the shaded black region. \textbf{Lower Panel:} The temperature profile evolution (normalized by $\mathrm{T_{vir}(t)}$) for both the fiducial sample and M8-h. For the fiducial sample, we calculate the median profile in $\mathrm{\sim0.5\,Gyr}$ bins across all 10 halos. In both the fiducial sample and M8-h, it is clear that by the present-day, the inner CGM is heated to above $\mathrm{T_{vir}}$.   
    \label{fig:temperature_evolution}}
\end{figure*}

We selected this snapshot ($\mathrm{z=0.28}$) specifically to highlight SNe-driven shock propagation on both small scales (near the star-forming disk) and large scales (near the halo boundary). More generally, we expect the CGM properties of our sample to evolve over time, since each galaxy undergoes a unique merger history and resides in a different environment. To understand this evolution and to better place M8-h at this particular snapshot (Figure \ref{fig:temp_slice_two}) in the context of our full galaxy sample, we include the time evolution of the gas temperature in Figure \ref{fig:temperature_evolution}. In particular, we calculate the volume-weighted temperature of the ISM ($r < 0.1\, \mathrm{R_{vir}}$), CGM ($0.1\, \mathrm{R_{vir}} < r <  1\, \mathrm{R_{vir}}$), and IGM ($1\, \mathrm{R_{vir}} < r < 2\, \mathrm{R_{vir}}$). We note that this is \emph{not} the mass-weighted temperature and therefore the ISM temperature will be much higher than typical dense, star-forming gas in the ISM. However, the volume-weighted temperature is better able to capture the hot SNe-driven outflows from the ISM, which is the main focus of this work. 

In the three upper-left panels in Figure~\ref{fig:temperature_evolution}, we show the median fiducial sample temperature evolution in the ISM, CGM, and IGM. In the CGM, we define three different mean temperature regions, averaged over different regions of the halo: (1) $\mathrm{T_{CGM,1.0}}$ ($\mathrm{0.1-1\,R_{vir}}$), (2) $\mathrm{T_{CGM,0.5}}$ ($\mathrm{0.1-0.5\,R_{vir}}$), and (3) $\mathrm{T_{CGM,0.25}}$ ($\mathrm{0.1-0.25\,R_{vir}}$). 
The median profiles are useful for understanding the typical temperature behavior; however, in order to see the temporal behavior of the temperature in a single halo, we also show, in the upper-right panels of Figure~\ref{fig:temperature_evolution}, the temperature evolution of the M8-h halo as the solid blue lines and its $\mathrm{T_{vir}(t)}$ as the dashed black lines. 
 
During the Figure~\ref{fig:temp_slice_two} outflow, the M8-h ISM temperature in Figure~\ref{fig:temperature_evolution} shows two peaks just before ${\sim}10\,\mathrm{Gyr}$, the second reaching $\sim10^7\,\mathrm{K}$, before declining 
rapidly. This SNe event then propagates outward into the CGM and with a time delay of ${\sim}0.6\,\mathrm{Gyr}$, corresponding to a shock velocity of $\sim98\,\mathrm{km\,s^{-1}}$ (assuming 
$\mathrm{R}_\mathrm{vir} \sim 60\,\mathrm{kpc}$), consistent with the radial velocities shown in Figure~\ref{fig:temp_slice_two}. The IGM temperature peaks precisely at the vertical line, reflecting the coincidence of its inner boundary with the outer shock at $\sim1\,\mathrm{R}_\mathrm{vir}$.

We find that the ISM temperature in M8-h is sustained near $\mathrm{\sim T_{vir}(t)}$. We note again that this is the volume-weighted temperature, which is dominated by volume-filling gas near the outer ISM boundary ($\sim0.1\,\mathrm{R_{vir}}$). The mass-weighted temperature is below $\sim10^4\,\mathrm{K}$, since most of the mass is in the cold phase. 

The interpretation of the temperature being sustained at $\mathrm{\sim T_{vir}}$ is less straightforward in the CGM. Although the CGM temperature reaches $\sim10^6\,\mathrm{K}$ during strong SNe bursts (Figure \ref{fig:temp_slice_two} and Figure \ref{fig:temperature_evolution}), the CGM gradually cools (adiabatically and radiatively) just below $\mathrm{T_{vir}(t)}$. This is most clear during the time period $\mathrm{t=[7,10]\,Gyr}$ in M8-h, where the volume-weighted CGM temperature falls to $\sim3\times10^4\,\mathrm{K}$. This apparent cooling is, however, a consequence of the strong (negative) radial temperature dependence at late times \citep[a result which is seen generically in CDM halos, e.g.][]{Loken2002, Vikhlinin2005, Vikhlinin2006, Pratt2007}. From the fiducial sample and M8-h temperature profile in the lower panels, we find that the inner CGM is sustained above $\mathrm{\sim T_{vir}(t)}$ at $\mathrm{z\sim0}$, while the outer CGM is $\mathrm{\lesssim T_{vir}(t)}$. Since volume weighting is biased toward the outer CGM volume, $\mathrm{T_{CGM,1.0}}$ (solid blue) will underestimate the temperature of the inner halo, motivating the inclusion of $\mathrm{T_{CGM,0.5}}$ (dashed blue) and $\mathrm{T_{CGM,0.25}}$ (dotted blue) to better represent the \emph{inner} CGM temperature.  The volume-averaged $\mathrm{T_{CGM,1.0}}$ is often below $\mathrm{T_{vir}}$, yet the inner $\mathrm{\sim45\%}$ ($\mathrm{T_{CGM,0.5}}$) and $\mathrm{\sim15\%}$ ($\mathrm{T_{CGM,0.25}}$) of the CGM are sustained at $\mathrm{T_{vir}}$ even between large SNe outbursts. The M8-h IGM similarly exceeds $\mathrm{T_{vir}}$ for extended periods, falling below only after $\mathrm{\sim 1\,Gyr}$ or until the next SNe-induced outflow. The CGM and IGM also cool more slowly than the ISM following a feedback event, likely reflecting the longer adiabatic and radiative cooling times at larger radii. 

We find that M8-h undergoes $\sim5$ major SNe-induced outbursts, marked by a sudden temperature peak well above the expected $\mathrm{T_{vir}}$ from merger-driven shocks. These temperature enhancements are particularly prominent at early times ($\lesssim5-6\,\mathrm{Gyr}$), where there are $\sim4$ large increases in CGM temperature, and $\sim2$ since that time. Interestingly, at $\sim2.5\,\mathrm{Gyr}$ there is a sudden temperature increase in the CGM and IGM but \emph{not} in the ISM. Upon further inspection, we find that this temperature anomaly is the result of a satellite galaxy undergoing feedback within the CGM of the main halo, M8-h. However, we do not expect this satellite SNe feedback behavior to be the main temperature driver of the central halo. 

Finally, we return to the full dwarf galaxy sample in the three upper-left panels of Figure \ref{fig:temperature_evolution}. We find that the overall picture is similar to the M8-h temperature evolution. In the fiducial sample ISM, ongoing star-formation drives frequent (but relatively small) temperature fluctuations, punctuated by occasional large SNe outbursts that raise the temperature to $\sim10^6-10^7\,\mathrm{K}$. The cadence of these events is difficult to discern in Figure \ref{fig:temperature_evolution}, since we show only the sample median and full range. However, upon individual inspection (not shown), the temperature evolution of each individual halo is similar to  M8-h.

In the CGM (middle panel), the sample also shows an enhanced volume-weighted temperature before $\mathrm{t\sim}\,6\,\mathrm{Gyr}$. The initial CGM temperature is low ($\mathrm{<2\,Gyr}$), likely due to efficient cooling during this epoch. However, there is a sudden temperature increase between $\mathrm{\sim2\,Gyr}$ and $\mathrm{\sim6\,Gyr}$ before declining after $\sim5-6\,\mathrm{Gyr}$. A similar increase in $\mathrm{T_{IGM}}$ is also seen (lower panel) during this time period. We discuss the origin of this transition in more detail in Section \ref{subsec:the physical picture}.  

\subsubsection{Temperature Profile Evolution} \label{subsubsec:temperature profile}

The lower panels in Figure \ref{fig:temperature_evolution} show the time evolution of the temperature profile of the fiducial sample (lower-left) and M8-h (lower-right). In particular, for M8-h, we calculate the median temperature profile normalized by $\mathrm{T_{vir}(t)}$ and $\mathrm{R_{vir}(t)}$ in a bin with a $\mathrm{\sim0.5\,Gyr}$ width. The calculation for the fiducial sample is the same, except we take an additional median of the ten temperature profiles in each bin. 

At early times in the fiducial sample (red; lower-left), the temperature profile increases from the center until $\mathrm{r/R_{vir}\sim0.5-0.75}$, where the volume-weighted temperature begins to decrease. In fact, the outer CGMs of these early halos were quite hot. The volume-weighted temperature reached a factor of $3-4$ more than $\mathrm{T_{vir}}$. This is consistent with $\mathrm{T_{CGM,1.0}}$, which is clearly elevated in both the fiducial sample and M8-h before $\mathrm{\sim5-6\,Gyr}$ and often exceeds $\mathrm{T_{CGM,0.5}}$ and $\mathrm{T_{CGM,0.25}}$ during this period, indicating that the outer CGM ($\mathrm{\gtrsim0.5\,R_{vir}}$) was preferentially hotter at early times.

By the present-day (blue lines; lower left), the temperature profile inverts: the inner halo ($\mathrm{r/R_{vir}\lesssim0.5}$) is sustained above $\mathrm{T_{vir}}$ while the outer halo falls below it, reflected in $\mathrm{T_{CGM,1.0}}$ dropping below $\mathrm{T_{CGM,0.5}}$ and $\mathrm{T_{CGM,0.25}}$ after $\mathrm{\sim5-6\,Gyr}$. We find that the M8-h temperature profile evolution (lower-right) is noisier but consistent with the fiducial sample. 

\subsection{Flow of Energy} \label{subsec:flow of mass and energy}

We have thus far demonstrated that the CGM of our low-mass halo sample is sustained at $\sim \mathrm{T_{vir}(t)}$ through SNe-induced shock heating, despite being at the peak of the cooling curve. In \S\ref{sec:intro}, we suggested that, compared to a Lagrangian approach with fixed mass resolution, Eulerian-based codes (e.g., \textsc{enzo}) may better capture the propagation of high-specific-energy winds from the ISM into the CGM and beyond \citep[see][for more discussion]{Smith_2024a}. In this section, we quantify both the specific energy of the outflows in the ISM boundary and more generally, the flow of energy between the ISM, CGM, and IGM. 

\subsubsection{Specific Energy of Flows} \label{subsubsec:specific energy}

\begin{figure*}
    \centering
    \includegraphics[width=1\textwidth]{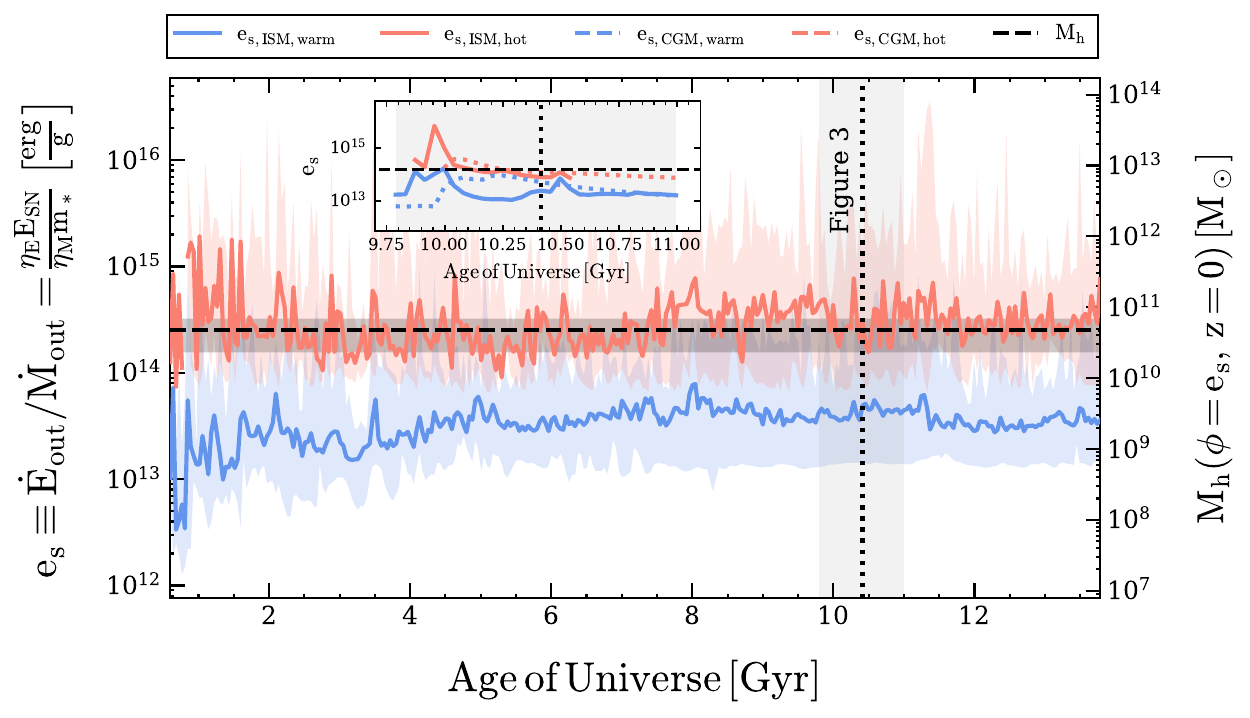} 
    \caption{The specific energy ($\mathrm{e_s}$) time evolution for the warm gas (blue; $\mathrm{T \leq 3 \times 10^5 K}$) and hot gas (red; $\mathrm{T > 3 \times 10^5 K}$). The solid line is the sample median and the shaded region is bounded by the minimum and maximum value. We also use a simple scaling relation (Equation \ref{equation:se_halo}; see \cite{Li_Bryan_2020}) to determine whether the outflow can escape the halo. This can be compared to the black dashed line, representing the median halo mass at z = 0 (shaded region is the minimum and maximum). This highlights the ability of the hot outflow to escape the halo and heat the surrounding medium, whereas the material within the warm outflow will likely return to the disk. The M8-h ISM and CGM $\mathrm{e_s}$ evolution near z = 0.28 (Figure \ref{fig:temp_slice_two}) is also included for reference (inset panel). We also include the scaling relation using the z = 0 M8-h halo mass (horizontal black dashed line).   
\label{fig:se_evolution}}
\end{figure*}

To quantify the specific energy of our SNe-induced outflows, we first separate the gas by temperature. We follow \cite{Li_Bryan_2020}, where we define warm gas as $\mathrm{T \leq 3 \times 10^5\,K}$ and hot gas as $\mathrm{T > 3 \times 10^5\,K}$. This warm gas definition will trace gas at $\mathrm{T_{vir}}$ and below. In comparison, the hot gas definition will only include gas well above $\mathrm{T_{vir}}$, which will be the result of SNe-driven feedback. We additionally require the gas to be outflowing ($\mathrm{v_r > 0\,km\,s^{-1}}$), where $\mathrm{v_r}$ is the bulk-velocity–corrected radial velocity relative to the halo center of mass. Since we are interested here in gas flowing out of the ISM, we further restrict our selection to gas at the ISM boundary, defined as a shell between $\mathrm{0.1-0.2\,R_{vir}}$.

The specific energy is calculated from the sum of the kinetic and thermal energy components of gas cells that satisfy these criteria, separated by temperature. The specific energy (Equation \ref{equation:specific_energy}) can be interpreted as the ratio of the energy outflow rate to the mass outflow rate, or equivalently the ratio of the energy and mass loading factors (since $\mathrm{E_{SN}/m_*}$ is constant in our simulation).

We calculate the median specific energy for both the warm and hot components at each simulation output. The results are shown in Figure \ref{fig:se_evolution}. The solid red line represents the median specific energy of the hot component, while the shaded red region is bound by the minimum and maximum value across the sample. The solid blue line and shaded blue region show the same quantities for the warm component.

Following \cite{Li_Bryan_2020}, we compare the wind specific energy to the gravitational potential of the dark matter halo to determine whether the outflow can escape. This comparison uses the scaling relation
\begin{gather} \label{equation:se_halo}
\mathrm{\phi = \frac{1}{2} (620\ km\ s^{-1})^2 \left(\frac{M_{halo}}{10^{12}M_\odot}\right)^{2/3}}
\end{gather}
A dashed black line shows the median halo potential at $\mathrm{z=0}$, with the shaded black region bound by the minimum and maximum $\mathrm{M_{vir}}$ in the sample.

We find that, for the sample median, the specific energy of the hot phase generally matches or exceeds the energy required to escape the halo potential ($\mathrm{M_h(\phi=e_s)}$). In contrast, the warm phase remains below the escape threshold. As a result, the warm gas will likely recycle within the CGM and eventually return to the ISM as fuel for future star-formation.
The specific energy of the hot phase is roughly an order of magnitude larger than that of the warm phase; later we will show that most of the outflow energy is carried by this phase.

We also include a vertical line at $\mathrm{z=0.28}$ corresponding to the M8-h snapshot from Figure \ref{fig:temp_slice_two}. The inset panel shows the M8-h specific energy evolution at both the ISM boundary (solid lines) and CGM boundary (dashed lines). 
The specific energy evolution of M8-h at $\mathrm{z = 0.28}$ is consistent with the M8-h temperature evolution (Figure \ref{fig:temperature_evolution}). There is a sharp increase in the ISM-boundary specific energy approximately $\sim500\,\mathrm{Myr}$ before the vertical line at $\mathrm{z=0.28}$. At the CGM boundary in the Figure \ref{fig:se_evolution} inset panel, the increase in specific energy is delayed and lower compared to the ISM, reflecting the outward propagation of the SNe-induced wind. In addition, the separation between the hot and warm phase is smaller in the CGM than at the ISM boundary. Nevertheless, the specific energy of the warm phase remains below $\mathrm{M_h(\phi=e_s)}$, indicating that this gas is likely to recycle within the halo. In contrast, the hot phase can escape the halo, if we assume the scaling relation in Equation \ref{equation:se_halo}. 

\subsubsection{Energy Flow Rate} \label{subsubsec:energy flow rate}

\begin{figure*}
    \centering
    \includegraphics[width=1\textwidth]{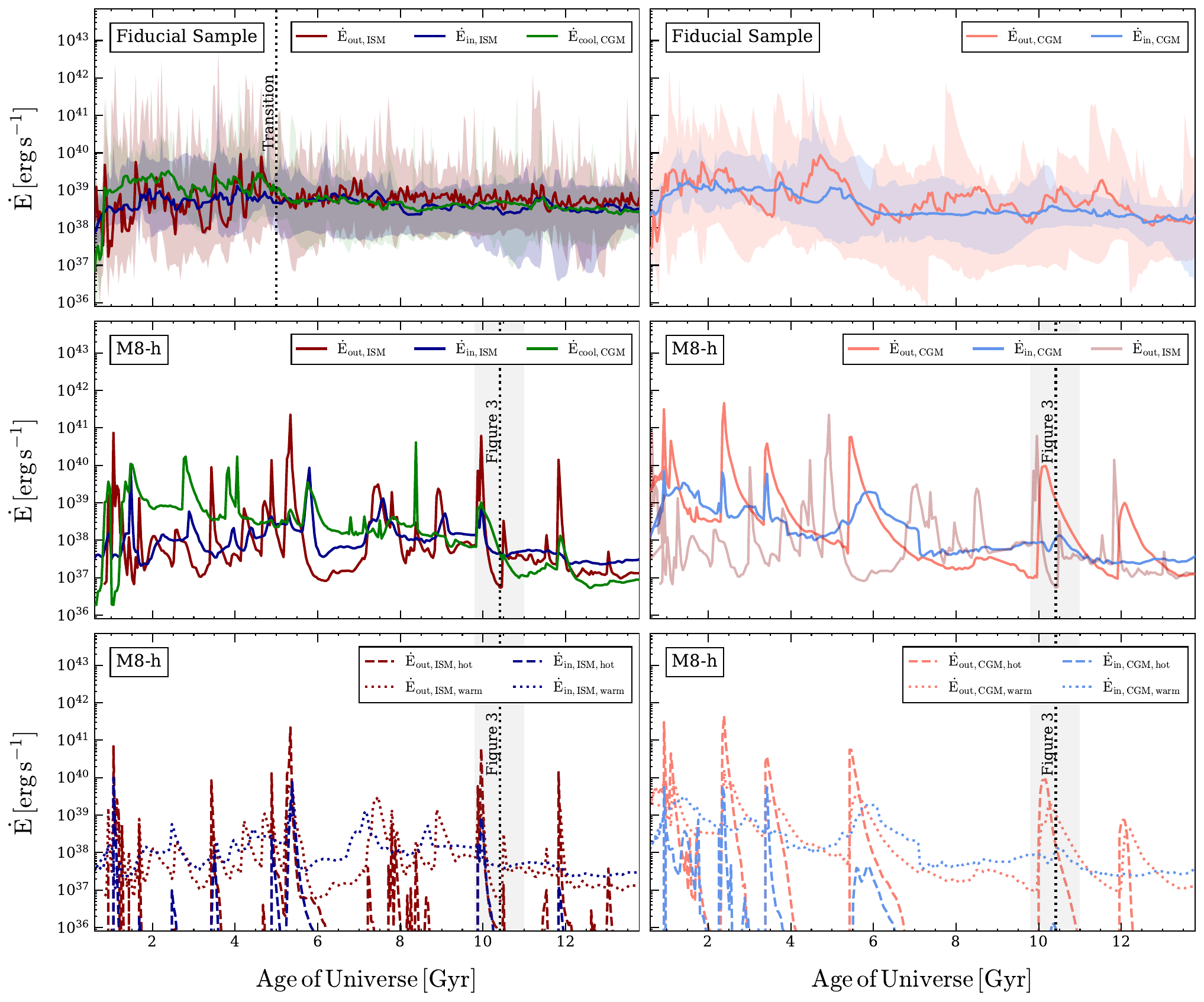} 
    \caption{\textbf{Upper Panels:} The median and full range (minimum to maximum) energy flow rate (Equation \ref{equation:energy flow}) in the fiducial sample at the ISM-scale (left) and CGM-scale (right). We also compare the ISM energy outflow rate (red) to the CGM cooling rate (green) in the upper right panel. \textbf{Middle Panels:} The same energy flow rate, but for M8-h. We also repeat $\mathrm{\dot{E}_{out,ISM}}$ on the right for better comparison with $\mathrm{\dot{E}_{out,CGM}}$. \textbf{Lower Panels:} The M8-h energy flow rate evolution as a function of the warm phase ($\mathrm{<3\times10^5\,K}$) and hot phase ($\mathrm{>3\times10^5\,K}$).       
\label{fig:energy_flow_evolution}}
\end{figure*}

Having quantified the specific energy of the flows, we now calculate the actual energy outflow and inflow rates at both the ISM and CGM boundaries, as well as the CGM cooling rate. These five rates should provide an accounting of how the energy content of the CGM is shaped. Similar to the definition presented in \cite{Pandya_2021}, we calculate the energy outflow rate using
\begin{gather} \label{equation:energy flow}
\mathrm{\dot{E} = \frac{A \sum E_i v_{r,i} (>0)}{\sum V_i}}
\end{gather}
where $\mathrm{A}$ is the surface area of the ISM or CGM boundary (at $\mathrm{0.15\,R_{vir}}$ and $\mathrm{1\,R_{vir}}$, respectively), and $\mathrm{E_i}$, $\mathrm{v_{r,i}}$, and $\mathrm{V_i}$ are the energy (kinetic plus thermal component), radial velocity, and volume of each cell, respectively.
We do not impose a lower velocity cutoff for winds; instead, we define outflow as $\mathrm{v_r > 0\,km\,s^{-1}}$ and inflow as $\mathrm{v_r < 0\,km\,s^{-1}}$. The inflow rate is calculated similarly but with the opposite radial velocity ($\mathrm{v_{r,i} < 0}$). 

We also compute the CGM cooling rate, $\mathrm{\dot{E}_{cool,CGM} = \sum \frac{E_i}{t_{cool,i}}}$, where $\mathrm{t_{cool}}$ is tracked for each gas cell in the simulation using the \textsc{grackle} cooling tables described in \S\ref{sec:methodology}. $\mathrm{E_i}$ refers to only the thermal energy of each \textsc{enzo} gas cell. The cooling time is the ratio between the thermal energy content and the radiative energy loss rate, $\mathrm{t_{cool} \equiv \frac{\rho \varepsilon}{\mathscr{C}}}$
where $\mathrm{\varepsilon = \frac{1}{\gamma - 1} \frac{k_B T}{\mu m_p}}$ and $\mathscr{C}$ is the volumetric cooling rate \citep{Mo_2010}. To avoid artificially large cooling rates during mergers, we remove very dense gas ($\mathrm{n>0.1\,cm^{-3}}$), which typically corresponds to star-forming gas within satellite ISM regions. 

The results are shown in Figure \ref{fig:energy_flow_evolution}. We begin with the median of the fiducial sample (top two plots) before exploring the time variation for a single halo (bottom four plots). In the upper-left panel, we compare the fiducial sample median $\mathrm{\dot{E}_{out,ISM}}$ (red; energy flowing from ISM $\rightarrow$ CGM) with $\mathrm{\dot{E}_{cool,CGM}}$ (green). If star formation heating balanced radiative cooling one might expect these two quantities to roughly balance in order to regulate the CGM. We also include $\mathrm{\dot{E}_{in,ISM}}$ (blue; CGM $\rightarrow$ ISM). The shaded region for each $\mathrm{\dot{E}}$ term is bounded by the minimum and maximum fiducial sample value. At early times ($\mathrm{\lesssim5\,Gyr}$), $\mathrm{\dot{E}_{out,ISM}}$ is characteristically bursty, with individual outbursts exceeding $\mathrm{\sim10^{42}\,erg\,s^{-1}}$ within a single timestep before immediately declining. In comparison, after $\mathrm{\sim5\,Gyr}$, the outbursts are smaller and more steady, with $\mathrm{\dot{E}_{out,ISM}}$ driven by lower-energy SNe-induced outflows that rarely exceed $\mathrm{\sim10^{42}\,erg\,s^{-1}}$.

Before $\mathrm{\sim5\,Gyr}$, the median $\mathrm{\dot{E}_{cool,CGM}}$ exceeds $\mathrm{\dot{E}_{out,ISM}}$ for a larger fraction of time. However, as we show in \S\ref{subsubsec:cumulative energy}, infrequent yet intense SNe outbursts are sufficient to dominate the cumulative energy balance within the first $\mathrm{\sim5\,Gyr}$. After $\mathrm{\sim5\,Gyr}$, there is a gradual transition and $\mathrm{\dot{E}_{out,ISM}}$ becomes generally dominant over $\mathrm{\dot{E}_{cool,CGM}}$, with the median $\mathrm{\dot{E}_{in,ISM}}$ becoming comparable to $\mathrm{\dot{E}_{cool,CGM}}$. We mark this transition with a vertical dashed line.

The upper-right panel of Figure \ref{fig:energy_flow_evolution}  shows the median $\mathrm{\dot{E}_{out,CGM}}$ (CGM $\rightarrow$ IGM) and $\mathrm{\dot{E}_{in,CGM}}$ (IGM $\rightarrow$ CGM). These flows also exhibit a transition near $\sim5\,\mathrm{Gyr}$, where both decrease by nearly an order of magnitude. $\mathrm{\dot{E}_{out,CGM}}$ also shows a characteristic sawtooth pattern (seen in the minimum to maximum range), with rapid increases followed by gradual declines. This is similar to what was seen in the volume-weighted temperature evolution (Figure~\ref{fig:temperature_evolution}). 

We now turn, in the middle panels of Figure \ref{fig:energy_flow_evolution}, to the energy flows for a single halo (M8-h). 
This shows some features similar to the median profiles; for example, the middle-left panel shows a gradual decline in $\mathrm{\dot{E}_{cool,CGM}}$ over cosmic time such that, after $\mathrm{\sim5\,Gyr}$, $\mathrm{\dot{E}_{cool,CGM}}$ is more comparable to $\mathrm{\dot{E}_{in,ISM}}$.  
However, it also allows us to explore the causal connection between the energy flows, which may be obscured in the sample median. In particular, there are sharp jumps superimposed on the gradual decline of the cooling rate that occur during or shortly after bursts in $\mathrm{\dot{E}_{out,ISM}}$.  

The middle-right panel shows the M8-h energy flows at the CGM/IGM boundary. We also repeat $\mathrm{\dot{E}_{out,ISM}}$ from the left panel to demonstrate the impact of strong star formation events: large energy flows out of the ISM are typically followed, after a $\sim 100$ Myr delay, by an abrupt increase in the CGM energy outflow rate, due to the CGM over-pressurization (as seen in the enhanced $\mathrm{T_{CGM}}$ values seen in Figure~\ref{fig:temperature_evolution}). The enhanced CGM outflow rate slowly declines over $\sim 1$ Gyr as the CGM returns to its pre-outburst state. Some outflow events are followed by an increase in $\mathrm{\dot{E}_{in,CGM}}$, although an integrated accounting confirms that energy outflow dominates inflow over the long term. This short-term enhancement of the energy inflow from the IGM is likely the result of hot gas returning to its equilibrium position after the shock passage. This decelerated and reversed material is also seen at the ISM boundary and in the mass inflow rate at the ISM and CGM boundary, which we discuss in \S\ref{subsec:flow of mass}.

In the lower panels, we repeat the M8-h $\mathrm{\dot{E}}$ evolution, but now separated by gas phase. In the lower-left panel, the warm phase ($\mathrm{<3\times10^5\,K}$) is nearly always present in both the ISM inflow ($\mathrm{\dot{E}_{in,ISM,warm}}$) and outflow ($\mathrm{\dot{E}_{out,ISM,warm}}$). In comparison, the hot phase ($\mathrm{>3\times10^5\,K}$) outflow ($\mathrm{\dot{E}_{out,ISM,hot}}$) is only present during an episode of peak SNe-outflow (i.e., $\mathrm{\dot{E}_{out,ISM}>10^{38}\,erg\,s^{-1}}$; middle-left panel). At the halo-scale in the lower-right panel, the picture is similar: the warm phase is nearly always present in the inflow and outflow, but the hot phase is dominant during an episode of strong SNe feedback. There is also a sharp increase in $\mathrm{\dot{E}_{in,CGM,hot}}$ during an increase in $\mathrm{\dot{E}_{out,CGM,hot}}$, further demonstrating the deceleration of shocked material. This can also be seen in $\mathrm{\dot{E}_{in,ISM}}$, but the peak is very narrow since the outflow occurs within one timestep.     

To quantify the total energy contribution from each phase, we compute the cumulative ratio $\mathrm{R \equiv\Sigma\,\dot{E}_{hot}/\Sigma\,\dot{E}_{warm}}$ for the outflow and inflow at both boundaries since $\mathrm{\sim0.5\,Gyr}$. At the ISM boundary, we find $\mathrm{R_{ISM,out}\sim9.3}$ and $\mathrm{R_{ISM,in}\sim0.090}$ for the fiducial sample median ($\mathrm{R_{ISM,out}\sim5.4}$, $\mathrm{R_{ISM,in}\sim0.57}$ for M8-h), confirming that the hot phase dominates the outflow and the warm phase dominates the inflow for every halo in the sample. At the CGM boundary, the picture is similar: $\mathrm{R_{CGM,out}\sim3.4}$ and $\mathrm{R_{CGM,in}\sim0.030}$ for the fiducial sample ($\mathrm{R_{CGM,out}\sim6.3}$, $\mathrm{R_{CGM,in}\sim0.28}$ for M8-h). This hot-phase energy dominance is particularly striking given that the hot phase is present in outflows for only $\mathrm{52-75\%}$ of snapshots, compared to $\mathrm{96-100\%}$ for the warm phase, at both the ISM and CGM boundaries.        

\subsubsection{Cumulative Energy} \label{subsubsec:cumulative energy}

\begin{figure*}
    \centering
    \includegraphics[width=1\textwidth]{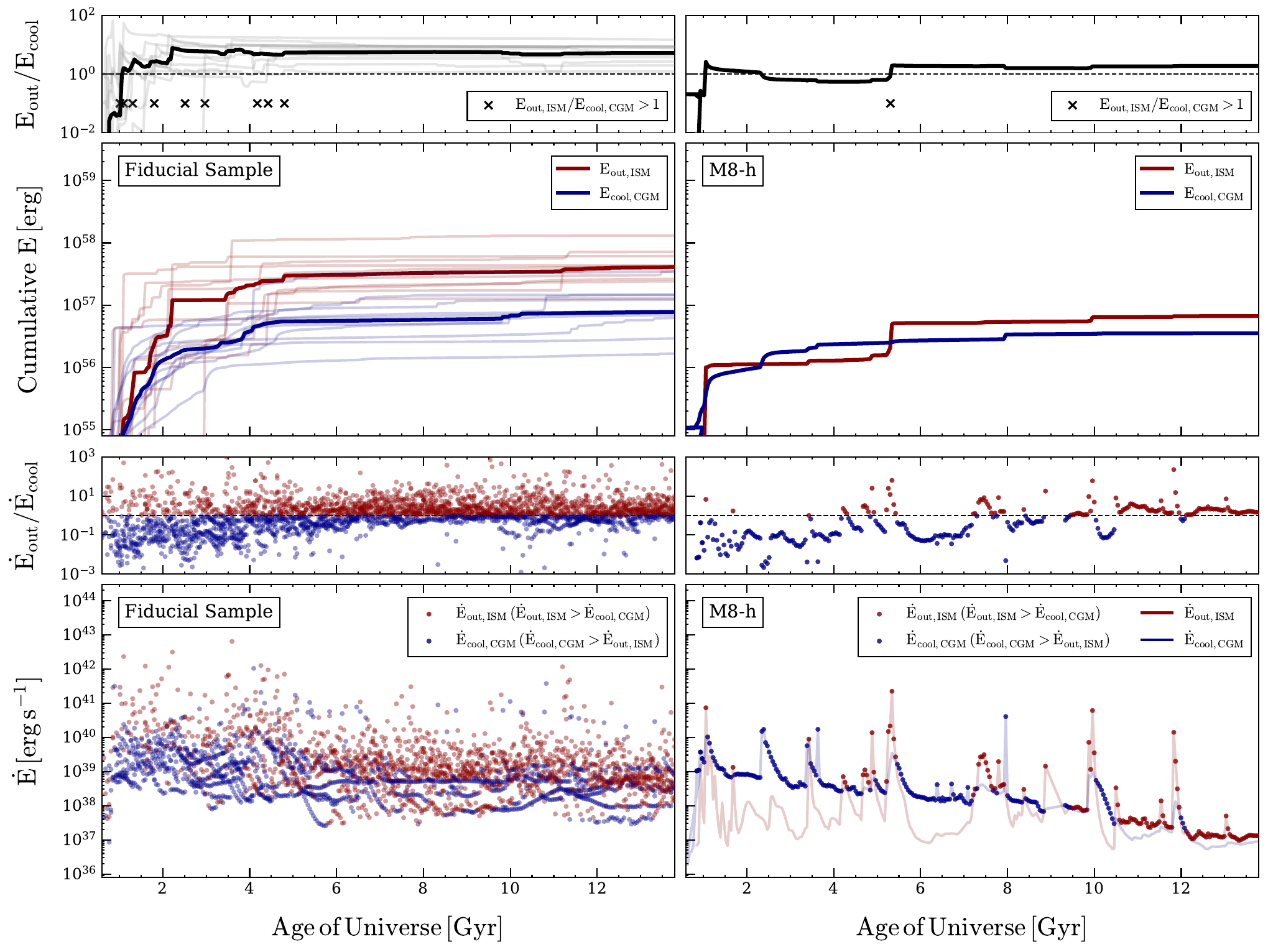} 
    \caption{\textbf{Upper Left:} The fiducial sample cumulative $\mathrm{E_{out,ISM}}$ (red)  and $\mathrm{E_{cool,CGM}}$ (blue). We include both the sample median and individual halo time evolution. In addition, the upper panel shows the cumulative ratio of both the sample median and individual halo time evolution. The \lq x' marker denotes the most recent timestep when an individual halo has its ratio $\mathrm{E_{out,ISM}}/\mathrm{E_{CGM,cool}}$ cross a value of 1. Importantly, by $\mathrm{\sim5\,Gyr}$, the cumulative SNe outflow rate is always dominant over the cumulative CGM cooling rate. \textbf{Lower Left:} The fiducial sample $\mathrm{\dot{E}}$ time evolution of either $\mathrm{\dot{E}_{out,ISM}}$ or $\mathrm{\dot{E}_{cool,CGM}}$, depending on which term is dominant at the given timestep. The upper panel is the ratio between $\mathrm{\dot{E}_{out,ISM}}$ and $\mathrm{\dot{E}_{cool,CGM}}$.  \textbf{Upper Right:} Similar to the upper left panel, we include the M8-h cumulative $\mathrm{E_{out,ISM}}$  and $\mathrm{E_{out,ISM}}$ time evolution and their ratio in the upper panel. \textbf{Lower Right:} Similar to the lower left panel, but for only M8-h. We also include the continuous $\mathrm{\dot{E}_{out,ISM}}$ and $\mathrm{\dot{E}_{cool,CGM}}$ time evolution (same as Figure~\ref{fig:energy_flow_evolution}) to better illustrate when one $\mathrm{\dot{E}}$ term is dominant over the other $\mathrm{\dot{E}}$ term.               
\label{fig:cumulative_energy}}
\end{figure*}

As shown in Figure \ref{fig:energy_flow_evolution}, $\mathrm{\dot{E}_{cool,CGM}}$ exceeds $\mathrm{\dot{E}_{out,ISM}}$ for a larger fraction of time at early epochs ($\mathrm{\lesssim5\,Gyr}$), with the opposite generally being true at late times ($\mathrm{\gtrsim5\,Gyr}$). However, a higher frequency of cooling dominance does not necessarily mean that the \emph{cumulative} energy is cooling-dominated — occasional SNe outbursts can drive $\mathrm{\dot{E}_{out,ISM}}$ far above $\mathrm{\dot{E}_{cool,CGM}}$, potentially tipping the cumulative balance. To address this, we show the time-integrated $\mathrm{E_{out,ISM}}$ and $\mathrm{E_{cool,CGM}}$ in Figure \ref{fig:cumulative_energy}.

The upper-left panel in Figure~\ref{fig:cumulative_energy} shows the cumulative energy for each halo in the fiducial sample and their median for both $\mathrm{E_{out,ISM}}$ (red) and $\mathrm{E_{cool,CGM}}$ (blue), while the panel immediately above shows their ratio $\mathrm{E_{out,ISM}}/\mathrm{E_{cool,CGM}}$\footnote{We caution that the integrated quantities we calculate are only estimates due to the finite time between snapshots. The integrated ISM energy flow is likely to be the most impacted by this due to its very bursty nature.}. We mark with an `x' (placed at an arbitrary ratio value of 0.1) the most recent timestep at which the ratio exceeded unity, indicating the timestep when cumulative $\mathrm{E_{out,ISM}}$ was last dominant. The upper-right panel shows the same cumulative energy and its ratio but for M8-h.

We find that the cumulative $\mathrm{E_{out,ISM}}$ exceeds the cumulative $\mathrm{E_{cool,CGM}}$ within the first $\mathrm{\sim 5\text{--}6\,Gyr}$ for both the fiducial sample and M8-h. By $\mathrm{z\sim0}$, the median cumulative $\mathrm{E_{out,ISM}}$ ($\mathrm{\sim4\times10^{57}\,erg}$) is a factor of $\mathrm{\sim5}$ larger than the median cumulative $\mathrm{E_{cool,CGM}}$ ($\mathrm{\sim8\times10^{56}\,erg}$), though with significant halo-to-halo scatter: some halos reach ratios exceeding $\sim10$ by $\mathrm{z\sim0}$, while others remain near unity. Similarly, the epoch at which the ratio first exceeds unity ranges from as early as $\mathrm{1-2\,Gyr}$ to as late as $\mathrm{\sim5.5\,Gyr}$.

The lower-left panel of Figure \ref{fig:cumulative_energy} resembles the upper-left panel of Figure \ref{fig:energy_flow_evolution}, but only shows $\mathrm{\dot{E}_{out,ISM}}$ or $\mathrm{\dot{E}_{cool,CGM}}$ if it is the dominant $\mathrm{\dot{E}}$ term at each timestep. The prevalence of blue points before $\mathrm{\sim5\,Gyr}$ confirms that $\mathrm{\dot{E}_{cool,CGM}}>\mathrm{\dot{E}_{out,ISM}}$ more frequently at early times, while the shift to red after $\mathrm{\sim5\,Gyr}$ reflects the transition to $\mathrm{\dot{E}_{out,ISM}}$ dominance, reinforced by the $\mathrm{E_{out,ISM}}/\mathrm{E_{cool,CGM}}$ ratio in the panel immediately above. This early CGM cooling dominance arises because $\mathrm{\dot{E}_{out,ISM}}$ drops an order of magnitude or more between infrequent SNe outbursts, whereas $\mathrm{\dot{E}_{cool,CGM}}$ remains relatively stable. During outbursts, however, $\mathrm{\dot{E}_{out,ISM}}$ can exceed $\mathrm{\dot{E}_{cool,CGM}}$ by an order of magnitude or more, and it is these rare peaks, not the time-averaged behavior, that tip the cumulative energy balance within the first $\mathrm{\sim5\,Gyr}$.

The lower-right panel shows the same quantities for M8-h, with full time series (solid lines, see Figure~\ref{fig:energy_flow_evolution}) overlaid on the dominant-term scatter points. Consistent with the fiducial sample, only $2-3$ early SNe events, each exceeding the CGM cooling rate by a factor of $\sim10-100$, suffice to offset the cumulative balance.

\subsection{Flow of Mass} \label{subsec:flow of mass}

We next track the flow of mass in our sample. As discussed in \S\ref{sec:intro}, the flow of energy and mass are complementary. For example, high resolution tall-box simulations find that the hot phase carries only a small fraction of the total mass but most of the energy (\S\ref{subsec:flow of mass and energy}). In addition, the flow of mass measured as different radii can indicate how much mass is \lq swept' up in outflows or how much gas is prevented from entering the halo. 

\subsubsection{Mass Flow Rate} \label{subsec:mass flow rate}

\begin{figure*}
    \centering
    \includegraphics[width=1\textwidth]{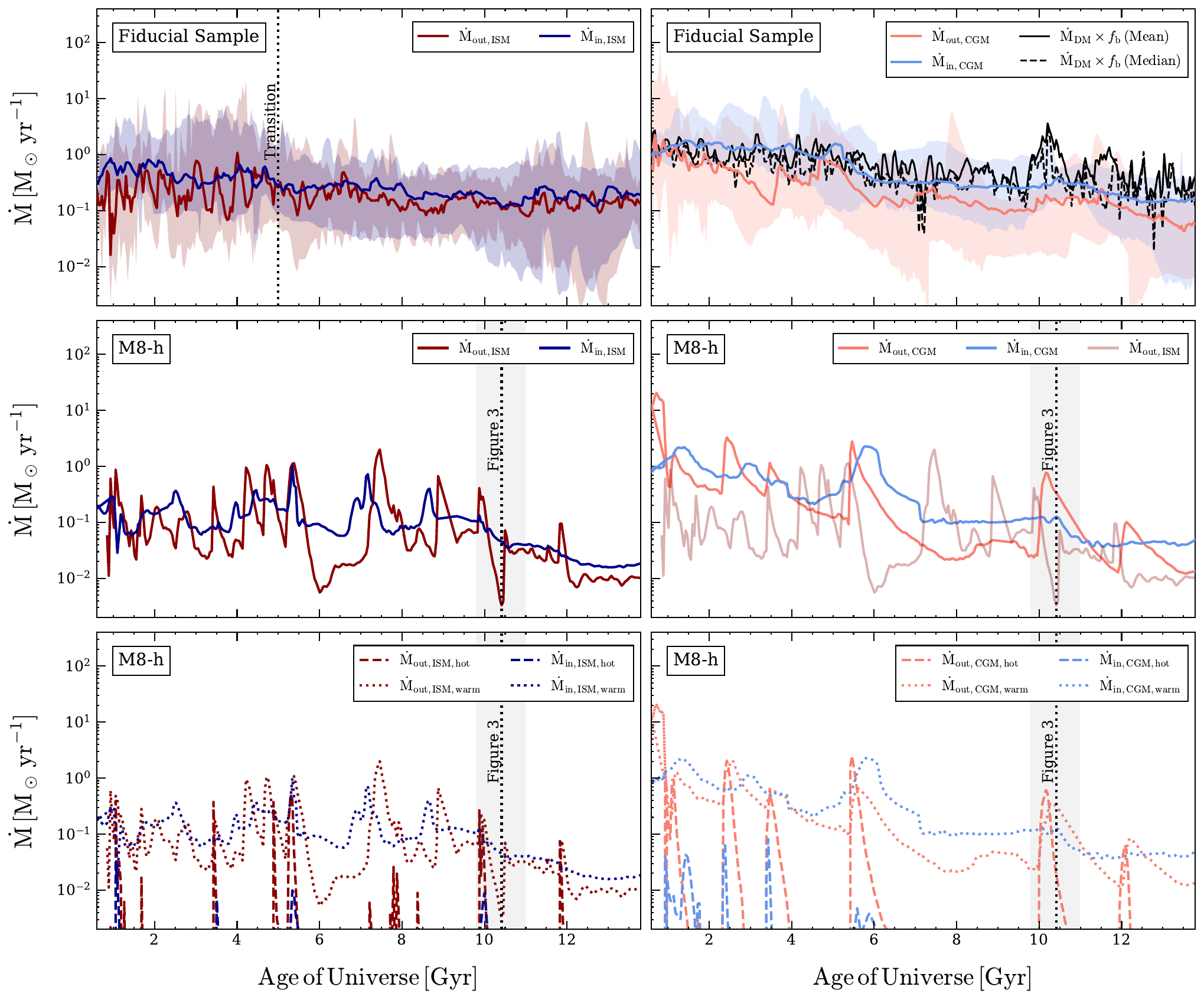} 
    \caption{\textbf{Upper Panels:} The median and full range (minimum to maximum) mass flow rate (Equation \ref{equation:energy flow}, where $\mathrm{E_i}$ is replaced with $\mathrm{M_i}$) in the fiducial sample at the ISM-scale (left) and CGM-scale (right). We also compare the CGM mass inflow rate (red) to the expected baryon accretion rate (black) in the upper left panel. \textbf{Middle Panels:} M8-h $\mathrm{\dot{M}}$ evolution, where we repeat $\mathrm{\dot{M}_{out,ISM}}$ on the right panel for comparison. \textbf{Lower Panels:} M8-h $\mathrm{\dot{M}}$ partitioned into the warm phase ($\mathrm{<3\times10^4\,K}$) and hot phase ($\mathrm{>3\times10^4\,K}$).  
}\label{fig:mass_flow_evolution}
\end{figure*}

To measure the mass flow rates, we use the same definition as the energy flow rate (Equation \ref{equation:energy flow}), but with $\mathrm{M_i}$, the cell mass, instead of $\mathrm{E_i}$. The result is shown in Figure \ref{fig:mass_flow_evolution} (with a format analogous to Figure \ref{fig:energy_flow_evolution}).

In general, the behavior of the mass flow evolution is similar to the energy flow (Figure \ref{fig:energy_flow_evolution}). In the upper-left panel, we see large variations in the CGM to ISM mass flow, with $\mathrm{\dot{M}_{out,ISM}}\sim0.1-1\,\mathrm{M_\odot\,yr^{-1}}$ for $\mathrm{t \lesssim5\,Gyr}$ before lowering to a more steady value of $\mathrm{\sim0.1\,M_\odot\,yr^{-1}}$ after $\mathrm{\sim5\,Gyr}$. In comparison, $\mathrm{\dot{M}_{in,ISM}}$ remains fairly steady, with a slow decline, and 
is nearly always higher than $\mathrm{\dot{M}_{out,ISM}}$ (although again we note that the comparison of medians should be interpreted with care). 

A similar behavior is  seen in $\mathrm{\dot{M}_{out,CGM}}$ and $\mathrm{\dot{M}_{in,CGM}}$ in the upper right-panel. For reference, we add the expected baryon accretion rate ($\mathrm{\dot{M}_{in,b}(t)}$) based on the mean and median dark matter accretion rate ($\mathrm{\dot{M}_{in,DM}(t)}$) of our fiducial sample. Since we track the dark matter halo mass, we approximate the time derivative as the difference between neighboring simulation output and ignore any timestep where the halo mass is lowered. We then multiply $\mathrm{\dot{M}_{in,DM}(t)}$ by $\mathrm{f_b\sim0.16}$. 

We find that, in the early Universe, the median $\mathrm{\dot{M}_{in,CGM}}$ and $\mathrm{\dot{M}_{in,b}(t)}$ trace each other quite closely. However, at $\mathrm{\sim5\,Gyr}$, there is a significant decrease in the median $\mathrm{\dot{M}_{in,CGM}}$ compared to $\mathrm{\dot{M}_{in,b}(t)}$. Therefore, the true baryon accretion rate \emph{does not} follow the expected baryon accretion rate after $\mathrm{\sim5-6\,Gyr}$. This is prevention in action. We calculate $f_\mathrm{prev}$, the fraction of the expected baryon accretion rate able to enter the halo and find that at early times there is significant scatter, but $f_\mathrm{prev}$ is close to unity. However, after $\mathrm{\sim 5\,Gyr}$, $f_\mathrm{prev}\sim0.25$, meaning that $\sim75\%$ of the expected baryon accretion rate is prevented.     

We next show the M8-h mass flow evolution in the middle panel. One interesting feature that cannot be seen in the fiducial sample is the increase in $\mathrm{\dot{M}_{in,ISM}}$ that is nearly always followed by an increase in $\mathrm{\dot{M}_{out,ISM}}$. 
This likely indicates that gas and satellite infall is triggering delayed SNe feedback.

At the CGM boundary, M8-h has a steady decrease in both $\mathrm{\dot{M}_{out,CGM}}$ and $\mathrm{\dot{M}_{in,CGM}}$. However, at early times ($\mathrm{\lesssim5\,Gyr}$), $\mathrm{\dot{M}_{out,CGM}}$ and $\mathrm{\dot{M}_{in,CGM}}$ trace each other much more closely than at late times ($\mathrm{\gtrsim5\,Gyr}$). Although we do not show $\mathrm{\dot{M}_{in,b}(t)}$, M8-h seems to exhibit a similar behavior as the fiducial sample median: after $\mathrm{\sim5-6\,Gyr}$, there is a decrease in $\mathrm{\dot{M}_{in,CGM}}$ compared to $\mathrm{\dot{M}_{in,b}(t)}$. We discuss the implication of this in \S\ref{subsubsec:mass_flow_rate}. 

Lastly, we separate the inflow and outflow by phase. This is shown for M8-h in the lower-left (ISM boundary) and lower-right (CGM boundary) panels in Figure \ref{fig:mass_flow_evolution}. Similar to $\mathrm{\dot{E}}$, there is a continuous inflow and outflow of warm gas ($\mathrm{\leq3\times10^5\,K}$) at both the ISM and CGM boundary. However, during a SNe-driven outburst, there is an increase in the contribution from the hot gas ($\mathrm{>3\times10^5\,K}$). During such outflows, $\mathrm{\dot{M}_{out,ISM,hot}}$ is often comparable to $\mathrm{\dot{M}_{out,ISM,warm}}$. 
However, there are other times when $\mathrm{\dot{M}_{out,ISM}}$ is dominated by the warm gas. For example, $\mathrm{\dot{M}_{out,ISM,warm}} \gg \mathrm{\dot{M}_{out,ISM,hot}}$ at $\mathrm{\sim7.5\,Gyr}$. In the lower-right panel, it can be seen that this large warm ISM outflow barely makes it to the CGM boundary -- there is only a gradual increase in $\mathrm{\dot{M}_{out,CGM}}$ at $\mathrm{\sim8.5\,Gyr}$. In comparison to the ISM, every significant $\mathrm{\dot{M}_{out,CGM}}$ is dominated by the hot phase, since it is only the hottest outflows that can escape the halo. This is directly related to the outflow specific energy (\S\ref{subsubsec:specific energy}).    

To better quantify the mass contribution from each phase, we perform a similar calculation to the energy flow (\S\ref{subsec:flow of mass and energy}), where we take the ratio of the cumulative mass in the warm and hot phase ($\mathrm{R \equiv\frac{\Sigma \,\dot{M}_{hot}}{\Sigma \,\dot{M}_{warm}}}$). We can therefore better estimate which phase is dominant in the inflow and outflow at both the ISM and CGM boundary. In the fiducial sample, $\mathrm{R_{ISM,out}\sim0.16}$ and $\mathrm{R_{ISM,in}\sim0.0035}$ ($\mathrm{R_{ISM,out}\sim0.080}$ and $\mathrm{R_{ISM,in}\sim0.0031}$ in M8-h). At the CGM boundary, $\mathrm{R_{CGM,out}\sim0.27}$ and $\mathrm{R_{CGM,in}\sim0.0020}$ ($\mathrm{R_{CGM,out}\sim0.080}$ and $\mathrm{R_{CGM,in}\sim0.0043}$ in M8-h). Together, these results demonstrate that the warm phase dominates the mass outflow and inflow at both $\sim0.1\,\mathrm{R_{vir}}$ and $\sim1\,\mathrm{R_{vir}}$. The mass flows are present for the same fraction of time as the energy flows (\S\ref{subsec:flow of mass and energy}), since these are only dependent on the presence of gas within the given temperature range.

\subsubsection{Mass Entrainment} \label{subsec:mass entrainment}

\begin{figure}
    \centering
    \includegraphics[width=0.47\textwidth]{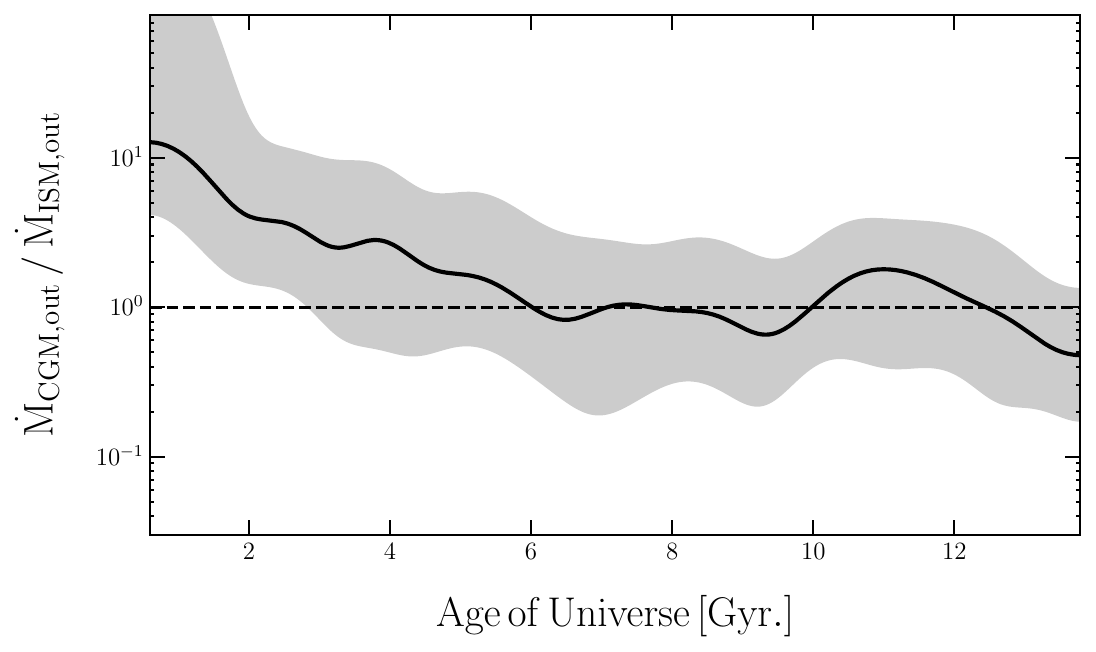} 
    \caption{The smoothened median of  $\mathrm{\dot{M}_{CGM,out}/\dot{M}_{ISM,out}}$ for both the fiducial sample and M8-h. There is clear evidence of mass entrainment in SNe-induced outflows at early times, since $\mathrm{\dot{M}_{CGM,out}>\dot{M}_{ISM,out}}$ before $\mathrm{\sim5\,Gyr}$. After $\mathrm{\sim5\,Gyr}$ $\mathrm{\dot{M}_{CGM,out}/\dot{M}_{ISM,out}}$ is more consistent with 1.      
}\label{fig:mass_entrainment}
\end{figure}

Turning to a measure of the amount of mass entrained as the outflows sweep over the CGM, we show in Figure~\ref{fig:mass_entrainment} the ratio of $\mathrm{\dot{M}_{out,CGM}}$ to $\mathrm{\dot{M}_{out,ISM}}$ for each halo in the fiducial sample and M8-h. In order to show the overall trend in these noisy data, we report the median and 16th/84th percentile $\mathrm{\dot{M}_{CGM,out}/\dot{M}_{ISM,out}}$, smoothed with a Gaussian filter of width $\mathrm{\sigma=10}$ timesteps ($\mathrm{\sim400\,Myr}$).
We find that, before $\mathrm{\sim5\,Gyr}$, for the typical halo in our sample, $\mathrm{\dot{M}_{CGM,out}/\dot{M}_{ISM,out}}>1$, suggesting that a large fraction of the mass in the CGM is entrained by the outflows. However, after $\mathrm{\sim5\,Gyr}$, $\mathrm{\dot{M}_{CGM,out}/\dot{M}_{ISM,out}}$ is more consistent with $\mathrm{\sim1}$. This is directly related to the CGM mass and baryon fraction, discussed in \S\ref{subsec:baryon fraction}. 

\subsection{Radiative Stability} \label{subsec:radiative stability}

The ability of gas to condense and accrete onto the galaxy is largely determined by the ratio of the local cooling time ($\mathrm{t_{cool}}$) to the free-fall time ($\mathrm{t_{ff}}$). In general, if $\mathrm{t_{cool} \gg t_{ff}}$ (where $\mathrm{t_{ff} \approx (G \rho)^{-1/2}}$), the system approaches hydrostatic equilibrium and radiative cooling is relatively unimportant. In contrast, if $\mathrm{t_{cool} < t_{ff}}$, the gas cools and collapses. In between, the gas can be in approximate hydrostatic equilibrium but be prone to precipitation \citep{McCourt_2012}. In galaxy clusters, it has been shown that feedback from the central black hole regulates the intracluster medium (ICM) such that the system remains radiatively stable such that $\mathrm{t_{cool}/t_{ff} \gtrsim 10}$ \citep[e.g.,][]{Voit_2017}. In the halo mass range $\mathrm{\sim10^{10}-10^{12}\,M_\odot}$, it has been suggested that SNe provide the primary source of heating \citep{Voit_2015}. However, as noted in \S\ref{sec:intro}, the CGM of  $\mathrm{\sim10^{10}-10^{11}\,M_\odot}$ halos are expected to cool on a short timescale due to their placement near the peak of the cooling curve \citep{White_Frenk_1991}. 

\begin{figure}
    \centering
    \includegraphics[width=0.48\textwidth]{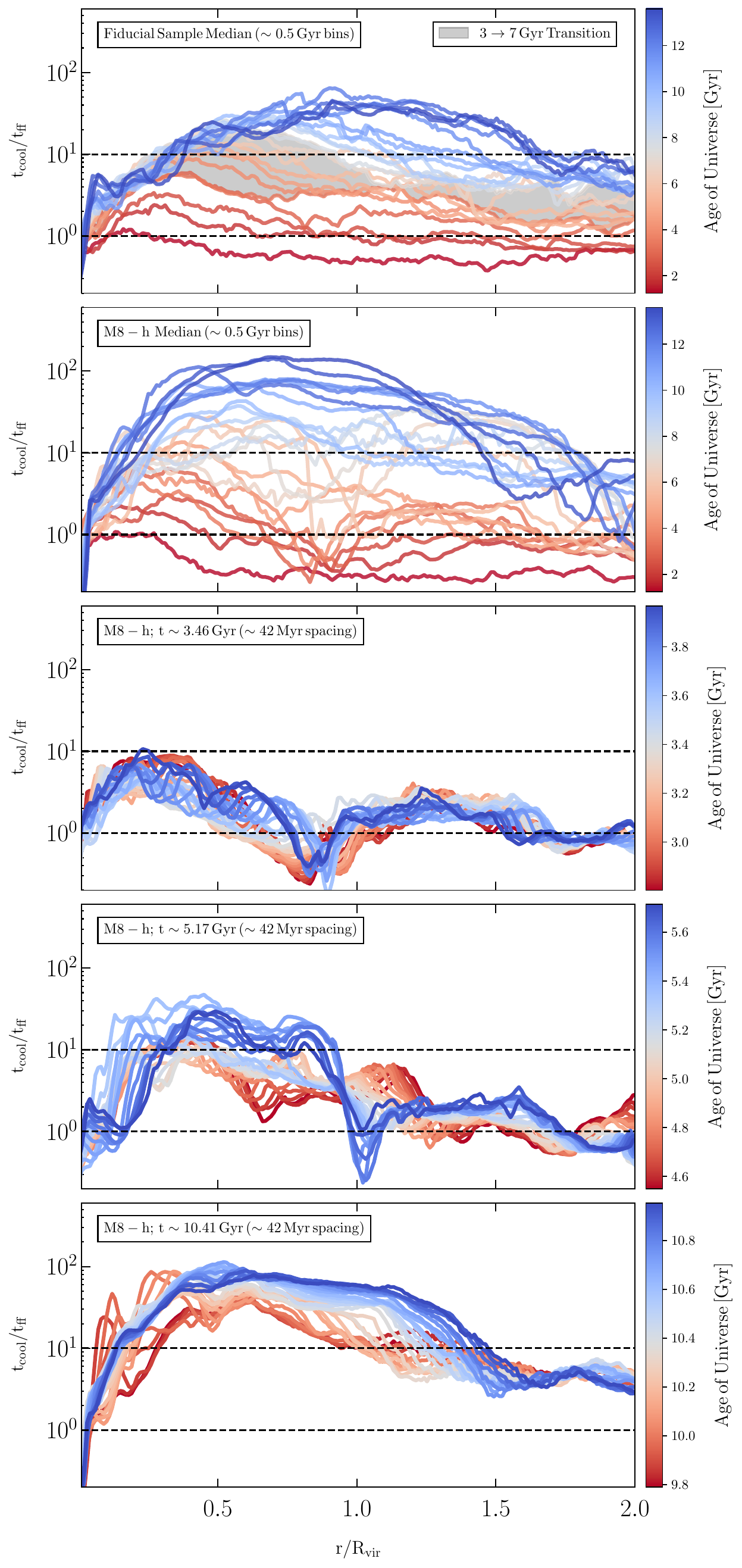} 
    \caption{\textbf{First Panel:} The fiducial sample median time evolution of $\mathrm{t_{cool}/t_{ff}}$, where we bin the profile every $\mathrm{\sim0.5\,Gyr}$. We include the approximate \lq transition' time period, where the outer CGM $\mathrm{t_{cool}/t_{ff}}$ profile exceeds the precipitation limit of $\mathrm{t_{cool}/t_{ff}\sim10}$. \textbf{Second Panel:} The same $\mathrm{t_{cool}/t_{ff}}$ profile time evolution, but for M8-h. \textbf{Third Panel:} The $\mathrm{t_{cool}/t_{ff}}$ time evolution for the M8-h outburst that occurred at $\mathrm{\sim3.46\,Gyr}$. We include every simulation output ($\sim42\,\mathrm{Myr}$ cadence) during the $\sim1\,\mathrm{Gyr}$ time period centered on the timestep where the outflow first passed through the ISM boundary ($\mathrm{\sim3.46\,Gyr}$).  \textbf{Fourth Panel:} The same analysis, but for a M8-h SNe outburst that occurred at $\mathrm{\sim5.17\,Gyr}$.  \textbf{Fifth Panel:} The outflow shock propagation in M8-h in the $\sim1\,\mathrm{Gyr}$ surrounding $\mathrm{\sim10.41\,Gyr}$ (Figure \ref{fig:temp_slice_two}).
    \label{fig:cooling_time_ratio}}
\end{figure}

We quantify the time evolution of $\mathrm{t_{cool}/t_{ff}}$ as a function of radius for our simulations in Figure~\ref{fig:cooling_time_ratio}. In the upper panel, we show the fiducial sample median $\mathrm{t_{cool}/t_{ff}}$ profile as a function of time. In particular, we calculate the median across all ten halos (excluding M8-h) in $\sim0.5\,\mathrm{Gyr}$ time bins from $\sim1\,\mathrm{Gyr}$ to the present day. Therefore, each median profile is calculated using approximately $\sim120$ $\mathrm{t_{cool}/t_{ff}}$ profiles. For reference, we include a horizontal line where $\mathrm{t_{cool} =t_{ff}}$ (also known as the cooling radius, $\mathrm{R_{cool}}$) and $\mathrm{t_{cool} = 10\,t_{ff}}$ (precipitation limit).     

In the upper panel, at the earliest time interval (dark red), $\mathrm{t_{cool}/t_{ff}\lesssim1}$ throughout much of the halo. This is most evident beyond $\mathrm{0.5\,R_{vir}}$.  $\mathrm{t_{cool}/t_{ff}\lesssim1}$ is indicative of rapid cooling of gas in the CGM onto the ISM. There is a modest enhancement of $\mathrm{t_{cool}/t_{ff}}$ at $\mathrm{0.1-0.2\,R_{vir}}$ ($\mathrm{t_{cool}/t_{ff}\sim1}$), likely due to early star formation and feedback that is heating the gas immediately outside the ISM. However, within $\mathrm{R_{cool}\sim0.1\,R_{vir}}$, $\mathrm{t_{cool}/t_{ff}}$ lowers, due to the cold and dense star-forming gas in the ISM. 

Prior to $\sim5\,\mathrm{Gyr}$, the CGM $\mathrm{t_{cool}/t_{ff}}$ profile tends to increase with time. There is also a slight radial evolution in the location of the peak value. For example, at $\sim1\,\mathrm{Gyr}$, the peak occurs near $\mathrm{0.1-0.2\,R_{vir}}$, whereas by $\sim5\,\mathrm{Gyr}$, the peak has shifted outward to $\sim0.5\,\mathrm{R_{vir}}$. The profile shape does not change at the center of the galaxy ($\mathrm{<0.05\,R_{vir}}$): $\mathrm{R_{cool}\sim0.02-0.05\,R_{vir}}$. 

At $\sim5\,\mathrm{Gyr}$, the CGM first begins to exceed $\mathrm{t_{cool}/t_{ff} \sim 10}$. We highlight the approximate time interval $\mathrm{t=[3,\,7]\,\mathrm{Gyr}}$ with a grey shaded region. During this $\mathrm{\sim4\,Gyr}$ time period, the CGM region bounded by $\mathrm{r/R_{vir} = [0.25,\,0.75]}$ begins to reach the $\mathrm{t_{cool}/t_{ff}\gtrsim10}$ criterion for radiative stability. Notably, this interval also coincides with the epoch in Figures~\ref{fig:energy_flow_evolution} and \ref{fig:cumulative_energy} when $\mathrm{\dot{E}_{ISM,out}}$ begins to dominate over $\mathrm{\dot{E}_{cool,CGM}}$. Note that this is the fiducial sample median, so we expect halo-to-halo scatter in the approximate time of the CGM transition to $\mathrm{t_{cool}/t_{ff}\gtrsim10}$.

After $\sim5-7\,\mathrm{Gyr}$, the main change in the $\mathrm{t_{cool}/t_{ff}}$ profile evolution is that $\mathrm{t_{cool}/t_{ff}}$ increases in the outer CGM ($\sim1\,\mathrm{R_{vir}}$). The peak value now occurs near the halo boundary and remains $\gtrsim10$ (though gradually decreasing with radius) out to $\sim1.5\,\mathrm{R_{vir}}$. By the present-day Universe, only the inner CGM ($\lesssim0.4\,\mathrm{R_{vir}}$) remains radiatively unstable. 

In the second panel of Figure~\ref{fig:cooling_time_ratio}, we show the same $\mathrm{t_{cool}/t_{ff}}$ evolution but for M8-h. The overall profile evolution is similar to the fiducial sample median, although several features become more apparent when examining a single system. For example, there are several broad depression-like features in the profile that likely correspond to infalling satellites containing their own star-forming ISM. Since these satellites remain within the halo for roughly $\sim0.5\,\mathrm{Gyr}$, the resulting decrease in $\mathrm{t_{cool}/t_{ff}}$ is broadened in radius. This effect is most apparent in the light-orange depression centered near $\sim1\,\mathrm{R_{vir}}$. In addition, we find that the $\mathrm{t_{cool}/t_{ff}}$ profile rises more steeply in the inner CGM compared to the fiducial sample median. In M8-h, only the inner $\lesssim0.2\,\mathrm{R_{vir}}$ remains radiatively unstable. 

In the lower three panels in Figure~\ref{fig:cooling_time_ratio}, we include a case study of the M8-h $\mathrm{t_{cool}/t_{ff}}$ profile during three different SNe-driven outflows. These occur at the ISM boundary at $\mathrm{\sim3.46\,Gyr}$, $\mathrm{\sim5.17\,Gyr}$, and $\mathrm{\sim10.41\,Gyr}$. The latter outflow is the same SNe event from Figure~\ref{fig:temp_slice_two}. Our choice of these three outflows enable us to explore the radiative stability of the halo before, during, and after the approximate transition at $\mathrm{\sim5\,Gyr}$, where there is a shift in the flow of energy and mass (\S\ref{subsec:flow of mass and energy} and \S\ref{subsec:flow of mass}).  Instead of plotting the binned time evolution, we show the profile at each simulation output ($\sim42\,\mathrm{Myr}$ resolution).

In the third panel in Figure~\ref{fig:cooling_time_ratio}, we show the $\mathrm{\sim1\,Gyr}$ timespan surrounding the SNe outburst at $\mathrm{t\sim 3.46\,Gyr}$. This SNe feedback event brought the volume-weighted CGM temperature ($\mathrm{T_{CGM,1.0}}$) to $\mathrm{\sim6\times10^5\,K}$ (Figure \ref{fig:temperature_evolution}). This snapshot also had $\mathrm{\dot{E}_{out,ISM}\sim10^{40}\,erg\,s^{-1}}$ (Figure \ref{fig:energy_flow_evolution}). Despite this seemingly significant SNe outflow, the $\mathrm{t_{cool}/t_{ff}}$ profile seems relatively stable during this time period. The only interesting feature can be seen in blue, where a satellite falls into the halo center after the SNe outflow (light blue to dark blue depression-like feature). In general, the entire CGM remains below $\mathrm{t_{cool}/t_{ff}}=10$, with the peak value of $\mathrm{t_{cool}/t_{ff}}\sim5$ occurring at $\mathrm{r/R_{vir}\sim0.25}$. At this early epoch, SNe feedback can't drive the CGM into the radiatively stable regime.   

The picture is slightly different in the fourth panel, which focuses on the time period surrounding the SNe outflow at $\mathrm{\sim5.17\,Gyr}$. Prior to the SNe outflow, the $\mathrm{t_{cool}/t_{ff}}$ profile has a similar shape but is slightly elevated compared to the SNe outflow at $\mathrm{\sim3.46\,Gyr}$ (third panel). However, there is a clear \lq lifting' of the $\mathrm{t_{cool}/t_{ff}}$ profile in the CGM \emph{after} the SNe outflow. At $\mathrm{\sim5.17\,Gyr}$, the $\mathrm{t_{cool}/t_{ff}}$ profile has a peak value of $\mathrm{t_{cool}/t_{ff}}\sim8-9$ at $\mathrm{r/R_{vir}\sim0.4}$. Compared to the earlier SNe profile, the peak value has shifted slightly outward. $\mathrm{T_{CGM,1.0}\sim4\times10^{6}\,K}$ and  $\mathrm{\dot{E}_{out,ISM}\sim10^{41}\,erg\,s^{-1}}$, which is a slightly higher $\mathrm{\dot{E}_{out,ISM}}$ compared to the previous SNe outflow. After $\mathrm{\sim5.17\,Gyr}$, the entire CGM is lifted and sustained at $\mathrm{t_{cool}/t_{ff}}\gtrsim10$. There is also a slight shift after the SNe outflow where the radius at which $\mathrm{t_{cool}/t_{ff}}\sim10$ shifts outward from $\mathrm{r/R_{vir}\sim0.1}$ to $\mathrm{r/R_{vir}\sim0.4}$.      

In the fifth panel in Figure~\ref{fig:cooling_time_ratio}, we show the propagation of the SNe-driven shock in M8-h corresponding to the $\mathrm{z=0.28}$ snapshot (Figure \ref{fig:temp_slice_two}). Before the SNe outflow, the $\mathrm{t_{cool}/t_{ff}}$ profile is already elevated. 
We can then see the shock propagate outward in the subsequent snapshots -- there are large peaks in the $\mathrm{t_{cool}/t_{ff}}$ profile corresponding to the region near the shock front. At $\mathrm{t}\, \mathrm{\sim10.41\,Gyr}$, $\mathrm{T_{CGM,1.0}\sim4\times10^{5}\,K}$ and  $\mathrm{\dot{E}_{out,ISM}\sim10^{41}\,erg\,s^{-1}}$. As the shock moves outward, the entire CGM is lifted well above the precipitation limit of $\mathrm{t_{cool}/t_{ff}}=10$ and even the region out to $\mathrm{r/R_{vir}\sim1.5}$ has $\mathrm{t_{cool}/t_{ff}}\sim10$.

\begin{figure}
    \centering
    \includegraphics[width=0.47\textwidth]{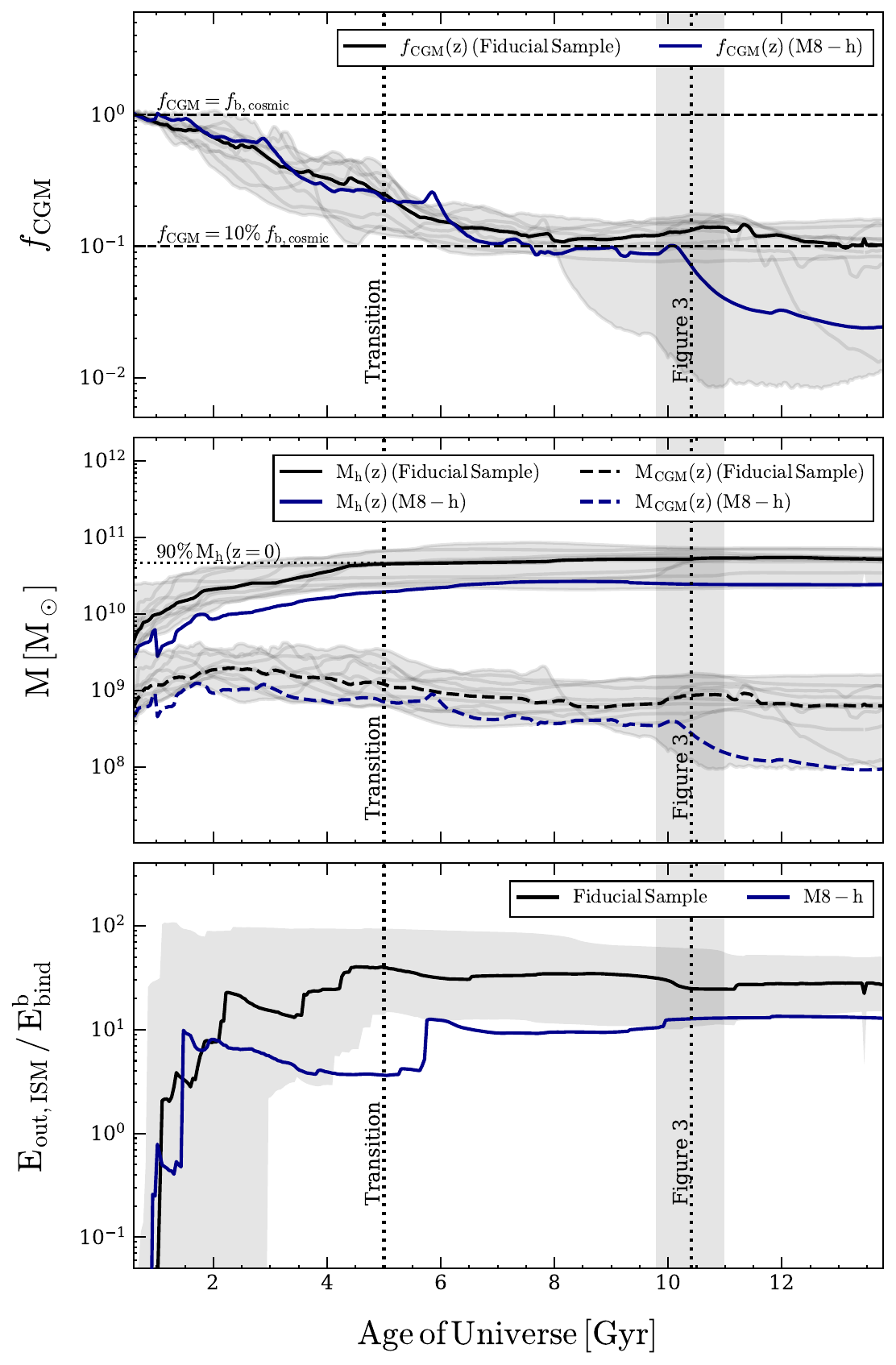} 
    \caption{\textbf{Upper Panel:} $f_\mathrm{CGM}$ evolution for the fiducial sample (solid black) and M8-h (solid dark blue). We also show the full sample range (minimum to maximum) and $f_\mathrm{CGM}$ for each individual halo. \textbf{Middle Panel:} The halo dark matter mass ($\mathrm{M_h}$; solid) and CGM gas mass ($\mathrm{M_{CGM}}$; dashed) growth history. We show this for both the fiducial sample (black) and M8-h (dark blue). \textbf{Lower Panel:} The ratio of the fiducial sample cumulative SNe energy outflow rate ($\mathrm{\dot{E}_{out,ISM}}$) to the baryon binding energy ($\mathrm{E_{bind}^b}$)). We show the fiducial sample median and full range (black) and M8-h (dark blue).       
    \label{fig:baryon_fraction}}
\end{figure}

\subsection{Baryon Fraction} \label{subsec:baryon fraction}

We now examine the CGM baryon fraction ($f_\mathrm{CGM}$) in our dwarf sample. Since the stellar mass in the CGM is negligible, $f_\mathrm{CGM}$ is merely the CGM gas fraction. Because the gravitational potentials of these low-mass halos are shallow, the CGM baryon fraction can serve as a measure of feedback efficiency. For example, weak feedback (i.e., slow winds, fixed $\mathrm{\eta_M}$) may produce a baryon-rich CGM, where gas is recycled and unable to leave the halo. In contrast, strong feedback (i.e., fast winds, fixed $\mathrm{\eta_M}$) may remove gas from the CGM, leaving behind a baryon-depleted halo. However, a similarly gas-depleted CGM can also arise in a preventive feedback scenario, where gas is prevented from accreting onto the halo in the first place. We define the CGM baryon fraction as the total gas mass between $0.1$--$1\,\mathrm{R_{vir}}$ divided by the halo mass, $\mathrm{M_{vir}}$, and normalize this value by the cosmic baryon fraction ($f_\mathrm{b}$). In the absence of feedback, we therefore expect $f_\mathrm{CGM}\sim1$, where the gas traces the dark matter.

In the upper panel of Figure~\ref{fig:baryon_fraction}, we show the time evolution of $f_\mathrm{CGM}$ for each halo in our sample. The solid black line shows the sample median, while the solid blue line corresponds to M8-h. On average, during the first $\sim8\,\mathrm{Gyr}$, the halos follow a similar evolutionary track. At early times, when star formation has just begun, $f_\mathrm{CGM}\sim1$. This value gradually declines to $\sim0.1$ by $\sim8\,\mathrm{Gyr}$. However, the evolution of $f_\mathrm{CGM}$ during this period is not smooth; several sharp jumps are present, indicating rapid episodes of gas accretion and removal at early times.

After $\sim5-8\,\mathrm{Gyr}$, $f_\mathrm{CGM}$ remains approximately constant at $\sim0.1$, an event that occurs at roughly the same time that the CGM becomes radiatively stable (\S\ref{subsec:radiative stability}). However, four halos exhibit a sudden decrease in their baryon fraction at later times. This behavior is evident in the solid blue line corresponding to M8-h, which shows a sharp drop in $f_\mathrm{CGM}$ at $\sim10\,\mathrm{Gyr}$. By the present day, $f_\mathrm{CGM}\sim0.01$--$0.02$. The feedback event shown in Figure~\ref{fig:temp_slice_two} appears to have stripped the CGM of most of its remaining gas content.

In the middle panel of Figure~\ref{fig:baryon_fraction}, we decompose the CGM baryon fraction into the halo mass and CGM gas mass. We find that halos in our sample acquire their present-day mass of $\mathrm{M_h}\sim10^{10}\,\mathrm{M_\odot}$ relatively rapidly: approximately $90\%$ of the final halo mass is assembled by our `transition' definition at $\sim5\,\mathrm{Gyr}$. In contrast, the CGM gas mass peaks earlier, at $\sim2\,\mathrm{Gyr}$, then gradually declines until $\sim8\,\mathrm{Gyr}$ before remaining approximately constant to the present day.

We lastly compare the baryon fraction to the time evolution of the ratio of the cumulative energy injected into the CGM from the ISM ($\mathrm{\dot{E}_{out,ISM}}$; Figure \ref{fig:energy_flow_evolution}) to the redshift dependent binding energy of the gas, given by 
\begin{gather} \label{equation:rate}
\mathrm{E_{bind}^b=\frac{1}{2}\mathit{f}_b\frac{G\,M_{vir}^2}{R_{vir}}}
\end{gather}
where G is the gravitational constant. We calculate this for both the fiducial sample and M8-h. In general, we find that the cumulative $\mathrm{E_{out,ISM}}$ nearly always exceeds $\mathrm{E_{bind}^b}$. In fact, $\mathrm{E_{out,ISM}}/ \mathrm{E_{bind}^b}\sim10-50$ once most of the present-day mass has been assembled. We discuss the implication of this result in \S\ref{subsubsec:cumulative_sne_energy}.

\subsection{Closure Radius} \label{subsec:closure radius}

\begin{figure*}
    \centering
    \includegraphics[width=1\textwidth]{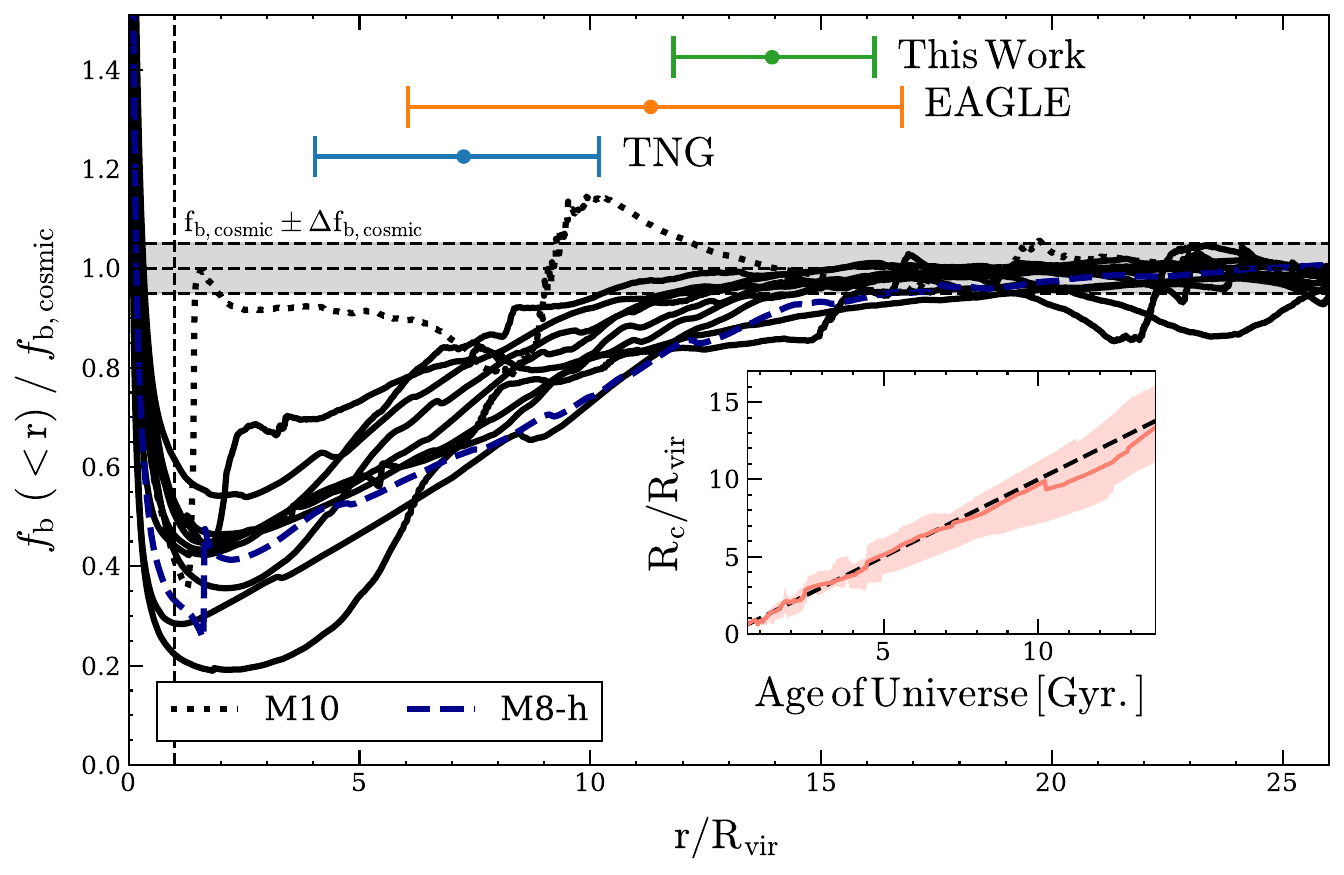} 
    \caption{The cumulative baryon fraction ($\mathrm{\mathit{f}_b(<r)}$) normalized by the cosmic baryon fraction ($\mathrm{\mathit{f}_{b,cosmic}}$) as a function of radius for each dwarf galaxy in our sample. There is one system where the presence of a nearby halo skews the $\mathrm{R_c}$ measurement (M10; dotted black). The M8-h profile is shown as the dashed dark blue line. The shaded region surrounding $\mathrm{\mathit{f}_{b,cosmic}} = 1$ represents the $\sim 5$ percent uncertainty in the \cite{Planck2018} measurement. We include a comparison to \cite{Ayromlou_2023}, where the closure radius range for a $\mathrm{M_{200c}\sim 5\times 10^{10}\ M_\odot}$ halo in TNG and EAGLE is shown (16th, 50th, 84th percentiles). We also include the median (and 16th, 84th percentiles) $\mathrm{R_c}$ as a function of time, which grows nearly linearly.       
    \label{fig:closure_radius}}
\end{figure*}

Having established that feedback drives $f_\mathrm{CGM}$ down to $\sim0.1$ by the present-day, a natural question is how far these depleted baryons have been redistributed. The closure radius ($\mathrm{R_c}$) quantifies this by measuring the distance at which the cumulative baryon fraction recovers to the cosmic mean.
We adopt the definition used in \cite{Ayromlou_2023}: 
\begin{gather}
\mathrm{f_b\ (< \mathrm{R_c}) = f_{b,cosmic} \pm \Delta f_{b,cosmic}}
\end{gather}
where $\Delta f_\mathrm{b,cosmic}$ is the uncertainty in the cosmic baryon fraction. We calculate the closure radius for each halo by calculating the ratio of the baryon mass profile and the total mass profile and find the approximate distance at which $\mathrm{f_\mathrm{b}(<r)/f_\mathrm{b,cosmic} \sim 0.95}$ (accounting for the uncertainty) for each halo. The result is shown in Figure \ref{fig:closure_radius}. The cumulative baryon fraction profile for each halo in the fiducial sample is represented by either a solid or dotted black line (the dotted black line is M10, where there is a nearby massive satellite at $\mathrm{z=0}$ that artificially boosts our estimate for that system). 

For reference, we include a vertical black line at $\mathrm{R_{vir}}$. The horizontal black lines denote the baryon fraction normalized to the cosmic baryon fraction (center) and normalized uncertainty, $\sim 5 \%$ (upper and lower). We also include the approximate median and 16th/84th percentile closure radius range for $\mathrm{M_{200c}\sim 5\times 10^{10}\ M_\odot}$ halos in TNG and EAGLE \citep[][]{Ayromlou_2023}.  

We find that at $\mathrm{1\, R_{vir}}$ the enclosed baryon fraction is well below the cosmic baryon fraction. The value presented here is different from Figure~\ref{fig:baryon_fraction} since the cumulative baryon fraction includes both stars and gas in the ISM, not just the CGM gas. We find that there is considerable scatter in the sample. For example, one halo has as little as $20 \%$ of the cosmic baryon fraction within $\mathrm{1\, R_{vir}}$. At the other extreme, another halo has nearly $60 \%$ of the cosmic baryon fraction within $\mathrm{1\, R_{vir}}$. Excluding M10, we find that the median $\mathrm{R_c}$ is $\sim15\,\mathrm{R_{vir}}$. Our sample is consistent with the upper 50 percent of EAGLE and is significantly larger than the IllustrisTNG distribution.

We also measure $\mathrm{R_c}$ as a function of time. This is shown in the inset panel in Figure \ref{fig:closure_radius} (solid red). We find that at the start of the simulation, $\mathrm{R_c\sim1}$, meaning that most of the expected baryons could be found in $\mathrm{R_{vir}}$. However, $\mathrm{R_c}$ appears to linearly grow with time. We also include the 16th and 84th percentiles (light red region), which also appear to become wider with time. 

\subsection{Ion Distribution} \label{subsec:metal distribution}

Lastly, we investigate the role of our feedback model on the distribution of metals in the CGM and IGM. In particular, we are interested in calculating the distribution of the same ions that are observed using HST/COS (e.g., {\sc Ovi}). Since we only use a six-species model and do not explicitly track e.g., O, we approximate the ion species using the open-source, hydrodynamical simulation post-processing tool \textsc{trident} \citep{Trident}. \textsc{trident} provides a look-up table of ion abundances for a given density and temperature assuming collisional ionization equilibrium and photoionization from a meta-galactic ultraviolet background. \textsc{cloudy} \citep{CLOUDY} and \textsc{roco} \citep{ROCO} are used to calculate this table using hydrogen density and temperature values that span the range found within our simulation. We use the same self-shielding ultraviolet background that our simulation was evolved with \citep{Haardt_Madau_2012}. We note that we do not include the contribution from the galaxy's radiation field. 

For a given gas cell hydrogen density and temperature, \textsc{trident} calculates $n_{X_i} = n_X f_{X_i}$,
where $n_X$ is is the total number density of the element and $f_{X_i}$ is the ionization fraction of the $i$'th ion of the element. Since the individual element abundance is not tracked in our simulation (except H, He), we use the solar abundance pattern. Therefore, \textsc{trident} calculates $n_X$ as $n_{X} = n_H Z \left( \frac{n_X}{n_H}\right)_\odot$,
where $n_H$ is the hydrogen number density, $Z$ is the metallicity, and $\left( n_X / n_H\right)_\odot$ is the solar abundance. 

\begin{figure*}
    \centering
    \includegraphics[width=1\textwidth]{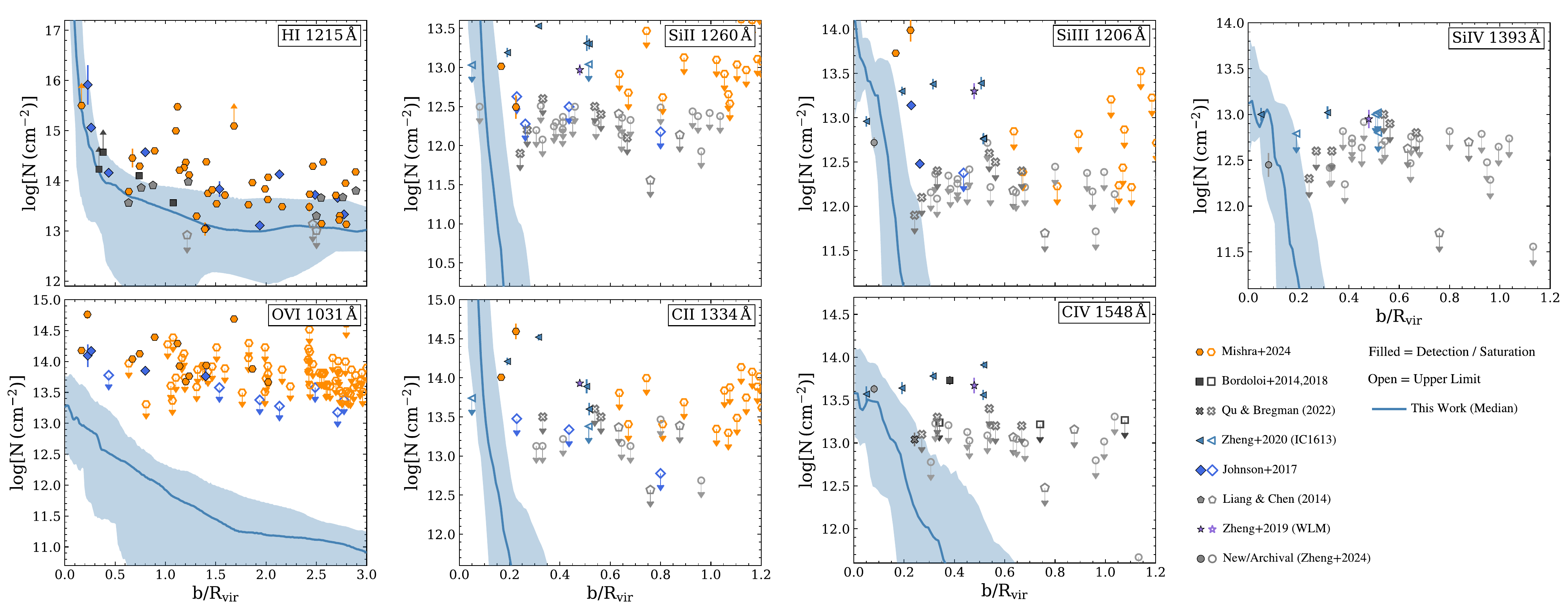} 
    \caption{
    The column density profiles of H~\textsc{i}, Si~\textsc{ii}, Si~\textsc{iii}, Si~\textsc{iv}, C~\textsc{ii}, C~\textsc{iv}, and O~\textsc{vi} as a function of projected impact parameter normalized by the virial radius ($\mathrm{b/R}_{\rm vir}$). For each simulated dwarf halo, we compute the column density map across a single random projection and extract a radial profile. Note that only H~\textsc{i} and O~\textsc{vi} go out to $\mathrm{b/R_{vir}=3}$ even though each ion projection is computed from a sphere centered on the halo out to $\mathrm{5\,R_{vir}}$. The solid blue line shows the median of the sample median profiles, and the blue shaded region spans the minimum to maximum range, reflecting halo-to-halo scatter. For comparison, observed column densities from the literature \citep[limited to $\mathrm{M_h\sim10^{10}-10^{11}\,M_\odot}$;][]{Zheng_2024} are shown as data points, where filled symbols indicate detections or saturated lines and open symbols indicate $3\sigma$ upper limits. In general, we find that our simulations agree with H~\textsc{i} and low ions but are well below the detected O~\textsc{vi} column densities reported in \cite{Johnson_2017} and \cite{Mishra_2024}.
    \label{fig:ions}}
\end{figure*}

We use the \textsc{trident} output to produce a column density map of each galaxy in our sample. To compare with recent observational work \citep{Zheng_2024,Mishra_2024}, we calculate $n_X$ for the following ions: H~\textsc{i}, Si~\textsc{ii}, Si~\textsc{iii}, Si~\textsc{iv}, C~\textsc{ii}, C~\textsc{iv}, and O~\textsc{vi}. We then use \textsc{yt} to produce a randomly projected column density image using gas out to $5\,\mathrm{R_{vir}}$ from the center of the galaxy. Each image has a pixel resolution of $1,000\times1,000$. We show the result in Figure \ref{fig:ions}, where we present both the median profile of our sample medians (solid blue) and their minimum to maximum range (shaded blue region).    

We also include the detections and non-detections reported in \cite{Zheng_2024}\footnote{We used the data and code made available at \url{https://github.com/yzhenggit/zheng_dwarfcgm_survey}. The data also included column density measurements reported in \cite{Liang_Chen2014,Bordoli_2014,Bordoloi_2018,Johnson_2017,Zheng_2019,Zheng_2020,Qu_Bregman2022}.} as well as the O~\textsc{vi} observations reported in \cite{Johnson_2017} and \cite{Mishra_2024}.
For both data sets, we only select halos consistent with our mass range ($\mathrm{10^{10}\,M_\odot \leq M_h\leq10^{11}\,M_\odot}$). We report detections and saturated lines using filled symbols and $3\sigma$ upper limits on non-detections using open symbols (see \cite{Zheng_2024} and \cite{Mishra_2024} for further details).  

We find that H~\textsc{i} and low ions generally agree with the literature.  H~\textsc{i} is consistent with its ubiquitous detection in observational work. The low H~\textsc{i} lower bound is a result of one of the halos undergoing a late-time merger and losing a substantial fraction of its CGM. The median H~\textsc{i} profiles of the other halos are more consistent with the solid blue line. The low ions (Si~\textsc{ii}, Si~\textsc{iii}, C\textsc{ii}) steeply drop off. Si~\textsc{iv} and  C~\textsc{iv} have a similar drop-off within $\mathrm{\sim0.5\,R_{vir}}$. Although O~\textsc{vi} has a more shallow profile, our median O~\textsc{vi} is still a factor of $10-100$ below detections in observational work.  We defer a more detailed comparison with observations and discussion to \S\ref{subsubsec:absorption discussion}.     

\section{Discussion} \label{sec:discussion}

Taken together, our results provide a coherent picture of energy and mass transport in dwarf galaxy halos, where SNe-driven feedback drives $f_\mathrm{CGM}$ down and brings the CGM into a radiatively stable regime. In this section, we first connect our results and discuss the physical picture that emerges and how it is related to the transition in a variety of properties around $\mathrm{\sim5\,Gyr}$ (\S\ref{subsec:the physical picture}). We then discuss our results in the context of other work, including a comparison to the outflow specific energy, temperature evolution, \lq inner CGM virialization', mass entrainment, and the baryon extent (\S\ref{subsec:comparison to other work}), before discussing potential caveats of our simulation suite (\S\ref{subsec:caveats}).

\subsection{The Physical Picture of the Transition} 
\label{subsec:the physical picture}

Bringing together all of our results, we find that there seems to be a significant transition in our halos around $\mathrm{\sim5\,Gyr}$. We re-state these physical elements more concisely here:

\begin{itemize}
    \item \textbf{Temperature:} In the first $\mathrm{\sim5\,Gyr}$, the inner CGM is cooler than the outer CGM, and the halos show frequent strong bursts of SNe feedback, visible in the $\mathrm{e_s}$, $\mathrm{\dot{E}}$, and $\mathrm{\dot{M}}$ evolution. After the transition, the temperature profile inverts and the inner CGM is sustained at  $\mathrm{T_{vir}}$.   

    \item \textbf{Energy:} By $\mathrm{\sim5\,Gyr}$, the cumulative SNe output exceeds CGM radiative losses. Although $\mathrm{\dot{E}_{cool,CGM}>\dot{E}_{out,ISM}}$ for a larger fraction of time, the infrequent yet large SNe-induced  $\mathrm{\dot{E}_{out,ISM}}$ dominate the total energy budget of the CGM. 

    \item \textbf{Mass:} At early times, $\mathrm{\dot{M}_{in,CGM}}$ is comparable to $\mathrm{\dot{M}_{in,b}}$, the expected baryon influx, but after $\mathrm{\sim5\,Gyr}$, it drops dramatically such that $\sim75\%$ of expected baryon accretion is prevented. $f_\mathrm{CGM}$ is lowered from unity to $\sim0.1$ by $\mathrm{\sim 5-8\,Gyr}$. This is largely driven by mass entrainment since we also find $\mathrm{M_{gas}}$ has been steadily decreasing since $\mathrm{\sim2\,Gyr}$.  

    \item \textbf{Halo Assembly:} For the dark matter halos in our sample, $\mathrm{\sim90\%}$ of the present-day mass had been acquired by $\mathrm{\sim5\,Gyr}$.   

    \item \textbf{Radiative Stability:} Before $\mathrm{\sim5\,Gyr}$, the entire halo is radiatively unstable,
    but by $\mathrm{\sim3-7\,Gyr}$ the CGM is lifted above the precipitation threshold, $\mathrm{t_{cool}/t_{ff}\gtrsim10}$. 
     
\end{itemize}

Together, these results point to a physical picture in which the CGM of halos in the mass range $\mathrm{M_h \sim 10^{10}-10^{11}\,M_\odot}$ undergo a fundamental transition at $\sim5\,\mathrm{Gyr}$, driven by the interplay between mass inflow, SNe feedback, entrainment, and radiative cooling.

At early times, rapid halo growth and frequent satellite and gas infall sustain a high cadence of SNe feedback, producing bursty outflows with high specific energy. These outflows do two things simultaneously: (1) deposit energy into and shock-heat the CGM and IGM and (2) entrain mass (i.e., remove mass from the CGM and nearby IGM). The energetic consequence is that, despite $\mathrm{\dot{E}_{cool,CGM} > \dot{E}_{out,ISM}}$ for a larger fraction of time, a small number of powerful outflow events come to dominate the cumulative energy budget -- the SNe energy input exceeds the total CGM cooling energy within the first $\sim5\,\mathrm{Gyr}$. The mass consequence is just as important: outflow-driven entrainment removes gas from the CGM faster than cosmological accretion can replenish it, as evidenced by $\mathrm{\dot{M}_{out,CGM} > \dot{M}_{out,ISM}}$ occurring preferentially before $\sim5\,\mathrm{Gyr}$ and the steady decline in $\mathrm{M_{gas}}$ from $\sim2\,\mathrm{Gyr}$ onward. Together, these processes gradually deplete the CGM baryon content, driving $f_\mathrm{CGM}$ from near unity down to $\sim0.1$ by $z\sim0$.

The thermal and structural state of the CGM reflects this evolution. In the early Universe, frequent SNe bursts heat the outer CGM while the inner CGM remains relatively cool, and the entire halo sits in a radiatively unstable regime ($\mathrm{t_{cool}/t_{ff} \lesssim 10}$). However, as outflows progressively remove gas and lower the CGM density, the cooling time increases relative to the free-fall time, pushing the halo toward radiative stability. By $\sim3$--$7\,\mathrm{Gyr}$, this manifests as the bulk of the CGM being \lq lifted' above the precipitation threshold ($\mathrm{t_{cool}/t_{ff} \gtrsim 10}$), with only the innermost regions ($\mathrm{r/R_{vir} \lesssim 0.40}$) remaining susceptible to cooling. 

Simultaneously, as halo growth slows after $\sim5\,\mathrm{Gyr}$ -- with $\sim90\%$ of the present-day mass already in place -- the frequency of merger-driven SNe bursts decreases, and $\mathrm{\dot{E}_{out,ISM}}$ begins to exceed $\mathrm{\dot{E}_{cool,CGM}}$ for a larger fraction of time. The inner CGM is sustained near $\mathrm{T_{vir}}$, baryon accretion from the IGM is suppressed by $\sim75\%$, and $f_\mathrm{CGM}$ has nearly plateaued -- the CGM is left in a hot, diffuse, and radiatively stable configuration that is in equilibrium at late times.

\subsection{Comparison to Previous Work}
\label{subsec:comparison to other work}

With this physical picture in mind, we now place our findings in the context of other work. We begin by comparing the temperature evolution of the CGM to other simulations (\S\ref{subsubsec:temperature_discussion}), followed by a comparison of energy flow (\S\ref{subsubsec:energy flow}, \S\ref{subsubsec:cumulative_sne_energy}), and mass flow  (\S\ref{subsubsec:mass_flow_rate}). We then discuss the radiative stability of our halos in the context of \lq inner CGM virialization' 
(\S\ref{subsubsec:radiative_stability_discussion}), and conclude with a discussion of the closure radius (\S\ref{subsubsec:closure radius discussion}) and comparison with recent observational work  (\S\ref{subsubsec:absorption discussion}).

\subsubsection{Temperature} \label{subsubsec:temperature_discussion}

In \S\ref{subsec:sustaining virial temperature}, we demonstrated that the CGM is continuously heated to $\mathrm{T_{vir}}$ by episodic SNe-induced shock-heating of the halo.  A similar behavior is seen in the IGM ($\mathrm{1-2\,R_{vir}}$), although the gas only temporarily exceeds $\mathrm{T_{vir}}$ before cooling on a $\mathrm{\sim Gyr}$ timescale. 

\cite{Fielding_2017} report the mass-weighted temperature time evolution of their idealized $\mathrm{M_h\sim10^{11}\,M_\odot}$ halo for both a low and high mass-loading ($\eta_\mathrm{M}$) feedback model, corresponding to high and low energy per unit mass, respectively. 
Although their halo is more massive than our sample, \cite{Fielding_2017} find that the low $\eta_\mathrm{M}$ model (characterized by low mass-loading and high wind velocity) is able to similarly sustain a larger fraction of the halo at $\mathrm{10^5-10^6\,K}$ compared to their high $\mathrm{\eta_M}$ model (high mass-loading and low wind velocity).

\cite{Fielding_2017} also report the radial temperature profile of their $\mathrm{M_h\sim10^{11}\,M_\odot}$ halo (see their Figure 7). In the high $\eta_\mathrm{M}$ case, the CGM is kept well below $\mathrm{T_{vir}}$ ($\mathrm{\sim1.2\times10^5\,K}$) and reaches a peak temperature at $\mathrm{\sim0.25\,R_{vir}}$. However, in the low $\eta_\mathrm{M}$ case, a better match to the high specific energy of our outflows, the temperature profile is flatter and is sustained near $\mathrm{T_{vir}}$ between $\mathrm{0.1-0.5\,R_{vir}}$. We find a qualitatively similar result in our work. In the volume-weighted temperature profile of our sample (Figure~\ref{fig:temperature_evolution}), the temperature peaks at $\mathrm{\sim0.25\,R_{vir}}$ at $\mathrm{z\sim2}$ (below $\mathrm{T_{vir}}$), and the peak gradually migrates inward to $\mathrm{\sim0.1\,R_{vir}}$ by $\mathrm{z\sim0}$ (above $\mathrm{T_{vir}}$). At $\mathrm{z\sim0}$, our temperature profile exceeds $\mathrm{T_{vir}}$ out to $\mathrm{\sim1\,R_{vir}}$, somewhat higher than \cite{Fielding_2017}, but note that \cite{Fielding_2017} averaged their temperature over $\mathrm{\sim6\,Gyr}$, whereas we report the median over $\mathrm{\sim0.5\,Gyr}$ bins.

The picture is quite different in the FIRE-2 simulation suite. In \cite{Stern2021}, the radial temperature profile of $\mathrm{\sim10^{10}-10^{11}\,M_\odot}$ halos peaks at $\mathrm{\sim0.5\,R_{vir}}$ at $\mathrm{z=0}$, with the peak volume-weighted temperature just below $\mathrm{T_{vir}}$ and the rest of the CGM remaining below $\mathrm{T_{vir}}$ (see their Figure 3). This is similar to the high $\eta_\mathrm{M}$ halo in \cite{Fielding_2017}. Interestingly, the time evolution of $\mathrm{T/T_{vir}}$ in our work more closely resembles the Milky Way-mass sample ($\mathrm{M_h\sim10^{12}\,M_\odot}$) in \cite{Stern2021} than their low-mass sample. For example, in the FIRE-2 m12i halo, the radial volume-weighted temperature peak occurs at $\mathrm{\sim0.5\,R_{vir}}$ below $\mathrm{T_{vir}}$ at $\mathrm{z\sim1}$ and migrates inward to $\mathrm{\sim0.1\,R_{vir}}$ above $\mathrm{T_{vir}}$ by $\mathrm{z\sim0}$ (see their Figure 2).\footnote{Here \lq m12' refers to $\mathrm{M_h\sim10^{12}\,M_\odot}$ at $\mathrm{z=0}$.} This inward migration of the temperature peak and heating of the inner CGM above $\mathrm{T_{vir}}$ is consistent with what we find in our lower mass galaxy sample (Figure~\ref{fig:temperature_evolution}).

We can also compare our results to IllustrisTNG and EAGLE. \cite{Wright_2024} report the median CGM temperature profile in the halo mass range $\mathrm{\log_{10}(M_{200c}/M_\odot)\in[10.75,11.75]}$, which overlaps with the upper end of our sample (Table~\ref{table:properties}). At $\mathrm{z\sim2}$, both IllustrisTNG and EAGLE have a CGM well below $\mathrm{T_{vir}}$, although IllustrisTNG is notably hotter than EAGLE at all radii within $\mathrm{\sim1\,R_{vir}}$ (see their Figure C1). By $\mathrm{z\sim0}$, the two simulations converge to a nearly identical temperature profile, with EAGLE being marginally hotter at the 84th percentile (see their Figure C2). This is consistent with the temperature evolution reported in EAGLE by \cite{Correa2018} (see their Figure 1). In the \textsc{arkenstone} model, \cite{Bennett2025} find that the mean CGM temperature is higher than the fiducial IllustrisTNG run in the halo mass range $\mathrm{M_{200}\gtrsim10^{11}\,M_\odot}$ at both $\mathrm{z\sim0}$ and $\mathrm{z\sim2}$ (see their Figures 10 and 15). However, at $\mathrm{z\sim0}$, the fiducial IllustrisTNG run is hotter than \textsc{arkenstone} in the halo mass range $\mathrm{M_{200}\lesssim5\times10^{10}\,M_\odot}$, and \cite{Bennett2025} find that the CGM temperature in this low-mass range increases with specific energy, yet still does not reach the fiducial IllustrisTNG result. The origin of this discrepancy at the lowest halo masses is unclear. We also note that \cite{Correa2018}, \cite{Wright_2024}, and \cite{Bennett2025} all report the mass-weighted temperature, which is likely biased toward the cold, dense gas at small radii ($\mathrm{\lesssim0.25\,R_{vir}}$), and therefore not directly comparable to our volume-weighted results.

Lastly, another related work is \cite{Cook2025}, who find that halos in a similar mass range have both mass- and volume-weighted CGM temperatures just below $\mathrm{T_{vir}}$, with the inner CGM notably cooler than what we find (see their Figures~3 and~4). Similarly, in a higher-resolution re-run of $\mathrm{M_h \sim 10^{10} - 10^{11}\,M_\odot}$ halos from IllustrisTNG, \cite{Tung2025} find that most CGM gas at $z \sim 0$ is below $10^5\,\mathrm{K}$ by mass, yet a substantial fraction of the CGM volume exceeds $10^5\,\mathrm{K}$ (see their Figure~5). These volume-filling hot outflows are qualitatively consistent with the hot outflows in our work.

\subsubsection{Specific Energy}
\label{subsubsec:specific_energy}

One of the key questions motivating this work is the role of high specific-energy outflows in driving the thermodynamic evolution of the CGM. In \S\ref{subsubsec:specific energy}, we demonstrated that the hot phase ($\mathrm{>3\times10^5\,K}$) has a large enough specific energy to escape the halo potential, therefore heating the surrounding CGM and IGM. In general, our results are broadly consistent with previous findings from high-resolution simulations, including \cite{Steinwandel_2024}, \cite{Kim_2020}, and \cite{Li_Bryan_2020}. In Figure~\ref{fig:se_evolution}, $\mathrm{e_{s,ISM,hot}\sim5\times10^{14}\,erg\,g^{-1}}$, which is consistent with around  half of the literature with a gas surface density of $\mathrm{\Sigma_{gas}\sim1-10\,M_\odot\,kpc^{-2}}$ (see \cite{Li_Bryan_2020} Figure 2). However, other work presented in \cite{Li_Bryan_2020} have $\mathrm{e_{s,ISM,hot}\sim10^{15}\,erg\,g^{-1}}$ in the same $\mathrm{\Sigma_{gas}}$ range \citep[e.g.,][]{Kim_2020}. This difference may be a lower limit due to where we measure $\mathrm{e_s}$ in the simulation.  

For example, the $\mathrm{e_s}$ reported in \cite{Li_Bryan_2020} are measured at $\mathrm{\sim0.5-1\,kpc}$ above the midplane, compared to our choice of $\mathrm{0.1\,R_{vir}}$ ($\mathrm{\sim10\,kpc}$ at $\mathrm{z=0}$). \cite{Steinwandel_2024} report $\mathrm{\dot{M}_{out}}$ and $\mathrm{\dot{E}_{out}}$ as a function of height above the midplane (see their Figures 8 and 9). If we use their hot wind component ($\mathrm{T>5\times10^5\,K}$), we find $\mathrm{e_s\sim8\times10^{14}\,erg\,g^{-1}}$ at $\mathrm{1\,kpc}$ and $\mathrm{e_s\sim1\times10^{15}\,erg\,g^{-1}}$ at $\mathrm{10\,kpc}$ above the disk. The increase in $\mathrm{e_s}$ with height is consistent with the notion that most mass recycling occurs close to the disk, which will drive down $\mathrm{e_s\equiv\dot{E}_{out}/\dot{M}_{out}}$ at a small height relative to our choice of $\mathrm{0.1\,R_{vir}}$. If this behavior is also present in the other work reported in \cite{Li_Bryan_2020}, the true discrepancy with our work may be larger than a factor of $\sim2$. We also note that the dwarf galaxy in \cite{Steinwandel_2024} is  $\mathrm{M_h\sim2\times10^{11}\,M_\odot}$, a factor of $\sim3$ more massive than our most massive halo, M1 (Table~\ref{table:properties}), so we caution against a direct comparison. 

In the fiducial low (high) $\mathrm{\eta_M}$ run in \cite{Fielding_2017}, the input specific energy is $\mathrm{\log_{10}e_s\sim14.9\,erg\,g^{-1}}$ ($\mathrm{15.2\,erg\,g^{-1}}$). It is important to note that this is the \emph{input} specific energy, meaning that this is the $\mathrm{e_s}$ placed into a cell at $\mathrm{0.025\,R_{vir}}$. Therefore, this cell will already have mass and a lower temperature and will likely have a much lower $\mathrm{e_s}$ (due to work against gravity and gas and radiative cooling) by the time the outflow reaches the boundary we measure $\mathrm{e_s}$ ($\mathrm{0.1-0.2\,R_{vir}}$). Note that our input $\mathrm{e_s}$ (along with the other works discussed above) is much higher, since it is simply given by $\mathrm{\log_{10}\frac{E_{SN}}{m_\ast}\sim15.70\,erg\,g^{-1}}$. The potentially low input $\mathrm{e_S}$ in \cite{Fielding_2017} may also explain the slight temperature profile discrepancy (and $\mathrm{t_{cool}/t_{ff}}$, see \S\ref{subsubsec:radiative_stability_discussion}).

We also compare our results to \cite{Pandya2023}, who measure $\mathrm{\dot{E}}$ and $\mathrm{\dot{M}}$ at both the ISM and CGM boundary in FIRE-2. Approximating their specific energy as $\mathrm{\dot{E}_\mathrm{wind}/\dot{M}_\mathrm{wind}}$ at the ISM and $\mathrm{\dot{E}_\mathrm{out,halo}/\dot{M}_\mathrm{out,halo}}$ at the CGM boundary (see their Figures~8 and 11) for halos with present-day mass 
$\sim10^{11}\,\mathrm{M_\odot}$, we find that our $\mathrm{e_s}$ values are a factor of $\sim2$ larger at both boundaries when no temperature cut is applied. 
In the same FIRE-2 sample, \cite{Pandya_2021} report the Bernoulli velocity for $\mathrm{T} > 10^5\,\mathrm{K}$ gas in FIRE-2, which is consistent with our hot phase reported in Figure~\ref{fig:se_evolution}. However, their intentional selection of the fastest outflows (using a Bernoulli velocity cut) means their results represent an upper limit on the FIRE-2 specific energy. Together, these comparisons suggest that outflows in FIRE-2 carry lower specific energy than in our simulations, consistent with their higher mass-loading factors \citep[see \S\ref{sec:intro};][]{Muratov2015}, therefore leading to a reduced ability to heat and deposit energy into the CGM and IGM.

\subsubsection{Energy Flow Rate}
\label{subsubsec:energy flow}

We now turn to the energy flow rate. In 
\S\ref{subsubsec:energy flow rate}, we demonstrated that although infrequent, SNe-induced outflows are able to drive $\mathrm{\dot{E}_{out,ISM}}$ to $\mathrm{10^{41}-10^{42}\,erg\,s^{-1}}$, with the largest and most frequent outbursts occurring at high redshift, likely due to an elevated merger rate. 

Turning to comparisons with other work, \cite{Pandya_2021,Pandya2023} compute $\mathrm{\dot{E}_{out,ISM}}$ and $\mathrm{\dot{E}_{out,CGM}}$ in FIRE-2, finding that, for their $\mathrm{M_h\sim10^{10}\,M_\odot}$ sample, $\mathrm{\dot{E}_{out,ISM}}$ and $\mathrm{\dot{E}_{out,CGM}}$ drop off very quickly \citep[see Figure 11 in ][]{Pandya2023}. In comparison, their $\mathrm{M_h\sim10^{11}\,M_\odot}$ sample is broadly consistent with our work in both the typical $\mathrm{\dot{E}}$ value and its bursty, redshift-dependent behavior. \cite{Oren2026} perform a similar calculation in TNG100, computing $\mathrm{\dot{E}_{out,ISM}}$ and $\mathrm{\dot{E}_{out,CGM}}$ as a function of $\mathrm{M_{200c}}$ (see their Figure 13); their median values are consistent with what we find though we refrain from a more direct comparison since their results aren't smoothed on a similar timescale.

In their regulator model, \cite{Carr_2023} report the $\mathrm{\dot{E}}$ time evolution for both a $\mathrm{M_h=10^{10}\,M_\odot}$ and $\mathrm{M_h=10^{11}\,M_\odot}$ halo, where $\mathrm{\dot{E}_{wind}}$, $\mathrm{\dot{E}_{out}}$, $\mathrm{\dot{E}_{in}}$, and $\mathrm{\dot{E}_{cool}}$ are equivalent to our $\mathrm{\dot{E}_{out,ISM}}$, $\mathrm{\dot{E}_{out,CGM}}$, $\mathrm{\dot{E}_{in,CGM}}$, and $\mathrm{\dot{E}_{cool,CGM}}$, respectively. Our results are significantly higher than their $\mathrm{M_h=10^{10}\,M_\odot}$ halo, so we restrict the comparison to their $\mathrm{M_h=10^{11}\,M_\odot}$ halo. There, our $\mathrm{\dot{E}_{out,ISM}}$, $\mathrm{\dot{E}_{out,CGM}}$, and $\mathrm{\dot{E}_{in,CGM}}$ fall below their $\mathrm{\dot{E}_{wind}}$, $\mathrm{\dot{E}_{out}}$, and $\mathrm{\dot{E}_{in}}$, respectively, suggesting that our outflows carry lower specific energy than assumed in the \cite{Carr_2023} model and therefore propagate less efficiently (see also \S\ref{subsubsec:mass_flow_rate}).

\subsubsection{Cumulative Energy}
\label{subsubsec:cumulative_sne_energy}

We demonstrated the significance of the cumulative energy in two ways. We first showed in \S\ref{subsubsec:cumulative energy} that, in the first $\mathrm{\sim5\,Gyr}$, the cumulative $\mathrm{E_{out,ISM}}$ exceeds $\mathrm{E_{cool,CGM}}$, highlighting the importance of infrequent yet energy-packed SNe outflows at high redshifts. A similar \lq bursty' nature of mass outflows has been characterized in other simulation work, including FIRE and FIRE-2 \citep{Muratov2015,Pandya_2021}. For example, \cite{Pandya_2021} calculated burst-averaged loading factors and found that the energy-loading in their $\mathrm{M_h\sim10^{10}-10^{11}\,M_\odot}$ halos is higher (and bursts are more frequent) at higher redshift (see their Figure 9). This result is qualitatively consistent with our finding. However, most literature doesn't calculate the cumulative $\mathrm{E_{out,ISM}}$ and $\mathrm{E_{cool,CGM}}$. \cite{Carr_2023} calculate $\mathrm{\dot{E}_{out,ISM}}$ and  $\mathrm{\dot{E}_{cool,CGM}}$ for their $\mathrm{M_h=10^{11}\,M_\odot}$ halo, but $\mathrm{\dot{E}_{cool,CGM}}>\mathrm{\dot{E}_{out,ISM}}$ at all model outputs. 

In \S\ref{subsec:baryon fraction}, we also showed a similar calculation, but compared the cumulative SNe energy ($\mathrm{E_{out,ISM}}$) to the binding energy of the halo baryon content ($\mathrm{E_{bind}^b}$). We found that the ratio $\mathrm{E_{out,ISM}}/\mathrm{E_{bind}^b}$ quickly exceeds 1 in the first few simulation outputs and remains at $\mathrm{\sim10-50}$ from $\mathrm{\sim5\,Gyr}$ until the present-day. A similar analysis was performed in \cite{Oppenheimer2020}, where they show that early-forming $L^\ast$ halos in EAGLE undergo rapid black hole growth (and therefore higher cumulative black hole energy) that suppress $f_\mathrm{CGM}$, producing a bimodality that may be observable via C~\textsc{iv} and O~\textsc{vi} absorption. Although their study looks at $L^\ast$ halos, they find that there is no correlation between the cumulative $\mathrm{E_{out,ISM}}$ and $f_\mathrm{CGM}$ and find $\mathrm{E_{out,ISM}}/\mathrm{E_{bind}^b}$ values consistent with our $\mathrm{z\sim0}$ result (Figure~\ref{fig:baryon_fraction}). However, with a larger sample, it could be interesting to test whether in the low-mass halo regime, a similar $f_\mathrm{CGM}$ bimodality can arise from a difference in the halo growth rate and subsequent SNe and outflows. For example, we found that by $\mathrm{\sim5\,Gyr}$, around $\mathrm{\sim90\%}$ of the present-day halo mass had already been acquired in the typical halo in our sample. It is likely that a different halo assembly history (e.g., late-time mergers) with the same present-day mass will show a similar bimodality that \cite{Oppenheimer2020} found.       

The other notable consequence of $\mathrm{E_{out,ISM}}/\mathrm{E_{bind}^b}$ exceeding $\mathrm{\sim1}$ within the first few Gyr is the ability for outflows to unbind and \lq sweep' up mass beyond $\mathrm{R_{vir}}$. This can also be clearly seen in the $\mathrm{M_{gas}}$ time evolution, where there is a turnover at $\mathrm{\sim2\,Gyr}$. This gradual decrease in $\mathrm{M_{gas}}$ is not seen in other simulations, including FIRE-2 \citep{Pandya_2021}. In EAGLE, \cite{Mitchell_2020} find that outflow mass entrainment arises from the over-pressurization compared to the ambient CGM \citep[also see][]{Wright_2024}. \cite{Mitchell_2020} argue that this is a direct consequence of injecting energy into a smaller amount of mass and therefore having higher specific energy outflows compared to other cosmological simulations (see their Section 5.2). Although it is not explicitly shown in other work, we expect that the $\mathrm{M_{gas}}$ time evolution in EAGLE will therefore be similar to our result.

\subsubsection{Mass Flow Rate}
\label{subsubsec:mass_flow_rate}

In \S\ref{subsec:flow of mass}, we reported the mass flow rate and found that there is a significant decrease in $\mathrm{\dot{M}_{in}}$ and $\mathrm{\dot{M}_{out}}$ (in both the ISM and CGM boundary) at $\mathrm{\sim5\,Gyr}$. We now compare these findings to other literature. As discussed in \cite{Wright_2024}, mass flow rates provide a more direct basis for comparison across the literature than mass-loading factors, which depend on the star formation rate and therefore vary with halo mass. \cite{Wright_2024} present a like-for-like comparison of inflow and outflow rates normalized by $\mathrm{M_{200c}\times\Omega_b/\Omega_m}$ as a function of radius, down to $\mathrm{M_{200c}\sim10^{10.5}\,M_\odot}$. For consistency, we adopt their $\mathrm{M_{200c}\sim10^{10.8}\,M_\odot}$ result and convert to $\mathrm{M_\odot\,yr^{-1}}$ (Figure~\ref{fig:mass_flow_evolution}). We find that our $\mathrm{\dot{M}_{in}}$ and $\mathrm{\dot{M}_{out}}$ results more closely resemble EAGLE than IllustrisTNG (in both the ISM and CGM boundary). 

In terms of the mass outflow rate, our median $\mathrm{\dot{M}_{out,ISM}\sim0.1\,M_\odot\,yr^{-1}}$ is similar to EAGLE and significantly lower than IllustrisTNG. 
However, unlike EAGLE, we do not find $\mathrm{\dot{M}_{out,CGM}>\dot{M}_{out,ISM}}$ at $\mathrm{z\sim0}$. \cite{Wright_2024} interpret this regime in EAGLE as evidence for swept-up material in outflows \citep[see also][]{Mitchell_2020}, whereas IllustrisTNG is characterized by strong, efficient recycling within the CGM. In our simulation suite, $\mathrm{\dot{M}_{out,CGM}>\dot{M}_{out,ISM}}$ only at early times ($\mathrm{\lesssim5\,Gyr}$; Figure~\ref{fig:mass_entrainment}), and their ratio is approximately unity afterwards. This suggests that entrainment is most efficient at early times in our work, consistent with the gradual decline in $\mathrm{M_{CGM}}$ that we measure (Figure~\ref{fig:baryon_fraction}). In addition, since so much material has been entrained in the outflows at early times, there simply isn't much CGM material left to sweep up after the first $\mathrm{\sim5\,Gyr}$. This will naturally result in a lower $\mathrm{\dot{M}}$ compared to a gas-rich CGM (e.g., IllustrisTNG). 

In terms of the mass inflow rate at the halo boundary, \cite{Wright_2024} find that at $\mathrm{z\sim0}$ IllustrisTNG prevents $\sim50\%$ of accretion ($f_\mathrm{prev}\sim0.5$), whereas EAGLE prevents $\sim90\%$ ($f_\mathrm{prev}\sim0.1$). \cite{Oren2026} find a similar result in their TNG100 analysis at the halo-boundary (see their Figure 8). In general, EAGLE shows strong suppression of the inflow rate at both CGM and ISM boundaries, while IllustrisTNG exhibits a higher inflow rate, especially at the ISM boundary where there is heavy recycling. In our work, we compute an analogous $f_\mathrm{prev}$ and find substantial scatter (although near unity) at early times, followed by convergence to $f_\mathrm{prev}\sim0.25$ after $\mathrm{\sim5\,Gyr}$, corresponding to $\sim75\%$ suppression. This is consistent with EAGLE. In FIRE-2, \cite{Pandya_2020} measure $\mathrm{\dot{M}_{in,gas}/\dot{M}_{in,DM}}$ at the halo boundary and also find $f_\mathrm{prev}\sim0.1$ (see their Figure 14). 

In their regulator model, \cite{Carr_2023} report the time evolution of $\mathrm{\dot{M}}$ for their $\mathrm{M_h=10^{11}\,M_\odot}$ halo, where $\mathrm{\dot{M}_{wind}}$, $\mathrm{\dot{M}_{out}}$, and $\mathrm{\dot{M}_{in}}$ correspond to our $\mathrm{\dot{M}_{out,ISM}}$, $\mathrm{\dot{M}_{out,CGM}}$, and $\mathrm{\dot{M}_{in,CGM}}$, respectively (see their Figure 7). At the ISM boundary, our $\mathrm{\dot{M}_{out,ISM}}$ is significantly higher than \cite{Carr_2023} for a large fraction of time. In their model, $\mathrm{\dot{M}_{wind}}$ peaks at $\mathrm{\sim0.1\,M_\odot\,yr^{-1}}$ around $\mathrm{\sim2\,Gyr}$ and declines to $\mathrm{\sim0.02\,M_\odot\,yr^{-1}}$ by $\mathrm{z=0}$, whereas our $\mathrm{\dot{M}_{out,ISM}}$ is bursty at early times ($\mathrm{\sim0.1-1\,M_\odot\,yr^{-1}}$ at $\mathrm{\lesssim5\,Gyr}$) before settling to $\mathrm{\sim0.1\,M_\odot\,yr^{-1}}$ at later times ($\mathrm{\gtrsim5\,Gyr}$). In contrast, at the CGM boundary, our $\mathrm{\dot{M}_{out,CGM}}$ does not exceed their $\mathrm{\dot{M}_{out}}$: it is comparable until $\mathrm{\sim5\,Gyr}$, after which it falls to a factor of $\sim5$ below. $\mathrm{\dot{M}_{in,CGM}}$ exceeds $\mathrm{\dot{M}_{in}}$ at early times and becomes slightly lower after $\mathrm{\sim5\,Gyr}$. \cite{Carr_2023} find that reproducing the SHMR for a $\mathrm{M_h=10^{11}\,M_\odot}$ halo requires a prevention factor of $f_\mathrm{prev}\sim0.4$, implying that $\sim60\%$ of the expected baryon accretion rate is suppressed. Together with the results from EAGLE and FIRE-2, these findings underline the importance of preventive feedback.

\subsubsection{Radiative Stability}
\label{subsubsec:radiative_stability_discussion}

In the upper two panels in Figure~\ref{fig:cooling_time_ratio}, we demonstrated that $\mathrm{t_{cool}/t_{ff}}$ exceeding $\sim10$ in the outer CGM occurs around $\mathrm{\sim3-7\,Gyr}$. \cite{Stern2021} identified a related process in their FIRE-2 $\mathrm{M_h\sim10^{12}-10^{13}\,M_\odot}$ sample, which they term \lq inner CGM virialization'. In their framework, the outer halo ($\mathrm{\sim0.5\,R_{vir}}$) is already heated to $\mathrm{\sim T_{vir}}$, followed by a relatively brief phase during which the inner CGM ($\mathrm{\sim0.1\,R_{vir}}$) is also raised to $\mathrm{\sim T_{vir}}$. During this transition, $\mathrm{t_{cool}}$ grows to exceed $\mathrm{t_{ff}}$ ($\mathrm{t_{cool}/t_{ff}\sim2-10}$), a large fraction of the inner CGM becomes subsonic, and the pressure distribution becomes more spherically symmetric. This transition is also associated with a shift from bursty to steady star-formation and the emergence of a rotation-dominated disk. In their FIRE-2 sample, m12i undergoes this transition at $\mathrm{z\sim0.3}$, m12b at $\mathrm{z\sim0.7}$, and m13A1 at $\mathrm{z\sim3.5}$.\footnote{Here \lq m13' refers to $\mathrm{M_h\sim10^{13}\,M_\odot}$ at $\mathrm{z=1}$.} We find that our results better resemble these halos than the $\mathrm{10^{10}-10^{11}\,M_\odot}$ halos in FIRE-2, in both the temperature and $\mathrm{t_{cool}/t_{ff}}$ evolution. 

\cite{Stern2021} conclude that no \lq inner CGM virialization' occurs in halos with $\mathrm{M_h\lesssim10^{12}\,M_\odot}$ (see their Figure 17). In their $\mathrm{z=0}$ sample, halos with $\mathrm{M_h\sim5\times10^{10}\,M_\odot}$ have $\mathrm{t_{cool}/t_{ff}\sim0.2-0.7}$ at $\mathrm{0.1\,R_{vir}}$, whereas we find $\mathrm{t_{cool}/t_{ff}\sim4}$ at the same radius (Figure~\ref{fig:cooling_time_ratio}). At $\mathrm{z\sim2}$, \cite{Stern2021} report $\mathrm{t_{cool}/t_{ff}\sim0.04-0.1}$, while we find $\mathrm{t_{cool}/t_{ff}\gtrsim1}$ since at least this time. \cite{Stern2021} suggest that their low $\mathrm{t_{cool}/t_{ff}}$ values may be the result of the FIRE-2 CGM gas mass being too high. For example, \cite{ElBadry2018} find a CGM mass exceeding the combined cold gas and stellar mass by factors of several, whereas in our sample the CGM mass is comparable to or smaller than the ISM gas mass (Table~\ref{table:properties}). Similarly, \cite{Hafen2019} report $f_\mathrm{CGM}\sim0.15-0.30$ at $\mathrm{z=0.25}$, compared to $\sim0.01-0.15$ in our simulations (Figure~\ref{fig:baryon_fraction}). 

\cite{Stern2021} further note that if \lq inner CGM virialization' is linked to disk formation, its occurrence in $\mathrm{\sim10^{10}-10^{11}\,M_\odot}$ halos could alleviate tensions between FIRE predictions and observed disk morphologies \citep{ElBadry2018,ElBadry2018b}, while also producing steadier star formation histories \citep{Sparre2017,Emami2019}. Although we do not quantify rotational support, all galaxies in our sample exhibit clear disk structures, present since $\mathrm{z\sim1-2}$.\footnote{See \url{https://www.youtube.com/watch?v=jfcwSDvQeJ0&list=PLXH_man2bw11ifqwursfffNUGMmOWadm4}.} This is earlier than the disk formation timescales reported for higher-mass halos, though it may also reflect overestimated stellar masses in our simulations. 

\cite{Correa2018} study hot halo formation in EAGLE for similar halo masses and find that most gas resides in $\mathrm{t_{cool}/t_{ff} < 1}$ at $z = 0$, with only a modest fraction exceeding unity. However, they report the mass-weighted distribution whereas the volume-weighted distribution (especially excluding $\mathrm{0.1-0.2\,R_{vir}}$) would likely be more consistent with our work (see their Figure 1). They also demonstrate that the metagalactic UV background can significantly increase $\mathrm{t_{cool}/t_{ff}}$ by suppressing the gas cooling rate. Although we include a time-dependent UV background \citep{Haardt_Madau_2012} with self-shielding in our simulations (see \S\ref{sec:methodology}), we do not explicitly investigate its contribution to the $\mathrm{t_{cool}/t_{ff}}$ structure, and its relative importance compared to SNe-driven heating warrants further investigation in future work.

\cite{Carr_2023} report $\mathrm{t_{cool}/t_{ff}\sim20-35}$ at $\mathrm{z=0}$ for $\mathrm{M_h\sim10^{10}-10^{11}\,M_\odot}$ halos, significantly higher than both FIRE-2 and our work. However, we note that the \cite{Carr_2023} regulator framework does not explicitly model thermal instabilities, so we caution over-interpreting this result. In \cite{Carr_2023}, the high $\mathrm{t_{cool}/t_{ff}}$ values  may also reflect their low CGM gas mass. At $\mathrm{M_h\sim5\times10^{10}\,M_\odot}$, they find $f_\mathrm{CGM}\sim0.01$ (rising to $\sim0.03$ in their most mass-loaded case). This low CGM density will therefore drive $\mathrm{t_{cool}}$ up.

\subsubsection{Closure Radius}\label{subsubsec:closure radius discussion}

The typical closure radius ($\mathrm{R_c/R_{vir}}$) in our sample extends to $\sim$15 by $z\sim0$, which is most consistent with EAGLE from the results of \cite{Ayromlou_2023}. This agreement is perhaps unsurprising given the low $f_\mathrm{CGM}$ in our sample (Figure~\ref{fig:baryon_fraction}), which more closely resembles EAGLE ($f_\mathrm{CGM}\sim0.2$) than TNG ($f_\mathrm{CGM}\sim0.6$) at $\mathrm{M_h\sim10^{11}\,M_\odot}$ \citep{Crain_VanDeVoort_2023,Wright_2024}. However, $\mathrm{R_c}$ provides additional diagnostic power beyond $f_\mathrm{CGM}$, as it captures the spatial extent to which the CGM and IGM are depleted of baryons. 

We also examine the time evolution of $\mathrm{R_c}$. The scatter in $\mathrm{R_c}$ grows with time, consistent with the behaviour seen in $f_\mathrm{CGM}$ (Figure~\ref{fig:baryon_fraction}), though the increased scatter in $f_\mathrm{CGM}$ beyond $\sim$8\,Gyr appears to be driven primarily by a single halo that underwent significant $\mathrm{M_{CGM}}$ loss during a late-time merger. We additionally find that $\mathrm{R_c/R_{vir}}$ grows nearly linearly with time, suggesting that feedback-driven energy propagates into the CGM at a roughly constant rate, consistent with a characteristic outflow velocity that remains approximately steady.

\subsubsection{Absorption Line Comparison}\label{subsubsec:absorption discussion}

 We calculate the column density profiles of ions in Figure~\ref{fig:ions} and find general agreement with the archival and new detections reported in \cite{Zheng_2024}. In addition, we find general agreement with the EAGLE predictions presented in \cite{Zheng_2024}, where there is detectable H\,\textsc{i} out to large radii and only detectable metal ion absorbers (Si\,\textsc{ii}, Si\,\textsc{iii}, Si\,\textsc{iv}, C\,\textsc{ii}, C\,\textsc{iii}, and C\,\textsc{iv}) at small radii. However, our {\sc Ovi} column density predictions do not agree with the recent findings of \cite{Mishra_2024}, where they report that the {\sc Ovi} covering fraction is $\sim$ 40 percent at $\mathrm{1-2\, R_{vir}}$ and a steep drop off beyond $\mathrm{2\, R_{vir}}$. There are a few possible reasons for this difference. First, we do not account for the radiation from the stars themselves, which will likely only affect {\sc Ovi} close to the disk. Second, the \cite{Mishra_2024} sample is star-forming out to $\mathrm{z\sim0.7}$, whereas our sample does not have this selection bias (and is at an overall lower star formation rate by $\mathrm{z\sim0.7}$). It is also possible that the high {\sc Ovi} covering fraction at large radii arise from the cooling of the SNe-induced shock propagation in our work that isn't captured in the $\mathrm{z\sim0}$ median profile for each halo in Figure~\ref{fig:ions}. In addition, we aren't able to resolve the {\sc Ovi} at the interface of clouds and the surrounding medium \citep[radiative turbulent mixing layers;][]{Fielding_Bryan2022}. We also assume ionization equilibrium in the \textsc{cloudy} output we use from \textsc{trident}. 
 
In a recent analysis, \cite{Piacitelli2025} found that their 64 isolated dwarfs from the Marvel-ous Dwarfs \citep{Munshi2021} and Marvelous Massive Dwarfs \citep[zoom-in ICs selected from Rolumus25;][]{Tremmel2017} simulations under-predict the {\sc Civ} and {\sc Ovi} column density from \cite{Mishra_2024} (see their Figure 5). Their low ions are consistent with the literature reported in \cite{Zheng_2024}. \cite{Piacitelli2025} suggest that the discrepancy may arise from under-predicting the warm and diffuse gas in the outer halo and the strength of the feedback model \citep[also see][]{Nelson2018,Sanchez2019}. This discrepancy is also seen in other cosmological zoom-in dwarf work, including the $\mathrm{M_h\sim10^{10}}$ sample in \cite{Shen2014,Mina2021,Baumschlager2025} and the Auriga project \citep{Cook2025}. \cite{Fielding_2017} find a higher {\sc Civ} and {\sc Ovi} column density in their idealized simulation setup, although they assume a constant metallicity ($\mathrm{\sim0.3\,Z_\odot}$), which likely does not capture the variability between low metallicity inflows and high metallicity outflows.
 
 Our comparison to recent observational work needs to be explored in more detail in future work to better understand the distribution of metals in the CGM and beyond and how it is related to the underlying feedback.  

 \subsection{Caveats of our Simulation Suite} \label{subsec:caveats}
 
There are several important caveats to note about our zoom-in simulation suite. The most significant is that, as discussed in \S\ref{sec:present day properties}, our stellar masses are too high compared to the observed SHMR \citep{Moster_2010,Behroozi_2019}. We find that $\mathrm{M_\ast/M_h\sim10^{-2}}$ ($\mathrm{z\sim0}$) at $z \sim 0$, roughly an order of magnitude above the \cite{Behroozi_2019} median for a $\mathrm{M_h \sim 5 \times 10^{10}\,M_\odot}$ halo. This excess stellar mass is formed at early times, pointing to inefficient ISM feedback at high redshift and the well-known overcooling problem \citep{Wise2012, Stinson2013}.
Although M8-h includes one additional level of initial CGM refinement, the ISM resolution remains fixed across all runs, and we find no significant improvement in $\mathrm{M_\ast/M_h}$ at higher CGM resolution. This suggests the overcooling originates at ISM rather than CGM scales, and future work should explore higher ISM resolution alongside improved early feedback prescriptions such as radiation pressure or early stellar winds.

While the high stellar mass implies a higher integrated SNe energy budget, the lower amount that flows into the CGM (after ISM cooling) is commensurate with the expected energy input of a correctly placed stellar component. Therefore, we argue that the key qualitative results -- shock heating of the CGM to $\sim \mathrm{T}_\mathrm{vir}$, hot-phase energy dominance, and the transition to radiative stability -- are robust to moderate changes in the star formation rate. Nevertheless, a simulation suite that correctly reproduces the SHMR will be needed to place our findings on a fully quantitative footing, and we defer this to future work.

A second caveat concerns our sample size of ten halos, which limits our ability to characterize halo-to-halo scatter and to identify systematic trends with mass or assembly history within the $\mathrm{M_h \sim 10^{10} - 10^{11}\,M_\odot}$ range. Finally, our OVI column densities fall short of recent observations. As discussed in \S\ref{subsubsec:absorption discussion}, this likely reflects a combination of missing stellar radiation fields, the assumption of ionization equilibrium in our \textsc{cloudy} tables, and the inability to resolve radiative turbulent mixing layers at the cloud-CGM interface \citep{Fielding_Bryan2022}. 

\section{Summary \& Conclusion} \label{sec:summary and conclusion}

In this work, we investigate the role of SNe-driven outflows in regulating the thermodynamic structure of the CGM in $\mathrm{M_h}\sim 10^{10}-10^{11}~\mathrm{M_\odot}$ halos. We run a suite of cosmological zoom-in simulations with the adaptive mesh refinement code \textsc{enzo}, combined with a discrete SNe feedback model, and simulate ten halos down to $z = 0$. We uniquely resolve and track the flow of energy and mass using high cadence outputs every $\mathrm{\sim42\,Myr}$ (332 total outputs for each zoom-in). This enabled us to capture the energy and mass flow between the ISM, CGM, and IGM and explore how this is directly related to other quantities, including the CGM cooling rate, $f_\mathrm{CGM}$, and the $\mathrm{t_{cool}/t_{ff}}$ profile.

Our main conclusions are as follows:
\begin{enumerate}

    \item \textbf{SNe wind-induced shock heating sustains the CGM of dwarf galaxies at $\boldsymbol{\mathrm{\sim T_{vir}}}$.} We find that episodic SNe outbursts drive shock fronts that propagate from the star-forming disk outward through the CGM and into the IGM. In particular, the inner CGM ($\mathrm{\lesssim 0.5~R_{vir}}$) is continuously maintained near $\mathrm{T_{vir}}$ by $\mathrm{z\sim0}$, while the outer CGM and IGM are temporarily heated before cooling  on $\sim$Gyr timescales (Figures~\ref{fig:temp_slice_two} and \ref{fig:temperature_evolution}).

    \item \textbf{Hot outflows dominate the energy budget; warm outflows dominate the mass budget.} We find that the hot outflows ($\mathrm{T > 3 \times 10^5\,K}$) have a specific energy $\sim 10\times$ larger than the warm phase ($\mathrm{T \leq 3 \times 10^5\,K}$) and consistently exceed the halo escape threshold, driving outflows and preventing inflows at both the CGM and IGM scale (Figure~\ref{fig:se_evolution}). Although hot outflows are present for a smaller fraction of time compared to the warm outflows, the cumulative energy of the hot outflows entering the CGM from the ISM is $\mathrm{\sim5-10\times}$ more than the warm outflows. However, the cumulative mass in warm outflows is $\mathrm{\sim5-10\times}$ more than the hot outflows (Figures~\ref{fig:energy_flow_evolution} and \ref{fig:mass_flow_evolution}).     

    \item \textbf{Early evolution dominated by infrequent yet large SNe outflows and mass entrainment.} Although $\mathrm{\dot{E}_{cool,CGM}}>\mathrm{\dot{E}_{out,ISM}}$ for a larger fraction of time at early times, the infrequent yet powerful $\mathrm{\dot{E}_{out,ISM}}$ outflows quickly dominate the cumulative energy placed into the CGM compared to the cumulative CGM cooling rate within the first $\mathrm{\sim5\,Gyr}$ (Figure~\ref{fig:cumulative_energy}). During this time period, we also find  $\mathrm{\dot{M}_{out,CGM}>\dot{M}_{out,ISM}}$, demonstrating that outflows are able to \lq sweep' up CGM and IGM material and gradually lower the CGM baryon mass, traced using either $\mathrm{M_{gas}}$ or $f_\mathrm{CGM}$ (Figures~\ref{fig:mass_entrainment} and \ref{fig:baryon_fraction}).     
    
    \item \textbf{SNe feedback drives the CGM toward a radiatively stable 
    regime after 
    $\boldsymbol{\sim 5~\mathrm{Gyr}}$.} At early times, the high $f_\mathrm{CGM}$ is able to maintain the CGM in the  
    $\mathrm{t_{cool}/t_{ff} \ll 10}$, despite ongoing SNe-induced shock heating. However, during the time period 
    $\sim 3-7\,\mathrm{Gyr}$, the decline in $f_\mathrm{CGM}$ and increase in $\mathrm{t_{cool}}$ 
    drive $\mathrm{t_{cool}/t_{ff}}$ to $\sim10$ in the outer CGM and nearby IGM ($\mathrm{r/R_{vir}\sim0.4-1.5}$). In contrast, the inner CGM ($\mathrm{r/R_{vir}\sim0.1-0.4}$) is maintained at $\mathrm{t_{cool}/t_{ff} \sim 1-10}$ (Figure~\ref{fig:cooling_time_ratio}).

    \item \textbf{There is a transition at $\boldsymbol{\mathrm{\sim5\,Gyr}}$ where the dwarf galaxy CGM enters an equilibrium.} We find that after $\mathrm{5-8\,Gyr}$, $f_\mathrm{CGM}$ is driven down from near unity to $\mathrm{\sim0.1}$ and $\sim75\%$ of the expected baryon accretion rate is prevented (Figure~\ref{fig:baryon_fraction}). This transition is also connected to heavy mass entrainment in the outflows (which inevitably drive down $f_\mathrm{CGM}$) and when the CGM becomes radiatively stable. Importantly, leading up to this transition is also when the cumulative SNe outflow energy rate begins to exceed the cumulative CGM cooling rate and the baryon binding energy (see \S\ref{subsec:the physical picture} for a more detailed discussion).       

    \item \textbf{H{\small I} and low ions  agree with observations but O{\small VI} is underestimated.} While our H{\small I} and low-ionization ion (e.g., Si{\small II}, Si{\small III}, Si{\small IV}, C{\small II}, C{\small IV}) column density 
    profiles broadly agree with observations \citep{Johnson_2017,Zheng_2024}, our predicted O{\small VI} covering fraction falls short of \cite{Mishra_2024} (Figure~\ref{fig:ions}).

\end{enumerate} 

In future work, it would be interesting to further investigate SNe-induced shock-heating in low mass halos and its role in driving the CGM to a radiatively stable regime. In particular, this would include incorporating tracer particles in \textsc{enzo} to better estimate the gas that is on first infall versus gas that has been recycled. In addition, it would be useful to increase the sample size to get a better sense of the halo-to-halo scatter, especially when comparing to IllustrisTNG and EAGLE.  

Lastly, it would be worthwhile to make a more direct comparison to previous simulation work. For example, examining how the $\mathrm{t_{cool}/t_{ff}}$ profile evolves as a function of time in e.g. IllustrisTNG, EAGLE, and FIRE-2. In addition, it would be insightful to make a more direct comparison with the bathtub model presented in \cite{Carr_2023} and investigate what combination of parameters best describes our findings. This comparison can also be performed in the more flexible \textsc{sapphire} model \citep{Pandya2026}, where galaxy regulation and the role of various astrophysical parameters can be understood from a semi-analytic perspective with unprecedented detail.
 
\begin{acknowledgments}
MM would like to thank Viraj Pandya, Mary Putman, and Mordecai-Mark Mac Low for the extensive feedback on the second-year report,
which was the basis of this project. In addition, M.M. would like to thank other members of his thesis committee for insightful conversations on the project, including Shy Genel and David Schiminovich. GLB acknowledges support from the NSF (AST-2108470, AST-2307419), NASA TCAN award 80NSSC21K1053, and the Simons Foundation through the Learning the Universe Collaboration. We acknowledge computing resources from NSF ACCESS Allocation PHY240043.
\end{acknowledgments}

\vspace{5mm}

\facilities{Anvil (Purdue RCAC), Frontera (TACC), Ginsburg \& Insomnia Cluster (Columbia University)}
            
\software{\textsc{enzo} \citep{Bryan_2014}, \textsc{music} \citep{Hahn_Abel_2011}, \textsc{yt} \citep{Turk_2011}, \textsc{trident} \citep{Trident}, Astropy \citep{astropy}}

\bibliography{bibliography}{}

@ARTICLE{Katz1996,
       author = {{Katz}, Neal and {Weinberg}, David H. and {Hernquist}, Lars},
        title = "{Cosmological Simulations with TreeSPH}",
      journal = {\apjs},
         year = 1996,
        month = jul,
       volume = {105},
        pages = {19},
          doi = {10.1086/192305},
archivePrefix = {arXiv},
       eprint = {astro-ph/9509107},
 primaryClass = {astro-ph},
       adsurl = {https://ui.adsabs.harvard.edu/abs/1996ApJS..105...19K}
}

@ARTICLE{Chen1998,
       author = {{Chen}, Hsiao-Wen and {Lanzetta}, Kenneth M. and {Webb}, John K. and {Barcons}, Xavier},
        title = "{The Gaseous Extent of Galaxies and the Origin of Ly{\ensuremath{\alpha}} Absorption Systems. III. Hubble Space Telescope Imaging of Ly{\ensuremath{\alpha}}-absorbing Galaxies at z < 1}",
      journal = {\apj},
         year = 1998,
        month = may,
       volume = {498},
       number = {1},
        pages = {77-94},
          doi = {10.1086/305554},
archivePrefix = {arXiv},
       eprint = {astro-ph/9710310},
 primaryClass = {astro-ph},
       adsurl = {https://ui.adsabs.harvard.edu/abs/1998ApJ...498...77C}
}

@ARTICLE{Gronke2020,
       author = {{Gronke}, Max and {Oh}, S. Peng},
        title = "{Is multiphase gas cloudy or misty?}",
      journal = {\mnras},
         year = 2020,
        month = may,
       volume = {494},
       number = {1},
        pages = {L27-L31},
          doi = {10.1093/mnrasl/slaa033},
archivePrefix = {arXiv},
       eprint = {1912.07808},
 primaryClass = {astro-ph.GA},
       adsurl = {https://ui.adsabs.harvard.edu/abs/2020MNRAS.494L..27G}
}

@ARTICLE{McCourt2018,
       author = {{McCourt}, Michael and {Oh}, S. Peng and {O'Leary}, Ryan and {Madigan}, Ann-Marie},
        title = "{A characteristic scale for cold gas}",
      journal = {\mnras},
         year = 2018,
        month = feb,
       volume = {473},
       number = {4},
        pages = {5407-5431},
          doi = {10.1093/mnras/stx2687},
archivePrefix = {arXiv},
       eprint = {1610.01164},
 primaryClass = {astro-ph.GA},
       adsurl = {https://ui.adsabs.harvard.edu/abs/2018MNRAS.473.5407M}
}

@ARTICLE{Abruzzo2024,
       author = {{Abruzzo}, Matthew W. and {Fielding}, Drummond B. and {Bryan}, Greg L.},
        title = "{Taming the TuRMoiL: The Temperature Dependence of Turbulence in Cloud{\textendash}Wind Interactions}",
      journal = {\apj},
         year = 2024,
        month = may,
       volume = {966},
       number = {2},
          eid = {181},
        pages = {181},
          doi = {10.3847/1538-4357/ad1e51},
archivePrefix = {arXiv},
       eprint = {2210.15679},
 primaryClass = {astro-ph.GA},
       adsurl = {https://ui.adsabs.harvard.edu/abs/2024ApJ...966..181A}
}

@ARTICLE{Tumlinson2013,
       author = {{Tumlinson}, Jason and {Thom}, Christopher and {Werk}, Jessica K. and {Prochaska}, J. Xavier and {Tripp}, Todd M. and {Katz}, Neal and {Dav{\'e}}, Romeel and {Oppenheimer}, Benjamin D. and {Meiring}, Joseph D. and {Ford}, Amanda Brady and {O'Meara}, John M. and {Peeples}, Molly S. and {Sembach}, Kenneth R. and {Weinberg}, David H.},
        title = "{The COS-Halos Survey: Rationale, Design, and a Census of Circumgalactic Neutral Hydrogen}",
      journal = {\apj},
         year = 2013,
        month = nov,
       volume = {777},
       number = {1},
          eid = {59},
        pages = {59},
          doi = {10.1088/0004-637X/777/1/59},
archivePrefix = {arXiv},
       eprint = {1309.6317},
 primaryClass = {astro-ph.CO},
       adsurl = {https://ui.adsabs.harvard.edu/abs/2013ApJ...777...59T}
}

@ARTICLE{Dekel_Silk_1986,
       author = {{Dekel}, A. and {Silk}, J.},
        title = "{The Origin of Dwarf Galaxies, Cold Dark Matter, and Biased Galaxy Formation}",
      journal = {\apj},
         year = 1986,
        month = apr,
       volume = {303},
        pages = {39},
          doi = {10.1086/164050},
       adsurl = {https://ui.adsabs.harvard.edu/abs/1986ApJ...303...39D}
}

@ARTICLE{White_Frenk_1991,
       author = {{White}, Simon D.~M. and {Frenk}, Carlos S.},
        title = "{Galaxy Formation through Hierarchical Clustering}",
      journal = {\apj},
         year = 1991,
        month = sep,
       volume = {379},
        pages = {52},
          doi = {10.1086/170483},
       adsurl = {https://ui.adsabs.harvard.edu/abs/1991ApJ...379...52W}
}

@ARTICLE{Mishra_2024,
       author = {{Mishra}, Nishant and {Johnson}, Sean D. and {Rudie}, Gwen C. and {Chen}, Hsiao-Wen and {Schaye}, Joop and {Qu}, Zhijie and {Zahedy}, Fakhri S. and {Boettcher}, Erin T. and {Cantalupo}, Sebastiano and {Chen}, Mandy C. and {Faucher-Gigu{\'e}re}, Claude-Andr{\'e} and {Greene}, Jenny E. and {Li}, Jennifer I.-Hsiu and {Liu}, Zhuoqi (Will) and {Lopez}, Sebastian and {Petitjean}, Patrick},
        title = "{The Cosmic Ultraviolet Baryon Survey (CUBS). IX. The Enriched Circumgalactic and Intergalactic Medium Around Star-forming Field Dwarf Galaxies Traced by O VI Absorption}",
      journal = {\apj},
         year = 2024,
        month = nov,
       volume = {976},
       number = {1},
          eid = {149},
        pages = {149},
          doi = {10.3847/1538-4357/ad7b0a},
archivePrefix = {arXiv},
       eprint = {2408.11151},
 primaryClass = {astro-ph.GA},
       adsurl = {https://ui.adsabs.harvard.edu/abs/2024ApJ...976..149M}
}

@ARTICLE{Tumlinson_2011,
       author = {{Tumlinson}, J. and {Thom}, C. and {Werk}, J.~K. and {Prochaska}, J.~X. and {Tripp}, T.~M. and {Weinberg}, D.~H. and {Peeples}, M.~S. and {O'Meara}, J.~M. and {Oppenheimer}, B.~D. and {Meiring}, J.~D. and {Katz}, N.~S. and {Dav{\'e}}, R. and {Ford}, A.~B. and {Sembach}, K.~R.},
        title = "{The Large, Oxygen-Rich Halos of Star-Forming Galaxies Are a Major Reservoir of Galactic Metals}",
      journal = {Science},
         year = 2011,
        month = nov,
       volume = {334},
       number = {6058},
        pages = {948},
          doi = {10.1126/science.1209840},
archivePrefix = {arXiv},
       eprint = {1111.3980},
 primaryClass = {astro-ph.CO},
       adsurl = {https://ui.adsabs.harvard.edu/abs/2011Sci...334..948T}
}

@ARTICLE{Werk_2014,
       author = {{Werk}, Jessica K. and {Prochaska}, J. Xavier and {Tumlinson}, Jason and {Peeples}, Molly S. and {Tripp}, Todd M. and {Fox}, Andrew J. and {Lehner}, Nicolas and {Thom}, Christopher and {O'Meara}, John M. and {Ford}, Amanda Brady and {Bordoloi}, Rongmon and {Katz}, Neal and {Tejos}, Nicolas and {Oppenheimer}, Benjamin D. and {Dav{\'e}}, Romeel and {Weinberg}, David H.},
        title = "{The COS-Halos Survey: Physical Conditions and Baryonic Mass in the Low-redshift Circumgalactic Medium}",
      journal = {\apj},
         year = 2014,
        month = sep,
       volume = {792},
       number = {1},
          eid = {8},
        pages = {8},
          doi = {10.1088/0004-637X/792/1/8},
archivePrefix = {arXiv},
       eprint = {1403.0947},
 primaryClass = {astro-ph.CO},
       adsurl = {https://ui.adsabs.harvard.edu/abs/2014ApJ...792....8W}
}

@ARTICLE{Johnson_2017,
       author = {{Johnson}, Sean D. and {Chen}, Hsiao-Wen and {Mulchaey}, John S. and {Schaye}, Joop and {Straka}, Lorrie A.},
        title = "{The Extent of Chemically Enriched Gas around Star-forming Dwarf Galaxies}",
      journal = {\apjl},
         year = 2017,
        month = nov,
       volume = {850},
       number = {1},
          eid = {L10},
        pages = {L10},
          doi = {10.3847/2041-8213/aa9370},
archivePrefix = {arXiv},
       eprint = {1710.06441},
 primaryClass = {astro-ph.GA},
       adsurl = {https://ui.adsabs.harvard.edu/abs/2017ApJ...850L..10J}
}

@ARTICLE{Zheng_2024,
       author = {{Zheng}, Yong and {Faerman}, Yakov and {Oppenheimer}, Benjamin D. and {Putman}, Mary E. and {McQuinn}, Kristen B.~W. and {Kirby}, Evan N. and {Burchett}, Joseph N. and {Telford}, O. Grace and {Werk}, Jessica K. and {Kim}, Doyeon A.},
        title = "{A Comprehensive Investigation of Metals in the Circumgalactic Medium of Nearby Dwarf Galaxies}",
      journal = {\apj},
         year = 2024,
        month = jan,
       volume = {960},
       number = {1},
          eid = {55},
        pages = {55},
          doi = {10.3847/1538-4357/acfe6b},
archivePrefix = {arXiv},
       eprint = {2301.12233},
 primaryClass = {astro-ph.GA},
       adsurl = {https://ui.adsabs.harvard.edu/abs/2024ApJ...960...55Z}
}

@ARTICLE{Zheng_2020,
       author = {{Zheng}, Yong and {Emerick}, Andrew and {Putman}, Mary E. and {Werk}, Jessica K. and {Kirby}, Evan N. and {Peek}, Joshua},
        title = "{Characterizing the Circumgalactic Medium of the Lowest-mass Galaxies: A Case Study of IC 1613}",
      journal = {\apj},
         year = 2020,
        month = dec,
       volume = {905},
       number = {2},
          eid = {133},
        pages = {133},
          doi = {10.3847/1538-4357/abc875},
archivePrefix = {arXiv},
       eprint = {2010.15645},
 primaryClass = {astro-ph.GA},
       adsurl = {https://ui.adsabs.harvard.edu/abs/2020ApJ...905..133Z}
}

@ARTICLE{Zheng_2019,
       author = {{Zheng}, Yong and {Putman}, Mary E. and {Emerick}, Andrew and {McQuinn}, Kristen B.~W. and {Werk}, Jessica K. and {Lockman}, Felix J. and {Oppenheimer}, Benjamin D. and {Fox}, Andrew J. and {Kirby}, Evan N. and {Burchett}, Joseph N.},
        title = "{Tentative detection of the circumgalactic medium of the isolated low-mass dwarf galaxy WLM}",
      journal = {\mnras},
         year = 2019,
        month = nov,
       volume = {490},
       number = {1},
        pages = {467-477},
          doi = {10.1093/mnras/stz2563},
archivePrefix = {arXiv},
       eprint = {1909.05407},
 primaryClass = {astro-ph.GA},
       adsurl = {https://ui.adsabs.harvard.edu/abs/2019MNRAS.490..467Z}
}

@ARTICLE{Bordoli_2014,
       author = {{Bordoloi}, Rongmon and {Tumlinson}, Jason and {Werk}, Jessica K. and {Oppenheimer}, Benjamin D. and {Peeples}, Molly S. and {Prochaska}, J. Xavier and {Tripp}, Todd M. and {Katz}, Neal and {Dav{\'e}}, Romeel and {Fox}, Andrew J. and {Thom}, Christopher and {Ford}, Amanda Brady and {Weinberg}, David H. and {Burchett}, Joseph N. and {Kollmeier}, Juna A.},
        title = "{The COS-Dwarfs Survey: The Carbon Reservoir around Sub-L* Galaxies}",
      journal = {\apj},
         year = 2014,
        month = dec,
       volume = {796},
       number = {2},
          eid = {136},
        pages = {136},
          doi = {10.1088/0004-637X/796/2/136},
archivePrefix = {arXiv},
       eprint = {1406.0509},
 primaryClass = {astro-ph.GA},
       adsurl = {https://ui.adsabs.harvard.edu/abs/2014ApJ...796..136B}
}

@ARTICLE{Birnboim_Dekel_2003,
       author = {{Birnboim}, Yuval and {Dekel}, Avishai},
        title = "{Virial shocks in galactic haloes?}",
      journal = {\mnras},
         year = 2003,
        month = oct,
       volume = {345},
       number = {1},
        pages = {349-364},
          doi = {10.1046/j.1365-8711.2003.06955.x},
archivePrefix = {arXiv},
       eprint = {astro-ph/0302161},
 primaryClass = {astro-ph},
       adsurl = {https://ui.adsabs.harvard.edu/abs/2003MNRAS.345..349B}
}

@ARTICLE{Dekel_Birnboim_2006,
       author = {{Dekel}, Avishai and {Birnboim}, Yuval},
        title = "{Galaxy bimodality due to cold flows and shock heating}",
      journal = {\mnras},
         year = 2006,
        month = may,
       volume = {368},
       number = {1},
        pages = {2-20},
          doi = {10.1111/j.1365-2966.2006.10145.x},
archivePrefix = {arXiv},
       eprint = {astro-ph/0412300},
 primaryClass = {astro-ph},
       adsurl = {https://ui.adsabs.harvard.edu/abs/2006MNRAS.368....2D}
}

@ARTICLE{Koudmani_2019,
       author = {{Koudmani}, Sophie and {Sijacki}, Debora and {Bourne}, Martin A. and {Smith}, Matthew C.},
        title = "{Fast and energetic AGN-driven outflows in simulated dwarf galaxies}",
      journal = {\mnras},
         year = 2019,
        month = apr,
       volume = {484},
       number = {2},
        pages = {2047-2066},
          doi = {10.1093/mnras/stz097},
archivePrefix = {arXiv},
       eprint = {1812.04629},
 primaryClass = {astro-ph.GA},
       adsurl = {https://ui.adsabs.harvard.edu/abs/2019MNRAS.484.2047K}
}

@ARTICLE{Barai_2019,
       author = {{Barai}, Paramita and {de Gouveia Dal Pino}, Elisabete M.},
        title = "{Intermediate-mass black hole growth and feedback in dwarf galaxies at high redshifts}",
      journal = {\mnras},
         year = 2019,
        month = aug,
       volume = {487},
       number = {4},
        pages = {5549-5563},
          doi = {10.1093/mnras/stz1616},
archivePrefix = {arXiv},
       eprint = {1807.04768},
 primaryClass = {astro-ph.GA},
       adsurl = {https://ui.adsabs.harvard.edu/abs/2019MNRAS.487.5549B}
}

@ARTICLE{Sharma_2020,
       author = {{Sharma}, Ray S. and {Brooks}, Alyson M. and {Somerville}, Rachel S. and {Tremmel}, Michael and {Bellovary}, Jillian and {Wright}, Anna C. and {Quinn}, Thomas R.},
        title = "{Black Hole Growth and Feedback in Isolated ROMULUS25 Dwarf Galaxies}",
      journal = {\apj},
         year = 2020,
        month = jul,
       volume = {897},
       number = {1},
          eid = {103},
        pages = {103},
          doi = {10.3847/1538-4357/ab960e},
archivePrefix = {arXiv},
       eprint = {1912.06646},
 primaryClass = {astro-ph.GA},
       adsurl = {https://ui.adsabs.harvard.edu/abs/2020ApJ...897..103S}
}

@ARTICLE{Nelson_2013,
       author = {{Nelson}, Dylan and {Vogelsberger}, Mark and {Genel}, Shy and {Sijacki}, Debora and {Kere{\v{s}}}, Du{\v{s}}an and {Springel}, Volker and {Hernquist}, Lars},
        title = "{Moving mesh cosmology: tracing cosmological gas accretion}",
      journal = {\mnras},
         year = 2013,
        month = mar,
       volume = {429},
       number = {4},
        pages = {3353-3370},
          doi = {10.1093/mnras/sts595},
archivePrefix = {arXiv},
       eprint = {1301.6753},
 primaryClass = {astro-ph.CO},
       adsurl = {https://ui.adsabs.harvard.edu/abs/2013MNRAS.429.3353N}
}

@ARTICLE{Nelson_2016,
       author = {{Nelson}, Dylan and {Genel}, Shy and {Pillepich}, Annalisa and {Vogelsberger}, Mark and {Springel}, Volker and {Hernquist}, Lars},
        title = "{Zooming in on accretion - I. The structure of halo gas}",
      journal = {\mnras},
         year = 2016,
        month = aug,
       volume = {460},
       number = {3},
        pages = {2881-2904},
          doi = {10.1093/mnras/stw1191},
archivePrefix = {arXiv},
       eprint = {1503.02665},
 primaryClass = {astro-ph.CO},
       adsurl = {https://ui.adsabs.harvard.edu/abs/2016MNRAS.460.2881N}
}

@ARTICLE{Wright_2024,
       author = {{Wright}, Ruby J. and {Somerville}, Rachel S. and {Lagos}, Claudia del P. and {Schaller}, Matthieu and {Dav{\'e}}, Romeel and {Angl{\'e}s-Alc{\'a}zar}, Daniel and {Genel}, Shy},
        title = "{The baryon cycle in modern cosmological hydrodynamical simulations}",
      journal = {\mnras},
         year = 2024,
        month = aug,
       volume = {532},
       number = {3},
        pages = {3417-3440},
          doi = {10.1093/mnras/stae1688},
archivePrefix = {arXiv},
       eprint = {2402.08408},
 primaryClass = {astro-ph.GA},
       adsurl = {https://ui.adsabs.harvard.edu/abs/2024MNRAS.532.3417W}
}

@ARTICLE{Moster_2010,
       author = {{Moster}, Benjamin P. and {Somerville}, Rachel S. and {Maulbetsch}, Christian and {van den Bosch}, Frank C. and {Macci{\`o}}, Andrea V. and {Naab}, Thorsten and {Oser}, Ludwig},
        title = "{Constraints on the Relationship between Stellar Mass and Halo Mass at Low and High Redshift}",
      journal = {\apj},
         year = 2010,
        month = feb,
       volume = {710},
       number = {2},
        pages = {903-923},
          doi = {10.1088/0004-637X/710/2/903},
archivePrefix = {arXiv},
       eprint = {0903.4682},
 primaryClass = {astro-ph.CO},
       adsurl = {https://ui.adsabs.harvard.edu/abs/2010ApJ...710..903M}
}

@ARTICLE{Behroozi_2019,
       author = {{Behroozi}, Peter and {Wechsler}, Risa H. and {Hearin}, Andrew P. and {Conroy}, Charlie},
        title = "{UNIVERSEMACHINE: The correlation between galaxy growth and dark matter halo assembly from z = 0-10}",
      journal = {\mnras},
         year = 2019,
        month = sep,
       volume = {488},
       number = {3},
        pages = {3143-3194},
          doi = {10.1093/mnras/stz1182},
archivePrefix = {arXiv},
       eprint = {1806.07893},
 primaryClass = {astro-ph.GA},
       adsurl = {https://ui.adsabs.harvard.edu/abs/2019MNRAS.488.3143B}
}

@ARTICLE{Crain_VanDeVoort_2023,
       author = {{Crain}, Robert A. and {van de Voort}, Freeke},
        title = "{Hydrodynamical Simulations of the Galaxy Population: Enduring Successes and Outstanding Challenges}",
      journal = {\araa},
         year = 2023,
        month = aug,
       volume = {61},
        pages = {473-515},
          doi = {10.1146/annurev-astro-041923-043618},
archivePrefix = {arXiv},
       eprint = {2309.17075},
 primaryClass = {astro-ph.GA},
       adsurl = {https://ui.adsabs.harvard.edu/abs/2023ARA&A..61..473C}
}

@ARTICLE{Schaye_2015,
       author = {{Schaye}, Joop and {Crain}, Robert A. and {Bower}, Richard G. and {Furlong}, Michelle and {Schaller}, Matthieu and {Theuns}, Tom and {Dalla Vecchia}, Claudio and {Frenk}, Carlos S. and {McCarthy}, I.~G. and {Helly}, John C. and {Jenkins}, Adrian and {Rosas-Guevara}, Y.~M. and {White}, Simon D.~M. and {Baes}, Maarten and {Booth}, C.~M. and {Camps}, Peter and {Navarro}, Julio F. and {Qu}, Yan and {Rahmati}, Alireza and {Sawala}, Till and {Thomas}, Peter A. and {Trayford}, James},
        title = "{The EAGLE project: simulating the evolution and assembly of galaxies and their environments}",
      journal = {\mnras},
         year = 2015,
        month = jan,
       volume = {446},
       number = {1},
        pages = {521-554},
          doi = {10.1093/mnras/stu2058},
archivePrefix = {arXiv},
       eprint = {1407.7040},
 primaryClass = {astro-ph.GA},
       adsurl = {https://ui.adsabs.harvard.edu/abs/2015MNRAS.446..521S}
}

@ARTICLE{Crain_2015,
       author = {{Crain}, Robert A. and {Schaye}, Joop and {Bower}, Richard G. and {Furlong}, Michelle and {Schaller}, Matthieu and {Theuns}, Tom and {Dalla Vecchia}, Claudio and {Frenk}, Carlos S. and {McCarthy}, Ian G. and {Helly}, John C. and {Jenkins}, Adrian and {Rosas-Guevara}, Yetli M. and {White}, Simon D.~M. and {Trayford}, James W.},
        title = "{The EAGLE simulations of galaxy formation: calibration of subgrid physics and model variations}",
      journal = {\mnras},
         year = 2015,
        month = jun,
       volume = {450},
       number = {2},
        pages = {1937-1961},
          doi = {10.1093/mnras/stv725},
archivePrefix = {arXiv},
       eprint = {1501.01311},
 primaryClass = {astro-ph.GA},
       adsurl = {https://ui.adsabs.harvard.edu/abs/2015MNRAS.450.1937C}
}

@ARTICLE{Pillepich_2018,
       author = {{Pillepich}, Annalisa and {Springel}, Volker and {Nelson}, Dylan and {Genel}, Shy and {Naiman}, Jill and {Pakmor}, R{\"u}diger and {Hernquist}, Lars and {Torrey}, Paul and {Vogelsberger}, Mark and {Weinberger}, Rainer and {Marinacci}, Federico},
        title = "{Simulating galaxy formation with the IllustrisTNG model}",
      journal = {\mnras},
         year = 2018,
        month = jan,
       volume = {473},
       number = {3},
        pages = {4077-4106},
          doi = {10.1093/mnras/stx2656},
archivePrefix = {arXiv},
       eprint = {1703.02970},
 primaryClass = {astro-ph.GA},
       adsurl = {https://ui.adsabs.harvard.edu/abs/2018MNRAS.473.4077P}
}

@ARTICLE{Nelson_2019,
       author = {{Nelson}, Dylan and {Pillepich}, Annalisa and {Springel}, Volker and {Pakmor}, R{\"u}diger and {Weinberger}, Rainer and {Genel}, Shy and {Torrey}, Paul and {Vogelsberger}, Mark and {Marinacci}, Federico and {Hernquist}, Lars},
        title = "{First results from the TNG50 simulation: galactic outflows driven by supernovae and black hole feedback}",
      journal = {\mnras},
         year = 2019,
        month = dec,
       volume = {490},
       number = {3},
        pages = {3234-3261},
          doi = {10.1093/mnras/stz2306},
archivePrefix = {arXiv},
       eprint = {1902.05554},
 primaryClass = {astro-ph.GA},
       adsurl = {https://ui.adsabs.harvard.edu/abs/2019MNRAS.490.3234N}
}

@ARTICLE{Mitchell_2020,
       author = {{Mitchell}, Peter D. and {Schaye}, Joop and {Bower}, Richard G. and {Crain}, Robert A.},
        title = "{Galactic outflow rates in the EAGLE simulations}",
      journal = {\mnras},
         year = 2020,
        month = may,
       volume = {494},
       number = {3},
        pages = {3971-3997},
          doi = {10.1093/mnras/staa938},
archivePrefix = {arXiv},
       eprint = {1910.09566},
 primaryClass = {astro-ph.GA},
       adsurl = {https://ui.adsabs.harvard.edu/abs/2020MNRAS.494.3971M}
}

@ARTICLE{McQuinn_2019,
       author = {{McQuinn}, Kristen. B.~W. and {van Zee}, Liese and {Skillman}, Evan D.},
        title = "{Galactic Winds in Low-mass Galaxies}",
      journal = {\apj},
         year = 2019,
        month = nov,
       volume = {886},
       number = {1},
          eid = {74},
        pages = {74},
          doi = {10.3847/1538-4357/ab4c37},
archivePrefix = {arXiv},
       eprint = {1910.04167},
 primaryClass = {astro-ph.GA},
       adsurl = {https://ui.adsabs.harvard.edu/abs/2019ApJ...886...74M}
}

@ARTICLE{Marasco_2023,
       author = {{Marasco}, A. and {Belfiore}, F. and {Cresci}, G. and {Lelli}, F. and {Venturi}, G. and {Hunt}, L.~K. and {Concas}, A. and {Marconi}, A. and {Mannucci}, F. and {Mingozzi}, M. and {McLeod}, A.~F. and {Kumari}, N. and {Carniani}, S. and {Vanzi}, L. and {Ginolfi}, M.},
        title = "{Shaken, but not expelled: Gentle baryonic feedback from nearby starburst dwarf galaxies}",
      journal = {\aap},
         year = 2023,
        month = feb,
       volume = {670},
          eid = {A92},
        pages = {A92},
          doi = {10.1051/0004-6361/202244895},
archivePrefix = {arXiv},
       eprint = {2209.02726},
 primaryClass = {astro-ph.GA},
       adsurl = {https://ui.adsabs.harvard.edu/abs/2023A&A...670A..92M}
}

@ARTICLE{KadoFong_2024,
       author = {{Kado-Fong}, Erin and {Geha}, Marla and {Mao}, Yao-Yuan and {de los Reyes}, Mithi A.~C. and {Wechsler}, Risa H. and {Asali}, Yasmeen and {Kallivayalil}, Nitya and {Nadler}, Ethan O. and {Tollerud}, Erik J. and {Weiner}, Benjamin},
        title = "{SAGAbg. I. A Near-unity Mass-loading Factor in Low-mass Galaxies via Their Low-redshift Evolution in Stellar Mass, Oxygen Abundance, and Star Formation Rate}",
      journal = {\apj},
         year = 2024,
        month = may,
       volume = {966},
       number = {1},
          eid = {129},
        pages = {129},
          doi = {10.3847/1538-4357/ad3042},
archivePrefix = {arXiv},
       eprint = {2401.16469},
 primaryClass = {astro-ph.GA},
       adsurl = {https://ui.adsabs.harvard.edu/abs/2024ApJ...966..129K}
}

@ARTICLE{Concas_2022,
       author = {{Concas}, Alice and {Maiolino}, Roberto and {Curti}, Mirko and {Hayden-Pawson}, Connor and {Cirasuolo}, Michele and {Jones}, Gareth C. and {Mercurio}, Amata and {Belfiore}, Francesco and {Cresci}, Giovanni and {Cullen}, Fergus and {Mannucci}, Filippo and {Marconi}, Alessandro and {Cappellari}, Michele and {Cicone}, Claudia and {Peng}, Yingjie and {Troncoso}, Paulina},
        title = "{Being KLEVER at cosmic noon: Ionized gas outflows are inconspicuous in low-mass star-forming galaxies but prominent in massive AGN hosts}",
      journal = {\mnras},
         year = 2022,
        month = jun,
       volume = {513},
       number = {2},
        pages = {2535-2562},
          doi = {10.1093/mnras/stac1026},
archivePrefix = {arXiv},
       eprint = {2203.11958},
 primaryClass = {astro-ph.GA},
       adsurl = {https://ui.adsabs.harvard.edu/abs/2022MNRAS.513.2535C}
}

@ARTICLE{Chisholm_2017,
       author = {{Chisholm}, John and {Tremonti}, Christy A. and {Leitherer}, Claus and {Chen}, Yanmei},
        title = "{The mass and momentum outflow rates of photoionized galactic outflows}",
      journal = {\mnras},
         year = 2017,
        month = aug,
       volume = {469},
       number = {4},
        pages = {4831-4849},
          doi = {10.1093/mnras/stx1164},
archivePrefix = {arXiv},
       eprint = {1702.07351},
 primaryClass = {astro-ph.GA},
       adsurl = {https://ui.adsabs.harvard.edu/abs/2017MNRAS.469.4831C}
}

@ARTICLE{Schneider_Robertson_2018,
       author = {{Schneider}, Evan E. and {Robertson}, Brant E.},
        title = "{Introducing CGOLS: The Cholla Galactic Outflow Simulation Suite}",
      journal = {\apj},
         year = 2018,
        month = jun,
       volume = {860},
       number = {2},
          eid = {135},
        pages = {135},
          doi = {10.3847/1538-4357/aac329},
archivePrefix = {arXiv},
       eprint = {1803.01008},
 primaryClass = {astro-ph.GA},
       adsurl = {https://ui.adsabs.harvard.edu/abs/2018ApJ...860..135S}
}

@ARTICLE{Kim_2020,
       author = {{Kim}, Chang-Goo and {Ostriker}, Eve C. and {Somerville}, Rachel S. and {Bryan}, Greg L. and {Fielding}, Drummond B. and {Forbes}, John C. and {Hayward}, Christopher C. and {Hernquist}, Lars and {Pandya}, Viraj},
        title = "{First Results from SMAUG: Characterization of Multiphase Galactic Outflows from a Suite of Local Star-forming Galactic Disk Simulations}",
      journal = {\apj},
         year = 2020,
        month = sep,
       volume = {900},
       number = {1},
          eid = {61},
        pages = {61},
          doi = {10.3847/1538-4357/aba962},
archivePrefix = {arXiv},
       eprint = {2006.16315},
 primaryClass = {astro-ph.GA},
       adsurl = {https://ui.adsabs.harvard.edu/abs/2020ApJ...900...61K}
}

@ARTICLE{Steinwandel_2024,
       author = {{Steinwandel}, Ulrich P. and {Kim}, Chang-Goo and {Bryan}, Greg L. and {Ostriker}, Eve C. and {Somerville}, Rachel S. and {Fielding}, Drummond B.},
        title = "{The Structure and Composition of Multiphase Galactic Winds in a Large Magellanic Cloud Mass Simulated Galaxy}",
      journal = {\apj},
         year = 2024,
        month = jan,
       volume = {960},
       number = {2},
          eid = {100},
        pages = {100},
          doi = {10.3847/1538-4357/ad09e1},
archivePrefix = {arXiv},
       eprint = {2212.03898},
 primaryClass = {astro-ph.GA},
       adsurl = {https://ui.adsabs.harvard.edu/abs/2024ApJ...960..100S}
}

@ARTICLE{Carr_2023,
       author = {{Carr}, Christopher and {Bryan}, Greg L. and {Fielding}, Drummond B. and {Pandya}, Viraj and {Somerville}, Rachel S.},
        title = "{Regulation of Star Formation by a Hot Circumgalactic Medium}",
      journal = {\apj},
         year = 2023,
        month = may,
       volume = {949},
       number = {1},
          eid = {21},
        pages = {21},
          doi = {10.3847/1538-4357/acc4c7},
archivePrefix = {arXiv},
       eprint = {2211.05115},
 primaryClass = {astro-ph.GA},
       adsurl = {https://ui.adsabs.harvard.edu/abs/2023ApJ...949...21C}
}

@ARTICLE{Donahue_Voit_2022,
       author = {{Donahue}, Megan and {Voit}, G. Mark},
        title = "{Baryon cycles in the biggest galaxies}",
      journal = {\physrep},
         year = 2022,
        month = aug,
       volume = {973},
        pages = {1-109},
          doi = {10.1016/j.physrep.2022.04.005},
archivePrefix = {arXiv},
       eprint = {2204.08099},
 primaryClass = {astro-ph.GA},
       adsurl = {https://ui.adsabs.harvard.edu/abs/2022PhR...973....1D}
}

@ARTICLE{Li_2015,
       author = {{Li}, Yuan and {Bryan}, Greg L. and {Ruszkowski}, Mateusz and {Voit}, G. Mark and {O'Shea}, Brian W. and {Donahue}, Megan},
        title = "{Cooling, AGN Feedback, and Star Formation in Simulated Cool-core Galaxy Clusters}",
      journal = {\apj},
         year = 2015,
        month = oct,
       volume = {811},
       number = {2},
          eid = {73},
        pages = {73},
          doi = {10.1088/0004-637X/811/2/73},
archivePrefix = {arXiv},
       eprint = {1503.02660},
 primaryClass = {astro-ph.GA},
       adsurl = {https://ui.adsabs.harvard.edu/abs/2015ApJ...811...73L}
}

@ARTICLE{Fielding_2017,
       author = {{Fielding}, Drummond and {Quataert}, Eliot and {McCourt}, Michael and {Thompson}, Todd A.},
        title = "{The impact of star formation feedback on the circumgalactic medium}",
      journal = {\mnras},
         year = 2017,
        month = apr,
       volume = {466},
       number = {4},
        pages = {3810-3826},
          doi = {10.1093/mnras/stw3326},
archivePrefix = {arXiv},
       eprint = {1606.06734},
 primaryClass = {astro-ph.GA},
       adsurl = {https://ui.adsabs.harvard.edu/abs/2017MNRAS.466.3810F}
}

@ARTICLE{Pandya_2020,
       author = {{Pandya}, Viraj and {Somerville}, Rachel S. and {Angl{\'e}s-Alc{\'a}zar}, Daniel and {Hayward}, Christopher C. and {Bryan}, Greg L. and {Fielding}, Drummond B. and {Forbes}, John C. and {Burkhart}, Blakesley and {Genel}, Shy and {Hernquist}, Lars and {Kim}, Chang-Goo and {Tonnesen}, Stephanie and {Starkenburg}, Tjitske},
        title = "{First Results from SMAUG: The Need for Preventative Stellar Feedback and Improved Baryon Cycling in Semianalytic Models of Galaxy Formation}",
      journal = {\apj},
         year = 2020,
        month = dec,
       volume = {905},
       number = {1},
          eid = {4},
        pages = {4},
          doi = {10.3847/1538-4357/abc3c1},
archivePrefix = {arXiv},
       eprint = {2006.16317},
 primaryClass = {astro-ph.GA},
       adsurl = {https://ui.adsabs.harvard.edu/abs/2020ApJ...905....4P}
}

@ARTICLE{Bryan_2014,
       author = {{Bryan}, Greg L. and {Norman}, Michael L. and {O'Shea}, Brian W. and {Abel}, Tom and {Wise}, John H. and {Turk}, Matthew J. and {Reynolds}, Daniel R. and {Collins}, David C. and {Wang}, Peng and {Skillman}, Samuel W. and {Smith}, Britton and {Harkness}, Robert P. and {Bordner}, James and {Kim}, Ji-hoon and {Kuhlen}, Michael and {Xu}, Hao and {Goldbaum}, Nathan and {Hummels}, Cameron and {Kritsuk}, Alexei G. and {Tasker}, Elizabeth and {Skory}, Stephen and {Simpson}, Christine M. and {Hahn}, Oliver and {Oishi}, Jeffrey S. and {So}, Geoffrey C. and {Zhao}, Fen and {Cen}, Renyue and {Li}, Yuan and {Enzo Collaboration}},
        title = "{ENZO: An Adaptive Mesh Refinement Code for Astrophysics}",
      journal = {\apjs},
         year = 2014,
        month = apr,
       volume = {211},
       number = {2},
          eid = {19},
        pages = {19},
          doi = {10.1088/0067-0049/211/2/19},
archivePrefix = {arXiv},
       eprint = {1307.2265},
 primaryClass = {astro-ph.IM},
       adsurl = {https://ui.adsabs.harvard.edu/abs/2014ApJS..211...19B}
}

@ARTICLE{Goldbaum_2015,
       author = {{Goldbaum}, Nathan J. and {Krumholz}, Mark R. and {Forbes}, John C.},
        title = "{Mass Transport and Turbulence in Gravitationally Unstable Disk Galaxies. I. The Case of Pure Self-gravity}",
      journal = {\apj},
         year = 2015,
        month = dec,
       volume = {814},
       number = {2},
          eid = {131},
        pages = {131},
          doi = {10.1088/0004-637X/814/2/131},
archivePrefix = {arXiv},
       eprint = {1510.08458},
 primaryClass = {astro-ph.GA},
       adsurl = {https://ui.adsabs.harvard.edu/abs/2015ApJ...814..131G}
}

@ARTICLE{Goldbaum_2016,
       author = {{Goldbaum}, Nathan J. and {Krumholz}, Mark R. and {Forbes}, John C.},
        title = "{Mass Transport and Turbulence in Gravitationally Unstable Disk Galaxies. II: The Effects of Star Formation Feedback}",
      journal = {\apj},
         year = 2016,
        month = aug,
       volume = {827},
       number = {1},
          eid = {28},
        pages = {28},
          doi = {10.3847/0004-637X/827/1/28},
archivePrefix = {arXiv},
       eprint = {1605.00646},
 primaryClass = {astro-ph.GA},
       adsurl = {https://ui.adsabs.harvard.edu/abs/2016ApJ...827...28G}
}

@ARTICLE{Planck2018,
       author = {{Planck Collaboration} and {Aghanim}, N. and {Akrami}, Y. and {Ashdown}, M. and {Aumont}, J. and {Baccigalupi}, C. and {Ballardini}, M. and {Banday}, A.~J. and {Barreiro}, R.~B. and {Bartolo}, N. and {Basak}, S. and {Battye}, R. and {Benabed}, K. and {Bernard}, J. -P. and {Bersanelli}, M. and {Bielewicz}, P. and {Bock}, J.~J. and {Bond}, J.~R. and {Borrill}, J. and {Bouchet}, F.~R. and {Boulanger}, F. and {Bucher}, M. and {Burigana}, C. and {Butler}, R.~C. and {Calabrese}, E. and {Cardoso}, J. -F. and {Carron}, J. and {Challinor}, A. and {Chiang}, H.~C. and {Chluba}, J. and {Colombo}, L.~P.~L. and {Combet}, C. and {Contreras}, D. and {Crill}, B.~P. and {Cuttaia}, F. and {de Bernardis}, P. and {de Zotti}, G. and {Delabrouille}, J. and {Delouis}, J. -M. and {Di Valentino}, E. and {Diego}, J.~M. and {Dor{\'e}}, O. and {Douspis}, M. and {Ducout}, A. and {Dupac}, X. and {Dusini}, S. and {Efstathiou}, G. and {Elsner}, F. and {En{\ss}lin}, T.~A. and {Eriksen}, H.~K. and {Fantaye}, Y. and {Farhang}, M. and {Fergusson}, J. and {Fernandez-Cobos}, R. and {Finelli}, F. and {Forastieri}, F. and {Frailis}, M. and {Fraisse}, A.~A. and {Franceschi}, E. and {Frolov}, A. and {Galeotta}, S. and {Galli}, S. and {Ganga}, K. and {G{\'e}nova-Santos}, R.~T. and {Gerbino}, M. and {Ghosh}, T. and {Gonz{\'a}lez-Nuevo}, J. and {G{\'o}rski}, K.~M. and {Gratton}, S. and {Gruppuso}, A. and {Gudmundsson}, J.~E. and {Hamann}, J. and {Handley}, W. and {Hansen}, F.~K. and {Herranz}, D. and {Hildebrandt}, S.~R. and {Hivon}, E. and {Huang}, Z. and {Jaffe}, A.~H. and {Jones}, W.~C. and {Karakci}, A. and {Keih{\"a}nen}, E. and {Keskitalo}, R. and {Kiiveri}, K. and {Kim}, J. and {Kisner}, T.~S. and {Knox}, L. and {Krachmalnicoff}, N. and {Kunz}, M. and {Kurki-Suonio}, H. and {Lagache}, G. and {Lamarre}, J. -M. and {Lasenby}, A. and {Lattanzi}, M. and {Lawrence}, C.~R. and {Le Jeune}, M. and {Lemos}, P. and {Lesgourgues}, J. and {Levrier}, F. and {Lewis}, A. and {Liguori}, M. and {Lilje}, P.~B. and {Lilley}, M. and {Lindholm}, V. and {L{\'o}pez-Caniego}, M. and {Lubin}, P.~M. and {Ma}, Y. -Z. and {Mac{\'\i}as-P{\'e}rez}, J.~F. and {Maggio}, G. and {Maino}, D. and {Mandolesi}, N. and {Mangilli}, A. and {Marcos-Caballero}, A. and {Maris}, M. and {Martin}, P.~G. and {Martinelli}, M. and {Mart{\'\i}nez-Gonz{\'a}lez}, E. and {Matarrese}, S. and {Mauri}, N. and {McEwen}, J.~D. and {Meinhold}, P.~R. and {Melchiorri}, A. and {Mennella}, A. and {Migliaccio}, M. and {Millea}, M. and {Mitra}, S. and {Miville-Desch{\^e}nes}, M. -A. and {Molinari}, D. and {Montier}, L. and {Morgante}, G. and {Moss}, A. and {Natoli}, P. and {N{\o}rgaard-Nielsen}, H.~U. and {Pagano}, L. and {Paoletti}, D. and {Partridge}, B. and {Patanchon}, G. and {Peiris}, H.~V. and {Perrotta}, F. and {Pettorino}, V. and {Piacentini}, F. and {Polastri}, L. and {Polenta}, G. and {Puget}, J. -L. and {Rachen}, J.~P. and {Reinecke}, M. and {Remazeilles}, M. and {Renzi}, A. and {Rocha}, G. and {Rosset}, C. and {Roudier}, G. and {Rubi{\~n}o-Mart{\'\i}n}, J.~A. and {Ruiz-Granados}, B. and {Salvati}, L. and {Sandri}, M. and {Savelainen}, M. and {Scott}, D. and {Shellard}, E.~P.~S. and {Sirignano}, C. and {Sirri}, G. and {Spencer}, L.~D. and {Sunyaev}, R. and {Suur-Uski}, A. -S. and {Tauber}, J.~A. and {Tavagnacco}, D. and {Tenti}, M. and {Toffolatti}, L. and {Tomasi}, M. and {Trombetti}, T. and {Valenziano}, L. and {Valiviita}, J. and {Van Tent}, B. and {Vibert}, L. and {Vielva}, P. and {Villa}, F. and {Vittorio}, N. and {Wandelt}, B.~D. and {Wehus}, I.~K. and {White}, M. and {White}, S.~D.~M. and {Zacchei}, A. and {Zonca}, A.},
        title = "{Planck 2018 results. VI. Cosmological parameters}",
      journal = {\aap},
         year = 2020,
        month = sep,
       volume = {641},
          eid = {A6},
        pages = {A6},
          doi = {10.1051/0004-6361/201833910},
archivePrefix = {arXiv},
       eprint = {1807.06209},
 primaryClass = {astro-ph.CO},
       adsurl = {https://ui.adsabs.harvard.edu/abs/2020A&A...641A...6P}
}

@ARTICLE{Spitzer_1956,
       author = {{Spitzer}, Lyman, Jr.},
        title = "{On a Possible Interstellar Galactic Corona.}",
      journal = {\apj},
         year = 1956,
        month = jul,
       volume = {124},
        pages = {20},
          doi = {10.1086/146200},
       adsurl = {https://ui.adsabs.harvard.edu/abs/1956ApJ...124...20S}
}

@ARTICLE{Eisenstein_Hu_1998,
       author = {{Eisenstein}, Daniel J. and {Hu}, Wayne},
        title = "{Baryonic Features in the Matter Transfer Function}",
      journal = {\apj},
         year = 1998,
        month = mar,
       volume = {496},
       number = {2},
        pages = {605-614},
          doi = {10.1086/305424},
archivePrefix = {arXiv},
       eprint = {astro-ph/9709112},
 primaryClass = {astro-ph},
       adsurl = {https://ui.adsabs.harvard.edu/abs/1998ApJ...496..605E}
}

@ARTICLE{Hahn_Abel_2011,
       author = {{Hahn}, Oliver and {Abel}, Tom},
        title = "{Multi-scale initial conditions for cosmological simulations}",
      journal = {\mnras},
         year = 2011,
        month = aug,
       volume = {415},
       number = {3},
        pages = {2101-2121},
          doi = {10.1111/j.1365-2966.2011.18820.x},
archivePrefix = {arXiv},
       eprint = {1103.6031},
 primaryClass = {astro-ph.CO},
       adsurl = {https://ui.adsabs.harvard.edu/abs/2011MNRAS.415.2101H}
}

@ARTICLE{Onorbe_2014,
       author = {{O{\~n}orbe}, Jose and {Garrison-Kimmel}, Shea and {Maller}, Ariyeh H. and {Bullock}, James S. and {Rocha}, Miguel and {Hahn}, Oliver},
        title = "{How to zoom: bias, contamination and Lagrange volumes in multimass cosmological simulations}",
      journal = {\mnras},
         year = 2014,
        month = jan,
       volume = {437},
       number = {2},
        pages = {1894-1908},
          doi = {10.1093/mnras/stt2020},
archivePrefix = {arXiv},
       eprint = {1305.6923},
 primaryClass = {astro-ph.CO},
       adsurl = {https://ui.adsabs.harvard.edu/abs/2014MNRAS.437.1894O}
}

@ARTICLE{Eisenstein_Hut_1998,
       author = {{Eisenstein}, Daniel J. and {Hut}, Piet},
        title = "{HOP: A New Group-Finding Algorithm for N-Body Simulations}",
      journal = {\apj},
         year = 1998,
        month = may,
       volume = {498},
       number = {1},
        pages = {137-142},
          doi = {10.1086/305535},
archivePrefix = {arXiv},
       eprint = {astro-ph/9712200},
 primaryClass = {astro-ph},
       adsurl = {https://ui.adsabs.harvard.edu/abs/1998ApJ...498..137E}
}

@ARTICLE{Turk_2011,
   author = {{Turk}, M.~J. and {Smith}, B.~D. and {Oishi}, J.~S. and {Skory}, S. and
     {Skillman}, S.~W. and {Abel}, T. and {Norman}, M.~L.},
    title = "{yt: A Multi-code Analysis Toolkit for Astrophysical Simulation Data}",
  journal = {The Astrophysical Journal Supplement Series},
archivePrefix = "arXiv",
   eprint = {1011.3514},
 primaryClass = "astro-ph.IM",
     year = 2011,
    month = jan,
   volume = 192,
      eid = {9},
    pages = {9},
      doi = {10.1088/0067-0049/192/1/9},
   adsurl = {https://ui.adsabs.harvard.edu/abs/2011ApJS..192....9T}
}

@ARTICLE{Kim_2016,
       author = {{Kim}, Ji-hoon and {Agertz}, Oscar and {Teyssier}, Romain and {Butler}, Michael J. and {Ceverino}, Daniel and {Choi}, Jun-Hwan and {Feldmann}, Robert and {Keller}, Ben W. and {Lupi}, Alessandro and {Quinn}, Thomas and {Revaz}, Yves and {Wallace}, Spencer and {Gnedin}, Nickolay Y. and {Leitner}, Samuel N. and {Shen}, Sijing and {Smith}, Britton D. and {Thompson}, Robert and {Turk}, Matthew J. and {Abel}, Tom and {Arraki}, Kenza S. and {Benincasa}, Samantha M. and {Chakrabarti}, Sukanya and {DeGraf}, Colin and {Dekel}, Avishai and {Goldbaum}, Nathan J. and {Hopkins}, Philip F. and {Hummels}, Cameron B. and {Klypin}, Anatoly and {Li}, Hui and {Madau}, Piero and {Mandelker}, Nir and {Mayer}, Lucio and {Nagamine}, Kentaro and {Nickerson}, Sarah and {O'Shea}, Brian W. and {Primack}, Joel R. and {Roca-F{\`a}brega}, Santi and {Semenov}, Vadim and {Shimizu}, Ikkoh and {Simpson}, Christine M. and {Todoroki}, Keita and {Wadsley}, James W. and {Wise}, John H. and {AGORA Collaboration}},
        title = "{The AGORA High-resolution Galaxy Simulations Comparison Project. II. Isolated Disk Test}",
      journal = {\apj},
         year = 2016,
        month = dec,
       volume = {833},
       number = {2},
          eid = {202},
        pages = {202},
          doi = {10.3847/1538-4357/833/2/202},
archivePrefix = {arXiv},
       eprint = {1610.03066},
 primaryClass = {astro-ph.GA},
       adsurl = {https://ui.adsabs.harvard.edu/abs/2016ApJ...833..202K}
}

@ARTICLE{Bryan_1995,
       author = {{Bryan}, Greg L. and {Norman}, Michael L. and {Stone}, James M. and {Cen}, Renyue and {Ostriker}, Jeremiah P.},
        title = "{A piecewise parabolic method for cosmological hydrodynamics}",
      journal = {Computer Physics Communications},
         year = 1995,
        month = aug,
       volume = {89},
       number = {1-3},
        pages = {149-168},
          doi = {10.1016/0010-4655(94)00191-4},
       adsurl = {https://ui.adsabs.harvard.edu/abs/1995CoPhC..89..149B}
}

@ARTICLE{Krumholz_Tan_2007,
       author = {{Krumholz}, Mark R. and {Tan}, Jonathan C.},
        title = "{Slow Star Formation in Dense Gas: Evidence and Implications}",
      journal = {\apj},
         year = 2007,
        month = jan,
       volume = {654},
       number = {1},
        pages = {304-315},
          doi = {10.1086/509101},
archivePrefix = {arXiv},
       eprint = {astro-ph/0606277},
 primaryClass = {astro-ph},
       adsurl = {https://ui.adsabs.harvard.edu/abs/2007ApJ...654..304K}
}

@ARTICLE{Krumholz_2012,
       author = {{Krumholz}, Mark R. and {Dekel}, Avishai and {McKee}, Christopher F.},
        title = "{A Universal, Local Star Formation Law in Galactic Clouds, nearby Galaxies, High-redshift Disks, and Starbursts}",
      journal = {\apj},
         year = 2012,
        month = jan,
       volume = {745},
       number = {1},
          eid = {69},
        pages = {69},
          doi = {10.1088/0004-637X/745/1/69},
archivePrefix = {arXiv},
       eprint = {1109.4150},
 primaryClass = {astro-ph.CO},
       adsurl = {https://ui.adsabs.harvard.edu/abs/2012ApJ...745...69K}
}

@ARTICLE{Haardt_Madau_2012,
       author = {{Haardt}, Francesco and {Madau}, Piero},
        title = "{Radiative Transfer in a Clumpy Universe. IV. New Synthesis Models of the Cosmic UV/X-Ray Background}",
      journal = {\apj},
         year = 2012,
        month = feb,
       volume = {746},
       number = {2},
          eid = {125},
        pages = {125},
          doi = {10.1088/0004-637X/746/2/125},
archivePrefix = {arXiv},
       eprint = {1105.2039},
 primaryClass = {astro-ph.CO},
       adsurl = {https://ui.adsabs.harvard.edu/abs/2012ApJ...746..125H}
}

@ARTICLE{McCourt_2012,
       author = {{McCourt}, Michael and {Sharma}, Prateek and {Quataert}, Eliot and {Parrish}, Ian J.},
        title = "{Thermal instability in gravitationally stratified plasmas: implications for multiphase structure in clusters and galaxy haloes}",
      journal = {\mnras},
         year = 2012,
        month = feb,
       volume = {419},
       number = {4},
        pages = {3319-3337},
          doi = {10.1111/j.1365-2966.2011.19972.x},
archivePrefix = {arXiv},
       eprint = {1105.2563},
 primaryClass = {astro-ph.CO},
       adsurl = {https://ui.adsabs.harvard.edu/abs/2012MNRAS.419.3319M}
}

@ARTICLE{Voit_2017,
       author = {{Voit}, G. Mark and {Meece}, Greg and {Li}, Yuan and {O'Shea}, Brian W. and {Bryan}, Greg L. and {Donahue}, Megan},
        title = "{A Global Model for Circumgalactic and Cluster-core Precipitation}",
      journal = {\apj},
         year = 2017,
        month = aug,
       volume = {845},
       number = {1},
          eid = {80},
        pages = {80},
          doi = {10.3847/1538-4357/aa7d04},
archivePrefix = {arXiv},
       eprint = {1607.02212},
 primaryClass = {astro-ph.GA},
       adsurl = {https://ui.adsabs.harvard.edu/abs/2017ApJ...845...80V}
}

@ARTICLE{Lochhaas_2021,
       author = {{Lochhaas}, Cassandra and {Tumlinson}, Jason and {O'Shea}, Brian W. and {Peeples}, Molly S. and {Smith}, Britton D. and {Werk}, Jessica K. and {Augustin}, Ramona and {Simons}, Raymond C.},
        title = "{Figuring Out Gas \& Galaxies In Enzo (FOGGIE). V. The Virial Temperature Does Not Describe Gas in a Virialized Galaxy Halo}",
      journal = {\apj},
         year = 2021,
        month = dec,
       volume = {922},
       number = {2},
          eid = {121},
        pages = {121},
          doi = {10.3847/1538-4357/ac2496},
archivePrefix = {arXiv},
       eprint = {2102.08393},
 primaryClass = {astro-ph.GA},
       adsurl = {https://ui.adsabs.harvard.edu/abs/2021ApJ...922..121L}
}

@ARTICLE{Voit_2015,
       author = {{Voit}, G. Mark and {Bryan}, Greg L. and {O'Shea}, Brian W. and {Donahue}, Megan},
        title = "{Precipitation-regulated Star Formation in Galaxies}",
      journal = {\apjl},
         year = 2015,
        month = jul,
       volume = {808},
       number = {1},
          eid = {L30},
        pages = {L30},
          doi = {10.1088/2041-8205/808/1/L30},
archivePrefix = {arXiv},
       eprint = {1505.03592},
 primaryClass = {astro-ph.GA},
       adsurl = {https://ui.adsabs.harvard.edu/abs/2015ApJ...808L..30V}
}

@ARTICLE{Kim_Ostriker_2015,
       author = {{Kim}, Chang-Goo and {Ostriker}, Eve C.},
        title = "{Momentum Injection by Supernovae in the Interstellar Medium}",
      journal = {\apj},
         year = 2015,
        month = apr,
       volume = {802},
       number = {2},
          eid = {99},
        pages = {99},
          doi = {10.1088/0004-637X/802/2/99},
archivePrefix = {arXiv},
       eprint = {1410.1537},
 primaryClass = {astro-ph.GA},
       adsurl = {https://ui.adsabs.harvard.edu/abs/2015ApJ...802...99K}
}

@ARTICLE{Martizzi_2015,
       author = {{Martizzi}, Davide and {Faucher-Gigu{\`e}re}, Claude-Andr{\'e} and {Quataert}, Eliot},
        title = "{Supernova feedback in an inhomogeneous interstellar medium}",
      journal = {\mnras},
         year = 2015,
        month = jun,
       volume = {450},
       number = {1},
        pages = {504-522},
          doi = {10.1093/mnras/stv562},
archivePrefix = {arXiv},
       eprint = {1409.4425},
 primaryClass = {astro-ph.GA},
       adsurl = {https://ui.adsabs.harvard.edu/abs/2015MNRAS.450..504M}
}

@ARTICLE{Kim_2011,
       author = {{Kim}, Chang-Goo and {Kim}, Woong-Tae and {Ostriker}, Eve C.},
        title = "{Regulation of Star Formation Rates in Multiphase Galactic Disks: Numerical Tests of the Thermal/Dynamical Equilibrium Model}",
      journal = {\apj},
         year = 2011,
        month = dec,
       volume = {743},
       number = {1},
          eid = {25},
        pages = {25},
          doi = {10.1088/0004-637X/743/1/25},
archivePrefix = {arXiv},
       eprint = {1109.0028},
 primaryClass = {astro-ph.GA},
       adsurl = {https://ui.adsabs.harvard.edu/abs/2011ApJ...743...25K}
}

@ARTICLE{Cioffi_1988,
       author = {{Cioffi}, Denis F. and {McKee}, Christopher F. and {Bertschinger}, Edmund},
        title = "{Dynamics of Radiative Supernova Remnants}",
      journal = {\apj},
         year = 1988,
        month = nov,
       volume = {334},
        pages = {252},
          doi = {10.1086/166834},
       adsurl = {https://ui.adsabs.harvard.edu/abs/1988ApJ...334..252C}
}

@ARTICLE{Li_Bryan_2020,
       author = {{Li}, Miao and {Bryan}, Greg L.},
        title = "{Simple Yet Powerful: Hot Galactic Outflows Driven by Supernovae}",
      journal = {\apjl},
         year = 2020,
        month = feb,
       volume = {890},
       number = {2},
          eid = {L30},
        pages = {L30},
          doi = {10.3847/2041-8213/ab7304},
archivePrefix = {arXiv},
       eprint = {1910.09554},
 primaryClass = {astro-ph.GA},
       adsurl = {https://ui.adsabs.harvard.edu/abs/2020ApJ...890L..30L}
}

@ARTICLE{Forbes_2016,
       author = {{Forbes}, John C. and {Krumholz}, Mark R. and {Goldbaum}, Nathan J. and {Dekel}, Avishai},
        title = "{Suppression of star formation in dwarf galaxies by photoelectric grain heating feedback}",
      journal = {\nat},
         year = 2016,
        month = jul,
       volume = {535},
       number = {7613},
        pages = {523-525},
          doi = {10.1038/nature18292},
archivePrefix = {arXiv},
       eprint = {1605.00650},
 primaryClass = {astro-ph.GA},
       adsurl = {https://ui.adsabs.harvard.edu/abs/2016Natur.535..523F}
}

@ARTICLE{Leitherer_1999,
       author = {{Leitherer}, Claus and {Schaerer}, Daniel and {Goldader}, Jeffrey D. and {Delgado}, Rosa M. Gonz{\'a}lez and {Robert}, Carmelle and {Kune}, Denis Foo and {de Mello}, Du{\'\i}lia F. and {Devost}, Daniel and {Heckman}, Timothy M.},
        title = "{Starburst99: Synthesis Models for Galaxies with Active Star Formation}",
      journal = {\apjs},
         year = 1999,
        month = jul,
       volume = {123},
       number = {1},
        pages = {3-40},
          doi = {10.1086/313233},
archivePrefix = {arXiv},
       eprint = {astro-ph/9902334},
 primaryClass = {astro-ph},
       adsurl = {https://ui.adsabs.harvard.edu/abs/1999ApJS..123....3L}
}

@ARTICLE{Vazquez_Leitherer_2005,
       author = {{V{\'a}zquez}, Gerardo A. and {Leitherer}, Claus},
        title = "{Optimization of Starburst99 for Intermediate-Age and Old Stellar Populations}",
      journal = {\apj},
         year = 2005,
        month = mar,
       volume = {621},
       number = {2},
        pages = {695-717},
          doi = {10.1086/427866},
archivePrefix = {arXiv},
       eprint = {astro-ph/0412491},
 primaryClass = {astro-ph},
       adsurl = {https://ui.adsabs.harvard.edu/abs/2005ApJ...621..695V}
}

@ARTICLE{Leitherer_2014,
       author = {{Leitherer}, Claus and {Ekstr{\"o}m}, Sylvia and {Meynet}, Georges and {Schaerer}, Daniel and {Agienko}, Katerina B. and {Levesque}, Emily M.},
        title = "{The Effects of Stellar Rotation. II. A Comprehensive Set of Starburst99 Models}",
      journal = {\apjs},
         year = 2014,
        month = may,
       volume = {212},
       number = {1},
          eid = {14},
        pages = {14},
          doi = {10.1088/0067-0049/212/1/14},
archivePrefix = {arXiv},
       eprint = {1403.5444},
 primaryClass = {astro-ph.GA},
       adsurl = {https://ui.adsabs.harvard.edu/abs/2014ApJS..212...14L}
}

@ARTICLE{Chabrier_2003,
       author = {{Chabrier}, Gilles},
        title = "{Galactic Stellar and Substellar Initial Mass Function}",
      journal = {\pasp},
         year = 2003,
        month = jul,
       volume = {115},
       number = {809},
        pages = {763-795},
          doi = {10.1086/376392},
archivePrefix = {arXiv},
       eprint = {astro-ph/0304382},
 primaryClass = {astro-ph},
       adsurl = {https://ui.adsabs.harvard.edu/abs/2003PASP..115..763C}
}

@ARTICLE{Smith_2024a,
       author = {{Smith}, Matthew C. and {Fielding}, Drummond B. and {Bryan}, Greg L. and {Kim}, Chang-Goo and {Ostriker}, Eve C. and {Somerville}, Rachel S. and {Stern}, Jonathan and {Su}, Kung-Yi and {Weinberger}, Rainer and {Hu}, Chia-Yu and {Forbes}, John C. and {Hernquist}, Lars and {Burkhart}, Blakesley and {Li}, Yuan},
        title = "{ARKENSTONE - I. A novel method for robustly capturing high specific energy outflows in cosmological simulations}",
      journal = {\mnras},
         year = 2024,
        month = jan,
       volume = {527},
       number = {1},
        pages = {1216-1243},
          doi = {10.1093/mnras/stad3168},
archivePrefix = {arXiv},
       eprint = {2301.07116},
 primaryClass = {astro-ph.GA},
       adsurl = {https://ui.adsabs.harvard.edu/abs/2024MNRAS.527.1216S}
}

@ARTICLE{Smith_2024b,
       author = {{Smith}, Matthew C. and {Fielding}, Drummond B. and {Bryan}, Greg L. and {Bennett}, Jake S. and {Kim}, Chang-Goo and {Ostriker}, Eve C. and {Somerville}, Rachel S.},
        title = "{ARKENSTONE - II. A model for unresolved cool clouds entrained in galactic winds in cosmological simulations}",
      journal = {\mnras},
         year = 2024,
        month = dec,
       volume = {535},
       number = {4},
        pages = {3550-3576},
          doi = {10.1093/mnras/stae2589},
archivePrefix = {arXiv},
       eprint = {2408.15321},
 primaryClass = {astro-ph.GA},
       adsurl = {https://ui.adsabs.harvard.edu/abs/2024MNRAS.535.3550S}
}

@BOOK{Mo_2010,
       author = {{Mo}, Houjun and {van den Bosch}, Frank C. and {White}, Simon},
        title = "{Galaxy Formation and Evolution}",
         year = 2010,
       adsurl = {https://ui.adsabs.harvard.edu/abs/2010gfe..book.....M}
}

@ARTICLE{Pandya_2021,
       author = {{Pandya}, Viraj and {Fielding}, Drummond B. and {Angl{\'e}s-Alc{\'a}zar}, Daniel and {Somerville}, Rachel S. and {Bryan}, Greg L. and {Hayward}, Christopher C. and {Stern}, Jonathan and {Kim}, Chang-Goo and {Quataert}, Eliot and {Forbes}, John C. and {Faucher-Gigu{\`e}re}, Claude-Andr{\'e} and {Feldmann}, Robert and {Hafen}, Zachary and {Hopkins}, Philip F. and {Kere{\v{s}}}, Du{\v{s}}an and {Murray}, Norman and {Wetzel}, Andrew},
        title = "{Characterizing mass, momentum, energy, and metal outflow rates of multiphase galactic winds in the FIRE-2 cosmological simulations}",
      journal = {\mnras},
         year = 2021,
        month = dec,
       volume = {508},
       number = {2},
        pages = {2979-3008},
          doi = {10.1093/mnras/stab2714},
archivePrefix = {arXiv},
       eprint = {2103.06891},
 primaryClass = {astro-ph.GA},
       adsurl = {https://ui.adsabs.harvard.edu/abs/2021MNRAS.508.2979P}
}

@ARTICLE{Ayromlou_2023,
       author = {{Ayromlou}, Mohammadreza and {Nelson}, Dylan and {Pillepich}, Annalisa},
        title = "{Feedback reshapes the baryon distribution within haloes, in halo outskirts, and beyond: the closure radius from dwarfs to massive clusters}",
      journal = {\mnras},
         year = 2023,
        month = oct,
       volume = {524},
       number = {4},
        pages = {5391-5410},
          doi = {10.1093/mnras/stad2046},
archivePrefix = {arXiv},
       eprint = {2211.07659},
 primaryClass = {astro-ph.GA},
       adsurl = {https://ui.adsabs.harvard.edu/abs/2023MNRAS.524.5391A}
}

@ARTICLE{Trident,
       author = {{Hummels}, Cameron B. and {Smith}, Britton D. and {Silvia}, Devin W.},
        title = "{Trident: A Universal Tool for Generating Synthetic Absorption Spectra from Astrophysical Simulations}",
      journal = {\apj},
         year = 2017,
        month = sep,
       volume = {847},
       number = {1},
          eid = {59},
        pages = {59},
          doi = {10.3847/1538-4357/aa7e2d},
archivePrefix = {arXiv},
       eprint = {1612.03935},
 primaryClass = {astro-ph.IM},
       adsurl = {https://ui.adsabs.harvard.edu/abs/2017ApJ...847...59H}
}

@ARTICLE{ROCO,
       author = {{Smith}, Britton and {Sigurdsson}, Steinn and {Abel}, Tom},
        title = "{Metal cooling in simulations of cosmic structure formation}",
      journal = {\mnras},
         year = 2008,
        month = apr,
       volume = {385},
       number = {3},
        pages = {1443-1454},
          doi = {10.1111/j.1365-2966.2008.12922.x},
archivePrefix = {arXiv},
       eprint = {0706.0754},
 primaryClass = {astro-ph},
       adsurl = {https://ui.adsabs.harvard.edu/abs/2008MNRAS.385.1443S}
}

@ARTICLE{CLOUDY,
       author = {{Ferland}, G.~J. and {Chatzikos}, M. and {Guzm{\'a}n}, F. and {Lykins}, M.~L. and {van Hoof}, P.~A.~M. and {Williams}, R.~J.~R. and {Abel}, N.~P. and {Badnell}, N.~R. and {Keenan}, F.~P. and {Porter}, R.~L. and {Stancil}, P.~C.},
        title = "{The 2017 Release Cloudy}",
      journal = {\rmxaa},
         year = 2017,
        month = oct,
       volume = {53},
        pages = {385-438},
          doi = {10.48550/arXiv.1705.10877},
archivePrefix = {arXiv},
       eprint = {1705.10877},
 primaryClass = {astro-ph.GA},
       adsurl = {https://ui.adsabs.harvard.edu/abs/2017RMxAA..53..385F}
}

@ARTICLE{Voit2024,
       author = {{Voit}, G. Mark and {Pandya}, Viraj and {Fielding}, Drummond B. and {Bryan}, Greg L. and {Carr}, Christopher and {Donahue}, Megan and {Oppenheimer}, Benjamin D. and {Somerville}, Rachel S.},
        title = "{Equilibrium States of Galactic Atmospheres. I. The Flip Side of Mass Loading}",
      journal = {\apj},
         year = 2024,
        month = dec,
       volume = {976},
       number = {2},
          eid = {150},
        pages = {150},
          doi = {10.3847/1538-4357/ad81d6},
archivePrefix = {arXiv},
       eprint = {2406.07631},
 primaryClass = {astro-ph.GA},
       adsurl = {https://ui.adsabs.harvard.edu/abs/2024ApJ...976..150V}
}

@ARTICLE{astropy,
       author = {{Astropy Collaboration} and {Robitaille}, Thomas P. and {Tollerud}, Erik J. and {Greenfield}, Perry and {Droettboom}, Michael and {Bray}, Erik and {Aldcroft}, Tom and {Davis}, Matt and {Ginsburg}, Adam and {Price-Whelan}, Adrian M. and {Kerzendorf}, Wolfgang E. and {Conley}, Alexander and {Crighton}, Neil and {Barbary}, Kyle and {Muna}, Demitri and {Ferguson}, Henry and {Grollier}, Fr{\'e}d{\'e}ric and {Parikh}, Madhura M. and {Nair}, Prasanth H. and {Unther}, Hans M. and {Deil}, Christoph and {Woillez}, Julien and {Conseil}, Simon and {Kramer}, Roban and {Turner}, James E.~H. and {Singer}, Leo and {Fox}, Ryan and {Weaver}, Benjamin A. and {Zabalza}, Victor and {Edwards}, Zachary I. and {Azalee Bostroem}, K. and {Burke}, D.~J. and {Casey}, Andrew R. and {Crawford}, Steven M. and {Dencheva}, Nadia and {Ely}, Justin and {Jenness}, Tim and {Labrie}, Kathleen and {Lim}, Pey Lian and {Pierfederici}, Francesco and {Pontzen}, Andrew and {Ptak}, Andy and {Refsdal}, Brian and {Servillat}, Mathieu and {Streicher}, Ole},
        title = "{Astropy: A community Python package for astronomy}",
      journal = {\aap},
         year = 2013,
        month = oct,
       volume = {558},
          eid = {A33},
        pages = {A33},
          doi = {10.1051/0004-6361/201322068},
archivePrefix = {arXiv},
       eprint = {1307.6212},
 primaryClass = {astro-ph.IM},
       adsurl = {https://ui.adsabs.harvard.edu/abs/2013A&A...558A..33A}
}

@ARTICLE{Bordoloi_2018,
       author = {{Bordoloi}, Rongmon and {Prochaska}, J. Xavier and {Tumlinson}, Jason and {Werk}, Jessica K. and {Tripp}, Todd M. and {Burchett}, Joseph N.},
        title = "{On the CGM Fundamental Plane: The Halo Mass Dependency of Circumgalactic H I}",
      journal = {\apj},
         year = 2018,
        month = sep,
       volume = {864},
       number = {2},
          eid = {132},
        pages = {132},
          doi = {10.3847/1538-4357/aad8ac},
archivePrefix = {arXiv},
       eprint = {1712.02348},
 primaryClass = {astro-ph.GA},
       adsurl = {https://ui.adsabs.harvard.edu/abs/2018ApJ...864..132B}
}

@ARTICLE{Voit2024b,
       author = {{Voit}, G. Mark and {Carr}, Christopher and {Fielding}, Drummond B. and {Pandya}, Viraj and {Bryan}, Greg L. and {Donahue}, Megan and {Oppenheimer}, Benjamin D. and {Somerville}, Rachel S.},
        title = "{Equilibrium States of Galactic Atmospheres. II. Interpretation and Implications}",
      journal = {\apj},
         year = 2024,
        month = dec,
       volume = {976},
       number = {2},
          eid = {151},
        pages = {151},
          doi = {10.3847/1538-4357/ad81d5},
archivePrefix = {arXiv},
       eprint = {2406.07632},
 primaryClass = {astro-ph.GA},
       adsurl = {https://ui.adsabs.harvard.edu/abs/2024ApJ...976..151V}
}

@ARTICLE{Behroozi2013,
       author = {{Behroozi}, Peter S. and {Wechsler}, Risa H. and {Wu}, Hao-Yi},
        title = "{The ROCKSTAR Phase-space Temporal Halo Finder and the Velocity Offsets of Cluster Cores}",
      journal = {\apj},
         year = 2013,
        month = jan,
       volume = {762},
       number = {2},
          eid = {109},
        pages = {109},
          doi = {10.1088/0004-637X/762/2/109},
archivePrefix = {arXiv},
       eprint = {1110.4372},
 primaryClass = {astro-ph.CO},
       adsurl = {https://ui.adsabs.harvard.edu/abs/2013ApJ...762..109B}
}

@ARTICLE{Behroozi2013b,
       author = {{Behroozi}, Peter S. and {Wechsler}, Risa H. and {Wu}, Hao-Yi and {Busha}, Michael T. and {Klypin}, Anatoly A. and {Primack}, Joel R.},
        title = "{Gravitationally Consistent Halo Catalogs and Merger Trees for Precision Cosmology}",
      journal = {\apj},
         year = 2013,
        month = jan,
       volume = {763},
       number = {1},
          eid = {18},
        pages = {18},
          doi = {10.1088/0004-637X/763/1/18},
archivePrefix = {arXiv},
       eprint = {1110.4370},
 primaryClass = {astro-ph.CO},
       adsurl = {https://ui.adsabs.harvard.edu/abs/2013ApJ...763...18B}
}

@ARTICLE{Stern2021,
       author = {{Stern}, Jonathan and {Faucher-Gigu{\`e}re}, Claude-Andr{\'e} and {Fielding}, Drummond and {Quataert}, Eliot and {Hafen}, Zachary and {Gurvich}, Alexander B. and {Ma}, Xiangcheng and {Byrne}, Lindsey and {El-Badry}, Kareem and {Angl{\'e}s-Alc{\'a}zar}, Daniel and {Chan}, T.~K. and {Feldmann}, Robert and {Kere{\v{s}}}, Du{\v{s}}an and {Wetzel}, Andrew and {Murray}, Norman and {Hopkins}, Philip F.},
        title = "{Virialization of the Inner CGM in the FIRE Simulations and Implications for Galaxy Disks, Star Formation, and Feedback}",
      journal = {\apj},
         year = 2021,
        month = apr,
       volume = {911},
       number = {2},
          eid = {88},
        pages = {88},
          doi = {10.3847/1538-4357/abd776},
archivePrefix = {arXiv},
       eprint = {2006.13976},
 primaryClass = {astro-ph.GA},
       adsurl = {https://ui.adsabs.harvard.edu/abs/2021ApJ...911...88S}
}

@ARTICLE{Stern2019,
       author = {{Stern}, Jonathan and {Fielding}, Drummond and {Faucher-Gigu{\`e}re}, Claude-Andr{\'e} and {Quataert}, Eliot},
        title = "{Cooling flow solutions for the circumgalactic medium}",
      journal = {\mnras},
         year = 2019,
        month = sep,
       volume = {488},
       number = {2},
        pages = {2549-2572},
          doi = {10.1093/mnras/stz1859},
archivePrefix = {arXiv},
       eprint = {1906.07737},
 primaryClass = {astro-ph.GA},
       adsurl = {https://ui.adsabs.harvard.edu/abs/2019MNRAS.488.2549S}
}

@ARTICLE{Stern2020,
       author = {{Stern}, Jonathan and {Fielding}, Drummond and {Faucher-Gigu{\`e}re}, Claude-Andr{\'e} and {Quataert}, Eliot},
        title = "{The maximum accretion rate of hot gas in dark matter haloes}",
      journal = {\mnras},
         year = 2020,
        month = mar,
       volume = {492},
       number = {4},
        pages = {6042-6058},
          doi = {10.1093/mnras/staa198},
archivePrefix = {arXiv},
       eprint = {1909.07402},
 primaryClass = {astro-ph.GA},
       adsurl = {https://ui.adsabs.harvard.edu/abs/2020MNRAS.492.6042S}
}

@ARTICLE{Pandya2023,
       author = {{Pandya}, Viraj and {Fielding}, Drummond B. and {Bryan}, Greg L. and {Carr}, Christopher and {Somerville}, Rachel S. and {Stern}, Jonathan and {Faucher-Gigu{\`e}re}, Claude-Andr{\'e} and {Hafen}, Zachary and {Angl{\'e}s-Alc{\'a}zar}, Daniel and {Forbes}, John C.},
        title = "{A Unified Model for the Coevolution of Galaxies and Their Circumgalactic Medium: The Relative Roles of Turbulence and Atomic Cooling Physics}",
      journal = {\apj},
         year = 2023,
        month = oct,
       volume = {956},
       number = {2},
          eid = {118},
        pages = {118},
          doi = {10.3847/1538-4357/acf3ea},
archivePrefix = {arXiv},
       eprint = {2211.09755},
 primaryClass = {astro-ph.GA},
       adsurl = {https://ui.adsabs.harvard.edu/abs/2023ApJ...956..118P}
}

@ARTICLE{Fielding_Bryan2022,
       author = {{Fielding}, Drummond B. and {Bryan}, Greg L.},
        title = "{The Structure of Multiphase Galactic Winds}",
      journal = {\apj},
         year = 2022,
        month = jan,
       volume = {924},
       number = {2},
          eid = {82},
        pages = {82},
          doi = {10.3847/1538-4357/ac2f41},
archivePrefix = {arXiv},
       eprint = {2108.05355},
 primaryClass = {astro-ph.GA},
       adsurl = {https://ui.adsabs.harvard.edu/abs/2022ApJ...924...82F}
}

@ARTICLE{Bennett2025,
       author = {{Bennett}, Jake S. and {Smith}, Matthew C. and {Fielding}, Drummond B. and {Bryan}, Greg L. and {Kim}, Chang-Goo and {Springel}, Volker and {Hernquist}, Lars and {Somerville}, Rachel S. and {Sommovigo}, Laura},
        title = "{Prevention is better than cure? Feedback from high specific energy winds in cosmological simulations with ARKENSTONE}",
      journal = {\mnras},
         year = 2025,
        month = oct,
       volume = {543},
       number = {2},
        pages = {1456-1478},
          doi = {10.1093/mnras/staf1440},
archivePrefix = {arXiv},
       eprint = {2410.12909},
 primaryClass = {astro-ph.GA},
       adsurl = {https://ui.adsabs.harvard.edu/abs/2025MNRAS.543.1456B}
}

@ARTICLE{Moser2022,
       author = {{Moser}, Emily and {Battaglia}, Nicholas and {Nagai}, Daisuke and {Lau}, Erwin and {Machado Poletti Valle}, Luis Fernando and {Villaescusa-Navarro}, Francisco and {Amodeo}, Stefania and {Angl{\'e}s-Alc{\'a}zar}, Daniel and {Bryan}, Greg L. and {Dave}, Romeel and {Hernquist}, Lars and {Vogelsberger}, Mark},
        title = "{The Circumgalactic Medium from the CAMELS Simulations: Forecasting Constraints on Feedback Processes from Future Sunyaev-Zeldovich Observations}",
      journal = {\apj},
         year = 2022,
        month = jul,
       volume = {933},
       number = {2},
          eid = {133},
        pages = {133},
          doi = {10.3847/1538-4357/ac70c6},
archivePrefix = {arXiv},
       eprint = {2201.02708},
 primaryClass = {astro-ph.CO},
       adsurl = {https://ui.adsabs.harvard.edu/abs/2022ApJ...933..133M}
}

@ARTICLE{Lau2025,
       author = {{Lau}, Erwin T. and {Nagai}, Daisuke and {Bogd{\'a}n}, {\'A}kos and {Medlock}, Isabel and {Oppenheimer}, Benjamin D. and {Battaglia}, Nicholas and {Angl{\'e}s-Alc{\'a}zar}, Daniel and {Genel}, Shy and {Ni}, Yueying and {Villaescusa-Navarro}, Francisco},
        title = "{X-Raying CAMELS: Constraining Baryonic Feedback in the Circumgalactic Medium with the CAMELS Simulations and eRASS X-Ray Observations}",
      journal = {\apj},
         year = 2025,
        month = may,
       volume = {984},
       number = {2},
          eid = {190},
        pages = {190},
          doi = {10.3847/1538-4357/adc450},
archivePrefix = {arXiv},
       eprint = {2412.04559},
 primaryClass = {astro-ph.GA},
       adsurl = {https://ui.adsabs.harvard.edu/abs/2025ApJ...984..190L}
}

@ARTICLE{Connor2025,
       author = {{Connor}, Liam and {Ravi}, Vikram and {Sharma}, Kritti and {Ocker}, Stella Koch and {Faber}, Jakob and {Hallinan}, Gregg and {Harnach}, Charlie and {Hellbourg}, Greg and {Hobbs}, Rick and {Hodge}, David and {Hodges}, Mark and {Kosogorov}, Nikita and {Lamb}, James and {Law}, Casey and {Rasmussen}, Paul and {Sherman}, Myles and {Somalwar}, Jean and {Weinreb}, Sander and {Woody}, David and {Konietzka}, Ralf M.},
        title = "{A gas-rich cosmic web revealed by the partitioning of the missing baryons}",
      journal = {Nature Astronomy},
         year = 2025,
        month = aug,
       volume = {9},
        pages = {1226-1239},
          doi = {10.1038/s41550-025-02566-y},
archivePrefix = {arXiv},
       eprint = {2409.16952},
 primaryClass = {astro-ph.CO},
       adsurl = {https://ui.adsabs.harvard.edu/abs/2025NatAs...9.1226C}
}

@ARTICLE{Muratov2015,
       author = {{Muratov}, Alexander L. and {Kere{\v{s}}}, Du{\v{s}}an and {Faucher-Gigu{\`e}re}, Claude-Andr{\'e} and {Hopkins}, Philip F. and {Quataert}, Eliot and {Murray}, Norman},
        title = "{Gusty, gaseous flows of FIRE: galactic winds in cosmological simulations with explicit stellar feedback}",
      journal = {\mnras},
         year = 2015,
        month = dec,
       volume = {454},
       number = {3},
        pages = {2691-2713},
          doi = {10.1093/mnras/stv2126},
archivePrefix = {arXiv},
       eprint = {1501.03155},
 primaryClass = {astro-ph.GA},
       adsurl = {https://ui.adsabs.harvard.edu/abs/2015MNRAS.454.2691M}
}

@ARTICLE{Angles-Alcazar2017,
       author = {{Angl{\'e}s-Alc{\'a}zar}, Daniel and {Faucher-Gigu{\`e}re}, Claude-Andr{\'e} and {Kere{\v{s}}}, Du{\v{s}}an and {Hopkins}, Philip F. and {Quataert}, Eliot and {Murray}, Norman},
        title = "{The cosmic baryon cycle and galaxy mass assembly in the FIRE simulations}",
      journal = {\mnras},
         year = 2017,
        month = oct,
       volume = {470},
       number = {4},
        pages = {4698-4719},
          doi = {10.1093/mnras/stx1517},
archivePrefix = {arXiv},
       eprint = {1610.08523},
 primaryClass = {astro-ph.GA},
       adsurl = {https://ui.adsabs.harvard.edu/abs/2017MNRAS.470.4698A}
}

@ARTICLE{Mitchell2020b,
       author = {{Mitchell}, Peter D. and {Schaye}, Joop and {Bower}, Richard G.},
        title = "{Galactic inflow and wind recycling rates in the EAGLE simulations}",
      journal = {\mnras},
         year = 2020,
        month = oct,
       volume = {497},
       number = {4},
        pages = {4495-4516},
          doi = {10.1093/mnras/staa2252},
archivePrefix = {arXiv},
       eprint = {2005.10262},
 primaryClass = {astro-ph.GA},
       adsurl = {https://ui.adsabs.harvard.edu/abs/2020MNRAS.497.4495M}
}

@ARTICLE{Mitchell2022,
       author = {{Mitchell}, Peter D. and {Schaye}, Joop},
        title = "{Baryonic mass budgets for haloes in the EAGLE simulation, including ejected and prevented gas}",
      journal = {\mnras},
         year = 2022,
        month = apr,
       volume = {511},
       number = {2},
        pages = {2600-2609},
          doi = {10.1093/mnras/stab3686},
archivePrefix = {arXiv},
       eprint = {2112.08244},
 primaryClass = {astro-ph.GA},
       adsurl = {https://ui.adsabs.harvard.edu/abs/2022MNRAS.511.2600M}
}

@ARTICLE{Oppenheimer2020,
       author = {{Oppenheimer}, Benjamin D. and {Davies}, Jonathan J. and {Crain}, Robert A. and {Wijers}, Nastasha A. and {Schaye}, Joop and {Werk}, Jessica K. and {Burchett}, Joseph N. and {Trayford}, James W. and {Horton}, Ryan},
        title = "{Feedback from supermassive black holes transforms centrals into passive galaxies by ejecting circumgalactic gas}",
      journal = {\mnras},
         year = 2020,
        month = jan,
       volume = {491},
       number = {2},
        pages = {2939-2952},
          doi = {10.1093/mnras/stz3124},
archivePrefix = {arXiv},
       eprint = {1904.05904},
 primaryClass = {astro-ph.GA},
       adsurl = {https://ui.adsabs.harvard.edu/abs/2020MNRAS.491.2939O}
}

@ARTICLE{ElBadry2018,
       author = {{El-Badry}, Kareem and {Quataert}, Eliot and {Wetzel}, Andrew and {Hopkins}, Philip F. and {Weisz}, Daniel R. and {Chan}, T.~K. and {Fitts}, Alex and {Boylan-Kolchin}, Michael and {Kere{\v{s}}}, Du{\v{s}}an and {Faucher-Gigu{\`e}re}, Claude-Andr{\'e} and {Garrison-Kimmel}, Shea},
        title = "{Gas kinematics, morphology and angular momentum in the FIRE simulations}",
      journal = {\mnras},
         year = 2018,
        month = jan,
       volume = {473},
       number = {2},
        pages = {1930-1955},
          doi = {10.1093/mnras/stx2482},
archivePrefix = {arXiv},
       eprint = {1705.10321},
 primaryClass = {astro-ph.GA},
       adsurl = {https://ui.adsabs.harvard.edu/abs/2018MNRAS.473.1930E}
}

@ARTICLE{ElBadry2018b,
       author = {{El-Badry}, Kareem and {Bradford}, Jeremy and {Quataert}, Eliot and {Geha}, Marla and {Boylan-Kolchin}, Michael and {Weisz}, Daniel R. and {Wetzel}, Andrew and {Hopkins}, Philip F. and {Chan}, T.~K. and {Fitts}, Alex and {Kere{\v{s}}}, Du{\v{s}}an and {Faucher-Gigu{\`e}re}, Claude-Andr{\'e}},
        title = "{Gas kinematics in FIRE simulated galaxies compared to spatially unresolved H I observations}",
      journal = {\mnras},
         year = 2018,
        month = jun,
       volume = {477},
       number = {2},
        pages = {1536-1548},
          doi = {10.1093/mnras/sty730},
archivePrefix = {arXiv},
       eprint = {1801.03933},
 primaryClass = {astro-ph.GA},
       adsurl = {https://ui.adsabs.harvard.edu/abs/2018MNRAS.477.1536E}
}

@ARTICLE{Emami2019,
       author = {{Emami}, Najmeh and {Siana}, Brian and {Weisz}, Daniel R. and {Johnson}, Benjamin D. and {Ma}, Xiangcheng and {El-Badry}, Kareem},
        title = "{A Closer Look at Bursty Star Formation with L $_{H{\ensuremath{\alpha}} }$ and L $_{UV}$ Distributions}",
      journal = {\apj},
         year = 2019,
        month = aug,
       volume = {881},
       number = {1},
          eid = {71},
        pages = {71},
          doi = {10.3847/1538-4357/ab211a},
archivePrefix = {arXiv},
       eprint = {1809.06380},
 primaryClass = {astro-ph.GA},
       adsurl = {https://ui.adsabs.harvard.edu/abs/2019ApJ...881...71E}
}

@ARTICLE{Sparre2017,
       author = {{Sparre}, Martin and {Hayward}, Christopher C. and {Feldmann}, Robert and {Faucher-Gigu{\`e}re}, Claude-Andr{\'e} and {Muratov}, Alexander L. and {Kere{\v{s}}}, Du{\v{s}}an and {Hopkins}, Philip F.},
        title = "{(Star)bursts of FIRE: observational signatures of bursty star formation in galaxies}",
      journal = {\mnras},
         year = 2017,
        month = apr,
       volume = {466},
       number = {1},
        pages = {88-104},
          doi = {10.1093/mnras/stw3011},
archivePrefix = {arXiv},
       eprint = {1510.03869},
 primaryClass = {astro-ph.GA},
       adsurl = {https://ui.adsabs.harvard.edu/abs/2017MNRAS.466...88S}
}

@ARTICLE{Hafen2019,
       author = {{Hafen}, Zachary and {Faucher-Gigu{\`e}re}, Claude-Andr{\'e} and {Angl{\'e}s-Alc{\'a}zar}, Daniel and {Stern}, Jonathan and {Kere{\v{s}}}, Du{\v{s}}an and {Hummels}, Cameron and {Esmerian}, Clarke and {Garrison-Kimmel}, Shea and {El-Badry}, Kareem and {Wetzel}, Andrew and {Chan}, T.~K. and {Hopkins}, Philip F. and {Murray}, Norman},
        title = "{The origins of the circumgalactic medium in the FIRE simulations}",
      journal = {\mnras},
         year = 2019,
        month = sep,
       volume = {488},
       number = {1},
        pages = {1248-1272},
          doi = {10.1093/mnras/stz1773},
archivePrefix = {arXiv},
       eprint = {1811.11753},
 primaryClass = {astro-ph.GA},
       adsurl = {https://ui.adsabs.harvard.edu/abs/2019MNRAS.488.1248H}
}

@article{Loken2002,
  author  = {Loken, Chris and Norman, Michael L. and Nelson, Erik and
             Burns, Jack and Bryan, Greg L. and Motl, Patrick},
  title   = {A Universal Temperature Profile for Galaxy Clusters},
  journal = {The Astrophysical Journal},
  year    = {2002},
  volume  = {579},
  pages   = {571--576},
  doi     = {10.1086/342825},
  eprint  = {astro-ph/0207095}
}

@article{Vikhlinin2005,
  author  = {Vikhlinin, A. and Markevitch, M. and Murray, S.~S. and
             Jones, C. and Forman, W. and {Van Speybroeck}, L.},
  title   = {Chandra Temperature Profiles for a Sample of Nearby
             Relaxed Galaxy Clusters},
  journal = {The Astrophysical Journal},
  year    = {2005},
  volume  = {628},
  pages   = {655--672},
  doi     = {10.1086/431142},
  eprint  = {astro-ph/0412306}
}

@article{Vikhlinin2006,
  author  = {Vikhlinin, A. and Kravtsov, A. and Forman, W. and
             Jones, C. and Markevitch, M. and Murray, S.~S. and
             {Van Speybroeck}, L.},
  title   = {Chandra Sample of Nearby Relaxed Galaxy Clusters:
             Mass, Gas Fraction, and Mass--Temperature Relation},
  journal = {The Astrophysical Journal},
  year    = {2006},
  volume  = {640},
  pages   = {691--709},
  doi     = {10.1086/500288},
  eprint  = {astro-ph/0507092}
}

@article{Pratt2007,
  author  = {Pratt, G.~W. and B{\"o}hringer, H. and Croston, J.~H. and
             Arnaud, M. and Borgani, S. and Finoguenov, A. and Temple, R.~F.},
  title   = {Temperature Profiles of a Representative Sample of Nearby
             {X}-ray Galaxy Clusters},
  journal = {Astronomy \& Astrophysics},
  year    = {2007},
  volume  = {461},
  pages   = {71--80},
  doi     = {10.1051/0004-6361:20065676},
  eprint  = {astro-ph/0609480}
}

@ARTICLE{Bryan_Norman1998,
       author = {{Bryan}, Greg L. and {Norman}, Michael L.},
        title = "{Statistical Properties of X-Ray Clusters: Analytic and Numerical Comparisons}",
      journal = {\apj},
         year = 1998,
        month = mar,
       volume = {495},
       number = {1},
        pages = {80-99},
          doi = {10.1086/305262},
archivePrefix = {arXiv},
       eprint = {astro-ph/9710107},
 primaryClass = {astro-ph},
       adsurl = {https://ui.adsabs.harvard.edu/abs/1998ApJ...495...80B}
}

@ARTICLE{Correa2018,
       author = {{Correa}, Camila A. and {Schaye}, Joop and {Wyithe}, J. Stuart B. and {Duffy}, Alan R. and {Theuns}, Tom and {Crain}, Robert A. and {Bower}, Richard G.},
        title = "{The formation of hot gaseous haloes around galaxies}",
      journal = {\mnras},
         year = 2018,
        month = jan,
       volume = {473},
       number = {1},
        pages = {538-559},
          doi = {10.1093/mnras/stx2332},
archivePrefix = {arXiv},
       eprint = {1709.01938},
 primaryClass = {astro-ph.GA},
       adsurl = {https://ui.adsabs.harvard.edu/abs/2018MNRAS.473..538C}
}

@ARTICLE{Pandya2026,
       author = {{Pandya}, Viraj and {Bryan}, Greg L. and {Makinen}, T. Lucas and {Gabrielpillai}, Austen and {Carr}, Christopher and {Fielding}, Drummond B. and {Hernquist}, Lars and {Ho}, Matthew and {Iyer}, Kartheik and {Jespersen}, Christian Kragh and {Koudmani}, Sophie and {Laska}, Marta and {Lemos}, Pablo and {Lovell}, Christopher C. and {Perez}, Lucia A. and {Robinson}, Jr., William F. and {Somerville}, Rachel S. and {Starkenburg}, Tjitske K. and {Stiskalek}, Richard and {Terrazas}, Bryan and {Voit}, G. Mark},
        title = "{Introducing sapphire: Towards Hybrid Physics-Informed, Data-Driven Modeling of Galaxy Formation}",
      journal = {arXiv e-prints},
         year = 2026,
        month = apr,
          eid = {arXiv:2604.06318},
        pages = {arXiv:2604.06318},
          doi = {10.48550/arXiv.2604.06318},
archivePrefix = {arXiv},
       eprint = {2604.06318},
 primaryClass = {astro-ph.GA},
       adsurl = {https://ui.adsabs.harvard.edu/abs/2026arXiv260406318P}
}

@ARTICLE{Piacitelli2025,
       author = {{Piacitelli}, Daniel R. and {Brooks}, Alyson M. and {Christensen}, Charlotte and {Sanchez}, N. Nicole and {Faerman}, Yakov and {Shen}, Sijing and {Cruz}, Akaxia and {Keller}, Ben and {Quinn}, Thomas R. and {Wadsley}, James},
        title = "{Marvelous Metals: Surveying the Circumgalactic Medium of Simulated Dwarf Galaxies}",
      journal = {\apj},
         year = 2025,
        month = nov,
       volume = {993},
       number = {2},
          eid = {230},
        pages = {230},
          doi = {10.3847/1538-4357/ae06a0},
archivePrefix = {arXiv},
       eprint = {2505.08861},
 primaryClass = {astro-ph.GA},
       adsurl = {https://ui.adsabs.harvard.edu/abs/2025ApJ...993..230P}
}

@ARTICLE{Sanchez2019,
       author = {{Sanchez}, N. Nicole and {Werk}, Jessica K. and {Tremmel}, Michael and {Pontzen}, Andrew and {Christensen}, Charlotte and {Quinn}, Thomas and {Cruz}, Akaxia},
        title = "{Not So Heavy Metals: Black Hole Feedback Enriches the Circumgalactic Medium}",
      journal = {\apj},
         year = 2019,
        month = sep,
       volume = {882},
       number = {1},
          eid = {8},
        pages = {8},
          doi = {10.3847/1538-4357/ab3045},
archivePrefix = {arXiv},
       eprint = {1810.12319},
 primaryClass = {astro-ph.GA},
       adsurl = {https://ui.adsabs.harvard.edu/abs/2019ApJ...882....8S}
}

@ARTICLE{Nelson2018,
       author = {{Nelson}, Dylan and {Kauffmann}, Guinevere and {Pillepich}, Annalisa and {Genel}, Shy and {Springel}, Volker and {Pakmor}, R{\"u}diger and {Hernquist}, Lars and {Weinberger}, Rainer and {Torrey}, Paul and {Vogelsberger}, Mark and {Marinacci}, Federico},
        title = "{The abundance, distribution, and physical nature of highly ionized oxygen O VI, O VII, and O VIII in IllustrisTNG}",
      journal = {\mnras},
         year = 2018,
        month = jun,
       volume = {477},
       number = {1},
        pages = {450-479},
          doi = {10.1093/mnras/sty656},
archivePrefix = {arXiv},
       eprint = {1712.00016},
 primaryClass = {astro-ph.GA},
       adsurl = {https://ui.adsabs.harvard.edu/abs/2018MNRAS.477..450N}
}

@ARTICLE{Wise2012,
       author = {{Wise}, John H. and {Abel}, Tom and {Turk}, Matthew J. and {Norman}, Michael L. and {Smith}, Britton D.},
        title = "{The birth of a galaxy - II. The role of radiation pressure}",
      journal = {\mnras},
         year = 2012,
        month = nov,
       volume = {427},
       number = {1},
        pages = {311-326},
          doi = {10.1111/j.1365-2966.2012.21809.x},
archivePrefix = {arXiv},
       eprint = {1206.1043},
 primaryClass = {astro-ph.CO},
       adsurl = {https://ui.adsabs.harvard.edu/abs/2012MNRAS.427..311W}
}

@ARTICLE{Stinson2013,
       author = {{Stinson}, G.~S. and {Brook}, C. and {Macci{\`o}}, A.~V. and {Wadsley}, J. and {Quinn}, T.~R. and {Couchman}, H.~M.~P.},
        title = "{Making Galaxies In a Cosmological Context: the need for early stellar feedback}",
      journal = {\mnras},
         year = 2013,
        month = jan,
       volume = {428},
       number = {1},
        pages = {129-140},
          doi = {10.1093/mnras/sts028},
archivePrefix = {arXiv},
       eprint = {1208.0002},
 primaryClass = {astro-ph.CO},
       adsurl = {https://ui.adsabs.harvard.edu/abs/2013MNRAS.428..129S}
}

@ARTICLE{Smith2021,
       author = {{Smith}, Matthew C. and {Bryan}, Greg L. and {Somerville}, Rachel S. and {Hu}, Chia-Yu and {Teyssier}, Romain and {Burkhart}, Blakesley and {Hernquist}, Lars},
        title = "{Efficient early stellar feedback can suppress galactic outflows by reducing supernova clustering}",
      journal = {\mnras},
         year = 2021,
        month = sep,
       volume = {506},
       number = {3},
        pages = {3882-3915},
          doi = {10.1093/mnras/stab1896},
archivePrefix = {arXiv},
       eprint = {2009.11309},
 primaryClass = {astro-ph.GA},
       adsurl = {https://ui.adsabs.harvard.edu/abs/2021MNRAS.506.3882S}
}

@ARTICLE{Porter2024,
       author = {{Porter}, Lori E. and {Orr}, Matthew E. and {Burkhart}, Blakesley and {Wetzel}, Andrew and {Kere{\v{s}}}, Du{\v{s}}an and {Faucher-Gigu{\`e}re}, Claude-Andr{\'e} and {Hopkins}, Philip F.},
        title = "{Any way the wind blows: quantifying superbubbles and their outflows in simulated galaxies across z ≍ 0-3}",
      journal = {\mnras},
         year = 2024,
        month = dec,
       volume = {535},
       number = {4},
        pages = {3451-3469},
          doi = {10.1093/mnras/stae2576},
archivePrefix = {arXiv},
       eprint = {2406.03535},
 primaryClass = {astro-ph.GA},
       adsurl = {https://ui.adsabs.harvard.edu/abs/2024MNRAS.535.3451P}
}

@ARTICLE{Oren2026,
       author = {{Oren}, Yossi and {Pandya}, Viraj and {Somerville}, Rachel S. and {Genel}, Shy and {Omoruyi}, Osase and {Sternberg}, Amiel},
        title = "{The Cosmic Baryon Cycle in IllustrisTNG: Flows of Mass, Energy, and Metals}",
      journal = {\apj},
         year = 2026,
        month = mar,
       volume = {999},
       number = {2},
          eid = {259},
        pages = {259},
          doi = {10.3847/1538-4357/ae41bc},
archivePrefix = {arXiv},
       eprint = {2510.23343},
 primaryClass = {astro-ph.GA},
       adsurl = {https://ui.adsabs.harvard.edu/abs/2026ApJ...999..259O}
}

@ARTICLE{Sullivan2026,
       author = {{Sullivan}, James M. and {Bryan}, Greg L. and {Smith}, Matthew C. and {Bennett}, Jake S. and {Fielding}, Drummond B. and {Terrazas}, Bryan A. and {Koudmani}, Sophie and {Somerville}, Rachel S. and {Hirschmann}, Michaela},
        title = "{ArkenstoneBH. A model for high-specific energy black hole feedback in cosmological simulations}",
      journal = {arXiv e-prints},
         year = 2026,
        month = may,
          eid = {arXiv:2605.03154},
        pages = {arXiv:2605.03154},
archivePrefix = {arXiv},
       eprint = {2605.03154},
 primaryClass = {astro-ph.GA},
       adsurl = {https://ui.adsabs.harvard.edu/abs/2026arXiv260503154S}
}

@ARTICLE{Arjona2024,
       author = {{Arjona-G{\'a}lvez}, Elena and {Di Cintio}, Arianna and {Grand}, Robert J.~J.},
        title = "{The role of active galactic nucleus feedback on the evolution of dwarf galaxies from cosmological simulations: Supermassive black holes suppress star formation in low-mass galaxies}",
      journal = {\aap},
         year = 2024,
        month = oct,
       volume = {690},
          eid = {A286},
        pages = {A286},
          doi = {10.1051/0004-6361/202449439},
archivePrefix = {arXiv},
       eprint = {2402.00929},
 primaryClass = {astro-ph.GA},
       adsurl = {https://ui.adsabs.harvard.edu/abs/2024A&A...690A.286A}
}

@ARTICLE{Baumschlager2025,
       author = {{Baumschlager}, Bernhard and {Shen}, Sijing and {Wadsley}, James W. and {Keller}, Benjamin and {Wissing}, Robert and {Mayer}, Lucio and {Madau}, Piero and {Munshi}, Ferah and {Brooks}, Alyson},
        title = "{The Seven Dwarfs illuminated. The impact of radiation on dwarf galaxies and their circumgalactic medium}",
      journal = {arXiv e-prints},
         year = 2025,
        month = aug,
          eid = {arXiv:2508.19396},
        pages = {arXiv:2508.19396},
          doi = {10.48550/arXiv.2508.19396},
archivePrefix = {arXiv},
       eprint = {2508.19396},
 primaryClass = {astro-ph.GA},
       adsurl = {https://ui.adsabs.harvard.edu/abs/2025arXiv250819396B}
}

@ARTICLE{Mina2021,
       author = {{Mina}, Mattia and {Shen}, Sijing and {Keller}, Benjamin Walter and {Mayer}, Lucio and {Madau}, Piero and {Wadsley}, James},
        title = "{The baryon cycle of Seven Dwarfs with superbubble feedback}",
      journal = {\aap},
         year = 2021,
        month = nov,
       volume = {655},
          eid = {A22},
        pages = {A22},
          doi = {10.1051/0004-6361/202039420},
archivePrefix = {arXiv},
       eprint = {2009.06646},
 primaryClass = {astro-ph.GA},
       adsurl = {https://ui.adsabs.harvard.edu/abs/2021A&A...655A..22M}
}

@ARTICLE{Shen2014,
       author = {{Shen}, Sijing and {Madau}, Piero and {Conroy}, Charlie and {Governato}, Fabio and {Mayer}, Lucio},
        title = "{The Baryon Cycle of Dwarf Galaxies: Dark, Bursty, Gas-rich Polluters}",
      journal = {\apj},
         year = 2014,
        month = sep,
       volume = {792},
       number = {2},
          eid = {99},
        pages = {99},
          doi = {10.1088/0004-637X/792/2/99},
archivePrefix = {arXiv},
       eprint = {1308.4131},
 primaryClass = {astro-ph.CO},
       adsurl = {https://ui.adsabs.harvard.edu/abs/2014ApJ...792...99S}
}

@ARTICLE{Qu_Bregman2022,
       author = {{Qu}, Zhijie and {Bregman}, Joel N.},
        title = "{Absorption Line Search through Three Local Group Dwarf Galaxy Halos}",
      journal = {\apj},
         year = 2022,
        month = mar,
       volume = {927},
       number = {2},
          eid = {228},
        pages = {228},
          doi = {10.3847/1538-4357/ac51df},
archivePrefix = {arXiv},
       eprint = {2203.08246},
 primaryClass = {astro-ph.GA},
       adsurl = {https://ui.adsabs.harvard.edu/abs/2022ApJ...927..228Q}
}

@ARTICLE{Liang_Chen2014,
       author = {{Liang}, Cameron J. and {Chen}, Hsiao-Wen},
        title = "{Mining circumgalactic baryons in the low-redshift universe}",
      journal = {\mnras},
         year = 2014,
        month = dec,
       volume = {445},
       number = {2},
        pages = {2061-2081},
          doi = {10.1093/mnras/stu1901},
archivePrefix = {arXiv},
       eprint = {1402.3602},
 primaryClass = {astro-ph.CO},
       adsurl = {https://ui.adsabs.harvard.edu/abs/2014MNRAS.445.2061L}
}

@ARTICLE{Munshi2021,
       author = {{Munshi}, Ferah and {Brooks}, Alyson M. and {Applebaum}, Elaad and {Christensen}, Charlotte R. and {Quinn}, T. and {Sligh}, Serena},
        title = "{Quantifying Scatter in Galaxy Formation at the Lowest Masses}",
      journal = {\apj},
         year = 2021,
        month = dec,
       volume = {923},
       number = {1},
          eid = {35},
        pages = {35},
          doi = {10.3847/1538-4357/ac0db6},
archivePrefix = {arXiv},
       eprint = {2101.05822},
 primaryClass = {astro-ph.GA},
       adsurl = {https://ui.adsabs.harvard.edu/abs/2021ApJ...923...35M}
}

@ARTICLE{Tremmel2017,
       author = {{Tremmel}, M. and {Karcher}, M. and {Governato}, F. and {Volonteri}, M. and {Quinn}, T.~R. and {Pontzen}, A. and {Anderson}, L. and {Bellovary}, J.},
        title = "{The Romulus cosmological simulations: a physical approach to the formation, dynamics and accretion models of SMBHs}",
      journal = {\mnras},
         year = 2017,
        month = sep,
       volume = {470},
       number = {1},
        pages = {1121-1139},
          doi = {10.1093/mnras/stx1160},
archivePrefix = {arXiv},
       eprint = {1607.02151},
 primaryClass = {astro-ph.GA},
       adsurl = {https://ui.adsabs.harvard.edu/abs/2017MNRAS.470.1121T}
}

@ARTICLE{Cook2025,
       author = {{Cook}, Andrew W.~S. and {van de Voort}, Freeke and {Pakmor}, R{\"u}diger and {Grand}, Robert J.~J.},
        title = "{The halo mass dependence of physical and observable properties in the circumgalactic medium at z = 0}",
      journal = {\mnras},
         year = 2025,
        month = oct,
       volume = {543},
       number = {2},
        pages = {1224-1238},
          doi = {10.1093/mnras/staf1537},
archivePrefix = {arXiv},
       eprint = {2409.05578},
 primaryClass = {astro-ph.GA},
       adsurl = {https://ui.adsabs.harvard.edu/abs/2025MNRAS.543.1224C}
}

@ARTICLE{Tung2025,
       author = {{Tung}, Pei-Cheng and {Chen}, Ke-Jung},
        title = "{Coevolution of Dwarf Galaxies and Their Circumgalactic Medium Across Cosmic Time}",
      journal = {\apj},
         year = 2025,
        month = jul,
       volume = {988},
       number = {1},
          eid = {127},
        pages = {127},
          doi = {10.3847/1538-4357/ade1d4},
archivePrefix = {arXiv},
       eprint = {2412.16440},
 primaryClass = {astro-ph.GA},
       adsurl = {https://ui.adsabs.harvard.edu/abs/2025ApJ...988..127T}
}

@ARTICLE{Applebaum2021,
       author = {{Applebaum}, Elaad and {Brooks}, Alyson M. and {Christensen}, Charlotte R. and {Munshi}, Ferah and {Quinn}, Thomas R. and {Shen}, Sijing and {Tremmel}, Michael},
        title = "{Ultrafaint Dwarfs in a Milky Way Context: Introducing the Mint Condition DC Justice League Simulations}",
      journal = {\apj},
         year = 2021,
        month = jan,
       volume = {906},
       number = {2},
          eid = {96},
        pages = {96},
          doi = {10.3847/1538-4357/abcafa},
archivePrefix = {arXiv},
       eprint = {2008.11207},
 primaryClass = {astro-ph.GA},
       adsurl = {https://ui.adsabs.harvard.edu/abs/2021ApJ...906...96A}
}

@ARTICLE{Tomaru2025,
       author = {{Tomaru}, Kazuki and {Oku}, Yuri and {Toyouchi}, Daisuke and {Nagamine}, Kentaro},
        title = "{CROCODILE-DWARF: Assembly and Kinematics of Field Dwarf Galaxies with GADGET4-OSAKA}",
      journal = {arXiv e-prints},
         year = 2025,
        month = oct,
          eid = {arXiv:2510.26513},
        pages = {arXiv:2510.26513},
          doi = {10.48550/arXiv.2510.26513},
archivePrefix = {arXiv},
       eprint = {2510.26513},
 primaryClass = {astro-ph.GA},
       adsurl = {https://ui.adsabs.harvard.edu/abs/2025arXiv251026513T}
}

@ARTICLE{Grackle2017,
       author = {{Smith}, Britton D. and {Bryan}, Greg L. and {Glover}, Simon C.~O. and {Goldbaum}, Nathan J. and {Turk}, Matthew J. and {Regan}, John and {Wise}, John H. and {Schive}, Hsi-Yu and {Abel}, Tom and {Emerick}, Andrew and {O'Shea}, Brian W. and {Anninos}, Peter and {Hummels}, Cameron B. and {Khochfar}, Sadegh},
        title = "{GRACKLE: a chemistry and cooling library for astrophysics}",
      journal = {\mnras},
         year = 2017,
        month = apr,
       volume = {466},
       number = {2},
        pages = {2217-2234},
          doi = {10.1093/mnras/stw3291},
archivePrefix = {arXiv},
       eprint = {1610.09591},
 primaryClass = {astro-ph.CO},
       adsurl = {https://ui.adsabs.harvard.edu/abs/2017MNRAS.466.2217S}
}

@ARTICLE{Sharma2012,
       author = {{Sharma}, Prateek and {McCourt}, Michael and {Quataert}, Eliot and {Parrish}, Ian J.},
        title = "{Thermal instability and the feedback regulation of hot haloes in clusters, groups and galaxies}",
      journal = {\mnras},
         year = 2012,
        month = mar,
       volume = {420},
       number = {4},
        pages = {3174-3194},
          doi = {10.1111/j.1365-2966.2011.20246.x},
archivePrefix = {arXiv},
       eprint = {1106.4816},
 primaryClass = {astro-ph.CO},
       adsurl = {https://ui.adsabs.harvard.edu/abs/2012MNRAS.420.3174S}
}
\bibliographystyle{aasjournal}



\end{document}